\documentclass[aastex, twocolumn]{aastex63}

\usepackage{mathtools}

\newcommand{\Msolyr}{\mbox{M\raisebox{-.6ex}{$\odot$} yr$^{-1}$}}
\newcommand{\Mdot}{\mbox{$\dot M$}}
\newcommand{\lMdot}{\mbox{$\log_{10}(\Mdot)$}}
\newcommand{\Lpk}{\mbox{$L_{\rm pk}$}}
\newcommand{\tpk}{\mbox{$t_{\rm pk}$}}

\newcommand{\ltpk}{\mbox{$\log_{10}(\tpk)$}}
\newcommand{\lLpk}{\mbox{$\log_{10}(\Lpk)$}}

\newcommand{\Lpkobs}{\mbox{$L_{\rm pk,obs}$}}
\newcommand{\tpkobs}{\mbox{$t_{\rm pk,obs}$}}
\newcommand{\ergsHz}{\mbox{erg~s$^{-1}$ Hz$^{-1}$}}

\newcommand{\kms}{\mbox{km s$^{-1}$}}

\newcommand{\ptot}{\mbox{$p_{\rm tot}$}}

\newcommand{\vwind}{\mbox{$v_{\rm wind}$}}
\newcommand{\vSSA}{\mbox{$v_{\rm SSA}$}}

\newcommand{\NSNe}{294}  % just to make updates easier; this is RSN_master v5.63

\newcommand{\newstuff}[1]{#1}

\begin{document}

\title[Radio Luminosity-Risetime Function of SNe]{The Radio
  Luminosity-Risetime Function of Core-Collapse Supernovae}

\correspondingauthor{M. F. Bietenholz}

\author{M. F. Bietenholz}
\affiliation{Hartebeesthoek Radio Astronomy Observatory, PO Box 443, Krugersdorp, 1740, South Africa}
\affiliation{Department of Physics and Astronomy, York University, Toronto, M3J~1P3, Ontario, Canada}

\author{N. Bartel}
\affiliation{Department of Physics and Astronomy, York University, Toronto, M3J~1P3, Ontario, Canada}

\author{M. Argo}
\affiliation{Jeremiah Horrocks Institute, University of Central Lancashire, Preston PR1 2HE, UK}

\author{R. Dua}
\affiliation{Birla Institute of Technology and Science, Pilani Campus, Pilani, India}

\author{S. Ryder}
\affiliation{Dept.\ of Physics \& Astronomy, Macquarie University, NSW 2109, Australia}
\affiliation{Astronomy, Astrophysics and Astrophotonics Research Centre, Macquarie University, Sydney, NSW 2109, Australia}

\author{A. Soderberg}
\affiliation{formerly at Harvard-Smithsonian Center for Astrophysics, 60 Garden
Street, Cambridge, MA 02138, USA}

%\revised{Version 7.5, \today}
\accepted{for publication in the {\em Atrophysical Journal}, \today}

\begin{abstract}
  We assemble a large set of 2--10 GHz radio flux density measurements
  and upper limits of \NSNe\ different supernovae (SNe), from the
  literature and our own and archival data.  Only 31\% of SNe were
  detected.  We characterize the SN radio lightcurves near the peak using a
  two-parameter model, with \tpk\ being the time to rise to a peak and
  \Lpk\ the spectral luminosity at that peak.  Over all SNe in our
  sample at $D<100$~Mpc, we find that $\tpk = 10^{1.7\pm0.9}$~d, and
  that $\Lpk = 10^{25.5\pm1.6}$~\ergsHz, and therefore that generally,
  50\% of SNe will have $\Lpk < 10^{25.5}$~\ergsHz.  These \Lpk\
  values are $\sim$30 times lower than those for only detected SNe.
  Types I b/c and II (excluding IIn's) have similar mean values of
  \Lpk\ but the former have a wider range, whereas Type IIn SNe have
  $\sim 10$ times higher values with
  \Lpk=$10^{26.5\pm1.1}$~\ergsHz\@. As for \tpk, Type I b/c have \tpk\
  of only $10^{1.1\pm0.5}$~d while Type II have \tpk=$10^{1.6\pm1.0}$
  and Type IIn the longest timescales with \tpk $ = 10^{3.1\pm0.7}$~d.
  We also estimate the distribution of progenitor mass-loss rates,
  \Mdot, and find the mean and standard deviation of
  $\log_{10}(\Mdot/[\Msolyr])$ are $-5.4\pm1.2$ (assuming
  \vwind=1000~\kms) for Type I~b/c SNe, and $-6.9\pm1.4$ (assuming
  \vwind = 10~\kms) for Type II SNe excluding Type IIn.
\end{abstract}


\section{Introduction}
\label{sintro}

Core collapse supernova (SNe) can produce bright radio emission.  The
chief source of this emission is the interaction of the rapidly
expanding ejecta with the circumstellar medium (CSM), which usually
consists of the stellar wind of the SN progenitor, but may also have a
significant contribution from mass-stripping in binary systems. Shocks
are formed in this interaction, which serve to accelerate particles to
relativistic velocities and amplify the magnetic field, resulting in
synchrotron radio emission.

The radio emission provides us with a probe of the CSM, as well as for
the outer, highest-velocity portion of the SN ejecta, for which few
other observational probes are available.  SNe are much less luminous
in the radio than in the optical, with typical radio luminosities
$<10^{-4}$ of those in the optical.
Compared to the thousands of SNe detected in the optical, only
$\sim$100 SNe have been detected in the radio.  Furthermore, only
core-collapse SNe have been detected to date, and as yet no Type Ia SN
\citep[for recent limits on the radio emission of Type Ia SNe,
see][]{Lundqvist+2020}.  In this paper, therefore, we consider only
core-collapse SNe, that is SN of Types Ib, Ic and II, and whenever we
use the term ``SN'' we are referring only to ones of the core-collapse
variety.

\begin{figure}
\centering
\includegraphics[width=\linewidth]{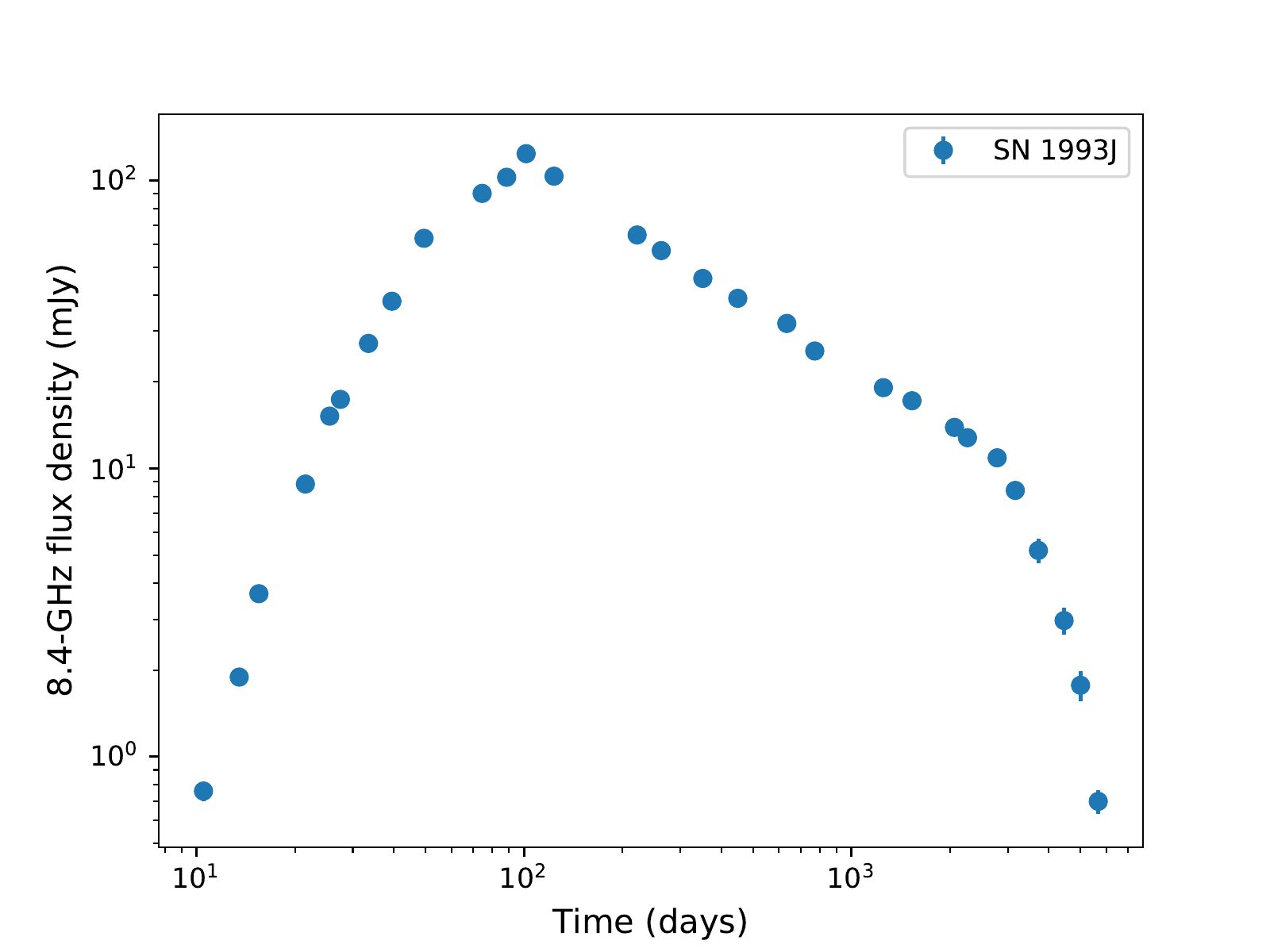}
\caption{An example radio lightcurve of a core-collapse SN\@.  We show
  the lightcurve of the Type~IIb SN~1993J (data from
  \citealt{SN93J-2}; see also \citealt{Marti-Vidal+2011a}), which is
  one of the most intensively observed radio SNe.  We plot the flux
  density, $S$, at 8.4~GHz against the time, $t$, since the explosion,
  with both axes being logarithmic.  The errorbars are mostly smaller
  than the plotted points.  The typical features are a rapid rise to a
  peak, which occurs at $\tpk \simeq 130$~d with
  $S_{\rm pk} = 123$~mJy, followed by a decline.  In the case of
  SN~1993J the decline is of an approximately power-law form
  ($S \propto t^{-\beta}$), until $t \simeq 2500$~d, at which time
  there is a distinct steepening of the slope of the logarithmic
  decline.  For our purposes here, we concentrate only on the region
  of the lightcurve near the peak and do not try to fit features such
  as the change in slope in the decay.}
\label{fsn93j}
\end{figure}

The radio emission from SNe is synchrotron emission.  It generally
displays a high brightness temperature, and a non-thermal spectrum.
Their radio lightcurves follow a general pattern with a rise to a
maximum, which can occur days to years after the SN explosion.  The
peak is followed by a decay, often of an approximately power-law form,
with $S_\nu \propto t^\beta$, where $S_\nu$ is the flux density at
frequency $\nu$, $t$ the time since the explosion, and $\beta$ is
usually in the range of $-1$ to $-3$.  As an illustration, we show the
8.4~GHz lightcurve of SN~1993J in Figure~\ref{fsn93j} \citep[data
from][and our own unpublished measurements]{SN93J-2}.  SN~1993J shows
the typical rise and then power-law decay, although in this case, there
is a distinct change in the slope of the decay after about 7 yr.

The radio lightcurves of SNe vary over a large range.  Although the
brightest SNe reach peak spectral luminosities, \Lpk\footnote{More
  formally, \Lpk\ should be denoted $L_{\nu,{\rm pk}}$ since the peak
  spectral luminosity will depend on the observing frequency, $\nu$\@.
  We omit the $\nu$ subscript on   $L_{\nu,{\rm pk}}$ and
   $L_{\nu,{\rm pk, obs}}$ for clarity.
  We expect in any case that the dependence on $\nu$ will not be large,
  since we restrict ourselves to frequencies, $4 < \nu  < 10$ GHz, with the
  exception of SN 1987A.  Indeed, \citet{Weiler+2002} found that the
  dependence of the \Lpk\ on $\nu$ was not large for a variety of SNe of
  Type Ib/c and II.}
$= 10^{29}$~\ergsHz\ (e.g., SN~1998bw, SN~2009bb), a considerable
fraction of even nearby SNe are never detected in the radio, and must
have \Lpk\ values at least 3 orders of magnitude lower, or
$< 10^{26}$~\ergsHz.  Indeed, the \Lpk\ of SN~1987A was another 2
orders of magnitude lower at $\lesssim 10^{24}$~\ergsHz.  Similarly,
the risetimes \tpk, have a very wide range.  Some SNe, such as
SN~1987A, have a very short $\tpk \simeq 1$~d, while others, such as
SN~1986J, can take several years to reach their peak.

A considerable number of radio flux density measurements of individual
SNe have been published over the years.  Much effort has also gone
into parameterizing and modeling the radio lightcurves for the subset
of SNe for which densely-sampled lightcurves are available \citep[see
e.g.,][]{Weiler+2002}.  However, there has been rather less
examination of the population as a whole.  In this paper, we will
explore, in a largely empirical way, the radio luminosity function of
supernovae, and attempt first to answer the questions: how bright do we
expect a core-collapse supernova to get in the radio, and how long
do we expect it to take to reach this peak?

Our approach is as follows: we will adopt a simple parameterization of
a supernova radio lightcurve, with only two parameters: \tpk, the time
between the explosion and \Lpk, the peak spectral luminosity at that
time.  The challenge is to find the values of \tpk\ and \Lpk\@. For a
lightcurve with many flux density measurements as depicted in
Figure~\ref{fsn93j} this can be done straightforwardly and relatively
unambiguously. However, if there is only a single flux density
measurement available, then the determination of \tpk\ and \Lpk\ is
ambiguous but as we will quantify later, the range of possible values
of \tpk\ and \Lpk\ is still well constrained.  In the case of only
upper limits on the flux density, the values of \tpk\ and \Lpk\ are
also ambiguous, but nonetheless still constrained, although generally
less so than in the case of a single measurement.  In this paper we
use all our measurements to derive statistically meaningful results.

Many SNe in fact show behavior more complex than assumed in our
simple model, with modulated lightcurves and anomalous rises at late
times (for example, SN~1993J, already shown in Figure~\ref{fsn93j}; but
also SN~1979C, \citealt{SN79C-shell}; SN~1986J, \citealt{SN86J-1};
SN~1987A, \citealt{Zanardo+2010, Cendes+2019}; SN~2001em
\citealt{SN2001em-1}; and SN~2001ig \citealt{Ryder+2004}). However, most
SNe do show an initial rise to a peak brightness and then a subsequent
decay, so our model should suffice for giving us some insight into the
population as a whole.  For those SNe, such as SN~1987A, which showed
a late-time rise in the radio emission, we use only the measurements
for the first rise and subsequent decay.

\newstuff{We divide SNe into different Types such as Types I b/c or II
  and determine the difference in the luminosity-risetime function for
  different SN Types.  We use the following three main
  classifications: Type I b/c, Type IIn, and then the remainder of the
  Type II's.  In what follows, when we mention Type II, we
  always mean Type II excluding the Type IIn.  In addition, we examine
  separately the subset of Type I b/c SNe which has broad optical
  lines, which we call ``BL'', and the Type IIb subset of Type II's.

  Type IIn SNe are those with narrow optical lines.  They constitute
  $\sim$12\% of all Type II SNe \citep{Smith+2011a}.  Examples are
  SN~1986J and SN~1998S\@.  These SNe are thought to be due to
  interaction with a dense CSM, which produces the narrow lines, and
  often strong radio emission. The radio evolution of Type IIn SNe is
  quite different from that of normal Type II SNe, which is why we
  treat then as a separate group.  Occasionally Type Ib SNe are also
  observed to have narrow lines, and classed as Type Ibn.  Our sample,
  however, contained only a single Type Ibn SN, SN 2015G, which was
  not detected, therefore we do not discuss the Ibn subtype
  separately.

  We also discuss the subset of Type I b/c SNe which have broad
  optical lines, indicating high ejection velocities, (BL) as a group.
  This subtype has been of special interest because it is associated
  with gamma-ray bursts \citep{WoosleyB2006, Cano+2017}.

  Finally, we also discuss the Type IIb subset of Type II SNe, of
  which SN~1993J is the most famous example.  These SNe initially have
  H in their spectra and are therefore classified as Type II, but
  transition subsequently to having He-dominated spectra more
  characteristic of Type Ib.  They constitute $\sim$14\% of all Type
  II's \citep{Smith+2011a}.

  An important caveat must be mentioned here.  The classification of
  SNe into Types is based on features in the optical spectrum.  Since
  such features can vary as the SN evolves, there is the possibility
  that a SN may appear as different Types at different stages in its
  evolution.  Indeed, we just mentioned the Type IIb SNe whose spectra
  changes from Type II to one resembling Type Ib.

  The classification of Type IIn SNe is also occasionally
  time-variable.  The interaction with the dense CSM giving rise to
  the narrow lines and the ``n'' characteristics can start only some
  time after the explosion, so some SNe might first appear to be
  normal Type I or II, and then develop the ``n'' characteristics.
  SN~2014C is a prominent example of this behavior, which started as a
  Type Ib but developed IIn characteristics after about 1~yr
  \citep{Milisavljevic+2015_SN2014C}. SN~2001em is the other example
  of this behaviour in our sample.  Since the dominant part of the
  radio lightcurve for both SN~2001em and SN~2014C occurs at later
  times, when the optical spectrum was of Type IIn, we classify
  both SNe as Type IIn.

  Given the possible time-variability of the spectral characteristics,
  and therefore the non-uniqueness of the SN Type classification, our
  division by the SN Types is not completely unique.  However, since
  only a small fraction of SNe show such time-variable spectral
  characteristics, our statistical results should not be greatly
  affected by their occurrence.}

The remainder of this paper is organized as follows.  First in
Section~\ref{sobs}, we briefly describe the observations and data
reduction for the new data in this paper.  Then, in
Section~\ref{sdata}, we describe our collection of radio measurements
of \NSNe\ SNe.  In Section~\ref{sfit} we describe the model of a SN
radio lightcurve we fit to our measurements, which is characterized by
only two parameters, \tpk\ and \Lpk\@.  For many SNe, the measurements
are not sufficient to uniquely determine the values of \tpk\ and \Lpk,
for example, if there is only a single measurement, or only upper
limits.  In Section~\ref{sdistrib}, we combine these constraints over
all our SNe, and determine the likelihood of different values of \tpk\
and \Lpk\ given our measurements.  We then parameterize the
distribution of \Lpk\ and \tpk, finding that lognormal form is the
most likely, and proceed to determine the particular lognormal
distributions for \tpk\ and \Lpk\ which are most compatible with our
measurements.  We also examine various SN subtypes, such as Type I b/c
and Type II, to ask whether the distribution of \tpk\ and \Lpk\
differs for different SN Types.  In Section~\ref{sMdot}, we use our
distrutions of \tpk\ and \Lpk\ to estimate the distribution of
mass-loss rates.  In Section~\ref{sdiscuss}, we discuss the
implications of our results, and finally in Section~\ref{sconclude} we
summarize them and give our conclusions.

\section{Observations and Data Reduction}
\label{sobs}

We discuss our complete data-set which includes both published and
previously unpublished values in the next section.  Here we give a
brief summary of the observations and data reduction of the 296
previously unpublished SN observations.

We re-reduced a number of archival observations of SNe from the Karl
G. Jansky Very Large Array (VLA). This was done in a standard manner,
using the Astronomical Image Processing System \citep[AIPS;][]{AIPS}
for observations from the VLA before about 2011, and Common Astronomy
Software Application \citep[CASA;][]{CASA}
The flux density calibration was done using observations of 3C~48,
3C~138 or 3C~286.  Phase self-calibration was done on the supernova
observations in cases where the signal-to-noise ratio was adequate,
but no amplitude self-calibration was done.  In most of the archival
data sets, the supernova was not detected, so no self-calibration was
done.

The flux densities were determined by fitting to the images elliptical
Gaussians, fixed to the dimensions of the restoring beam, along with a
zero-level to account for any extended emission from the host
galaxies.  The total uncertainties (in Table~\ref{tnewvals}) include a
5\% uncertainty on the flux-density calibration, and in some cases a
contribution from the uncertainty in separating the SN from the
background emission, added in quadrature to the image background rms.

All observations with the Australia Telescope Compact Array (ATCA)
used the 2 GHz bandwidth CABB system \citep{Wilson+2011} and were
processed and measured using the {\sc miriad} package
\citep{SaultTW1995}, as described in \citet{Bufano+2014}.  The primary
flux density calibrator was PKS~B1934-638, and no self-calibration was
applied.

Observations with the Multi-Element Radio-Linked Interferometer
Network (MERLIN) used the e-Merlin pipeline \citep{e-Merlin_Pipeline}
using 512 MHz bandwidth.  The primary flux density calibrator was
3C~286, and no self-calibration was done.

\begin{table*}  % needed to get single column here
  \caption{Table: Supernova flux densities or limits from radio observations}
  \label{tnewvals}
  
  \tablecomments{Table \ref{tnewvals} is published in its entirety in
    the machine-readable format.  Only a portion is shown here for
    guidance regarding its form and content.}
 
\footnotesize
\begin{verbatim}

Title: The Radio Luminosity-Risetime Function of Core-Collapse Supernovae
Authors: Bietenholz M.F., Bartel N., Argo M., Dua R., Ryder S., Soderberg A.
Table: Supernova flux densities or limits from radio observations
================================================================================
Byte-by-byte Description of file: datafile1.txt
--------------------------------------------------------------------------------
   Bytes Format Units   Label  Explanations
--------------------------------------------------------------------------------
   1-  9 A9     ---     ID     SN identifier
      11 A1     ---     Limit  [L] Limit flag on Flux (1)
  13- 16 I4     yr      Obs.Y  UT Year of observation midpoint
  18- 19 I2     month   Obs.M  UT Month of observation midpoint
  21- 25 F5.2   d       Obs.D  UT Day of observation midpoint
  27- 33 A7     ---     Tel    Telescope identifier (2)
  36- 40 F5.2   GHz     Freq   Observed frequency 
  42- 49 F8.4   mJy     Flux   Measured flux density at Freq (3)
  51- 56 F6.4   mJy   e_Flux   Uncertainty in Flux 
  58- 63 A6     ---     Com    Additional comment (4)
--------------------------------------------------------------------------------
Note (1): "L" indicates a limit, blank indicates a measured value.
Note (2): 
    VLA = Very Large Array, USA; if known, the VLA configuration 
          is appended,  e.g. VLA-A;
  MERLIN= the Multi-Element Radio-Linked Interferometer Network, UK;
    ATCA= Australia Telescope Compact Array, Australia.
Note (3): A negative value indicates a limit, with the magnitude of 
          the value being the 3-sigma upper limit
Note (4): "Weiler" indicates that this value was retrieved from the 
          website of the late Kurt Weiler.
--------------------------------------------------------------------------------
SN1980O   L 1988  2  1.32 VLA-AB    4.86  -0.360  0.120
SN1982F   L 1984  8 31.00 VLA-D     4.86  -1.160  0.390
SN1982F   L 1984 12 23.00 VLA-A     4.86  -0.180  0.060
SN1985F   L 1985  3 18.00 VLA       4.86  -0.189  0.064
SN1985F   L 1985  7 31.00 VLA       4.86  -0.330  0.110
SN1985G   L 1985  5  7.3  VLA       4.86  -0.212  0.071
SN1985G   L 1985  9  1.00 VLA       4.86  -0.675  0.225
SN1985G   L 1986 12 15.00 VLA       4.86  -0.623  0.208
....
SN1993N   L 1994  2 18.32 VLA       8.44  -0.110  0.037
SN1993N   L 1997  1 23.00 VLA       8.46  -0.186  0.062  Weiler
....
SN2010as    2010  4 16.7  ATCA      9.00   2.19   0.11
SN2010as    2010  4 25.5  ATCA      9.00   3.10   0.9
....
\end{verbatim}
\end{table*}

\section{The Data-set}
\label{sdata}

We avail ourselves of as many of the published results as possible,
taking care to include any published upper limits in the cases of
non-detection.  To keep our data-set as uniform as possible, we
restricted ourselves to measurements between 4 and 10 GHz since the
most commonly used observing frequencies are 4.8 and 8.4 GHz, making
an exception for SN~1987A, where only a very few measurements are
available in the first years at those frequencies and we therefore use
the more complete 2.3~GHz lightcurve. We add to the previously
published values a number of previously unpublished measurements,
which are listed in Table~\ref{tnewvals}.

Our previously unpublished values include new measurements
from the ATCA and MERLIN,
as well as a number of results from re-reduced data from the Karl
G. Jansky Very Large Array (VLA) available in the National Radio
Astronomy Observatory (NRAO)\footnote{The NRAO, is a facility of the
  National Science Foundation operated under cooperative agreement by
  Associated Universities, Inc.} data archive.  There are a
considerable number of such observations which were never published.
Many SNe, even relatively nearby ones, are never detected in the
radio.  Such non-detections are much less likely to be published,
therefore the sample of published values is likely to be biased
towards detections and thus higher radio luminosities.  We have
therefore re-reduced a significant number of unpublished archival
measurements, the majority of which are indeed non-detections.

Finally, we include values from the website of the late Kurt
W. Weiler.  Dr.\ Weiler obtained many radio observations of SNe during
his illustrious career --- see, for example, \citet{Weiler+2002}.
Some of these were made available for a time on his website at the
U. S. Naval Observatory, but were never formally published.  We had
retrieved some of those values from the website, which we now include
also in our data set and in Table~\ref{tnewvals}.

While the largest fraction of our assembled observations are from the
VLA and ATCA, we also have measurements from a number of other
telescopes including  MERLIN, the Westerbork Synthesis Radio
Telescope, the European VLBI Network, the Urumqi radio telescope and
the Parkes-Tidbinbilla Interferometer.

In total we have 1475 measurements the flux density, or upper limits
on it, for \NSNe\ SNe.  For well observed SNe, such as SN~1993J
(Figure~\ref{fsn93j}) or SN~1986J, we have very well sampled
lightcurves, with many measurements ($n = 29$ and 39 respectively),
allowing \tpk\ and \Lpk\ to be accurately determined.
For the majority of SNe, however, only one or two measurements are
available, which thus provide only weak constraints on \tpk\ or \Lpk.
In fact, in many cases, the observations yielded only upper limits on
the SN's flux density.

Of our \NSNe\ SNe, only 31\% ($n=90$) are detected.  For the remaining
69\% ($n=204$) we have only upper limits on the flux density.  The
average number of measurements or limits per SN, detected or not, is
5.0\@.  However, this number is skewed by the 9\% ($n=26$) of %PPAA (5.1)
well-observed SNe which have more than 12 measurements each.  In fact,
35\% ($n=104$) of our SNe have only a single measurement or limit.  At least
three observations are required to uniquely determine the peak of the
lightcurve (one near, one before and one after the peak).
Only 27\% ($n=79$) of our SNe have three or more measurements or
limits, although in many of those cases, they all occur after the
peak, so that the peak is not determined.

Given this relatively modest number of measurements, compared to what
is available in the optical, and the fact that our sample is of
necessity heterogeneous and incomplete, we cannot provide a definitive
radio luminosity function for supernovae.  Nonetheless, we have a
larger data set than has ever previously been assembled, and
sufficiently large that some reasonably robust inferences can be
drawn.  It is crucial for this purpose to consider the non-detections
as well as the published detections.

Table~\ref{tsne} gives some details of the SNe in our database.  In
order to determine luminosities, we need the distances, $D$, for our
SNe.  In most cases, we calculated $D$ from the recession velocity for
the parent galaxy from the NASA/IPAC Extragalactic Database
(NED)\footnote{\url{https://ned.ipac.caltech.edu}}, using the value
corrected for our motion with respect to the cosmic microwave
background, and infall to the Virgo cluster, to the Great Attractor
and to the Shapley supercluster \citep{Mould+2000}.

We use the latest values from the Planck collaboration, which are
$H_0 = 67.4$~\kms~Mpc$^{-1}$, $\Omega_{\rm matter} =0.315$ and
$\Omega_\Lambda = 0.685$ \citep{Planck+2018vi}.  Since our most
distant (SN 2010ay) is at $D \simeq 300$~Mpc, and most SNe (89\%) are at
$D < 100$~Mpc, the precise values adopted for the cosmological
parameters do not significantly affect our results. For SNe closer
than 30~Mpc, we use the mean of the redshift-independent distances
from NED when available in preference to those calculated from the
recession velocity.

\startlongtable
\begin{deluxetable*}{lllrlrcl}
  \tabletypesize{\scriptsize}
\tablecaption{Supernovae Observed in Radio\label{tsne}}
\tablewidth{0pt}
\tablehead{
  \colhead{SN Name} &
  \colhead{Type\tablenotemark{a}} &
  \colhead{Galaxy} &
  \colhead{Distance\tablenotemark{b}} &
  \colhead{Explosion} &
  \colhead{Number of} &
  \colhead{Detected} &
  \colhead{References\tablenotemark{e}}  \\
  & & & \colhead{($D$; Mpc)} & \colhead{date\tablenotemark{c}} & \colhead{measurements\tablenotemark{d}}
}
\startdata
SN 1979C   & IIL    & NGC 4321       & 16.2 & 1979 04 06 & 67 & Y & 1, 2 \\
SN 1980K   & IIb-L  & NGC 6946       &  5.5 & 1980 10 25 & 69 & Y & 3, 4 \\
SN 1980O   & II     & NGC 1255       & 17.9 & 1980 12 30 &  2 &   & 5, 6  \\
SN 1981A   & II     & NGC 1532       & 17.9 & 1981 02 28 &  1 &   & 5  \\
SN 1981K   & II     & NGC 4258       &  7.3 & 1981 07 31 & 30 & Y & 1 \\  %scale 10->5 GHz using alpha = -0.6
SN 1982F   & IIP    & NGC 4490       &  6.2 & 1982 02 24 &  2 &   & 6 \\ 
SN 1982aa  & ?      & NGC 6052       & 80.5 & 1979 08 16 & 11 & Y & 7 \\
SN 1983I   & Ic     & NGC 4051       & 13.7 & 1983 04 25 &  2 &   & 8 \\ % no other source found
SN 1983K   & II     & NGC 4699       & 19.7 & 1983 06 22 &  3 &   & 9 \\
SN 1983N   & Ib     & NGC 5236       &  4.9 & 1983 06 29 & 15 & Y & 10 \\
SN 1984E   & IIL    & NGC 3169       & 22.4 & 1984 03 29 &  4 &   & 9 \\
SN 1984L   & Ib     & NGC 991        &  8.8 & 1984 08 10 &  3 & Y & 11 \\
SN 1985F   & Ib/c   & NGC 4618       &  7.2 & 1984 03 30 &  2 &   & 6 \\ % mfb
SN 1985G   & IIP    & NGC 4451       & 20.9 & 1985 03 17 &  3 &   & 6 \\ % mfb
SN 1985H   & II     & NGC 3359       & 16.0 & 1985 04 12 &  2 &   & 6 \\ % mfb
SN 1985L   & IIL    & NGC 5033       & 16.5 & 1985 06 13 &  7 & Y & 12 \\
SN 1986E   & IIL    & NGC 4302       & 16.8 & 1986 03 28 &  7 & Y & 13 \\ % max light
SN 1986J   & IIn    & NGC 891        & 10.0 & 1983 03 14 & 39 & Y & 14, 15 \\
SN 1987A   & IIf    & LMC            & 0.051& 1997 02 23 &  8 & Y & 16 \\  %  AKK- lc graph-only for ~6 days in
SN 1987F   & IIn:   & NGC 4615       & 79.6 & 1987 03 22 &  4 &   & 6 \\ % mfb
SN 1987K   & IIb    & NGC 4651       & 16.5 & 1987 07 31 &  2 &   & 6 \\ % mfb; 8->5GHz alpha -1
SN 1988I   & IIn    & Leda 86944     & 178  & 1988 03 07 &  1 &   & 9 \\ 
SN 1988Z   & IIn    & MCG+03-28-22   & 111  & 1988 12 01 & 26 & Y & 6, 17 \\
SN 1989C   & IIP    & UGC 5249       & 32.1 & 1989 02 01 &  1 &   & 9 \\ 
SN 1989L   & II     & NGC 7339       & 22.0 & 1989 05 04 &  3 &   & 6 \\ % mfb
SN 1989R   & IIn    & UGC 2912       & 80.1 & 1989 09 15 &  1 &   & 9  \\ 
SN 1990B   & Ic     & NGC 4568       & 17.4 & 1990 01 18 &  8 & Y & 18 \\
SN 1990K   & II     & NGC 150        & 23.4 & 1990 05 14 &  2 &   & 6 \\ %mfb
SN 1991G   & IIP    & NGC 4088       & 13.9 & 1991 01 23 &  2 &   & 6 \\ %mfb 8->5GHz alpha = -1
SN 1991N   & Ic     & NGC 3310       & 18.1 & 1991 04 02 &  2 &   & 8, 19 \\
SN 1991ae  & IIn    & MCG+11-19-18   & 138  & 1991 05 15 &  2 &   & 6, 9 \\ % 8->5GHz alpha -1
SN 1991av  & IIn  & Anon J215601+0059& 288  & 1991 09 15 &  3 &   & 9 \\ 
SN 1992H   & II     & NGC 5377       & 35.1 & 1992 02 11 &  2 &   & 6 \\  % limit listed but no numbers in Weiler+2002
SN 1992ad  & II     & NGC 4411B      & 22.4 & 1992 06 30 &  5 & Y & 6, 20 \\
SN 1992bd  & II     & NGC 1097       & 16.9 & 1992 10 12 &  5 &   & 6 \\ %mfb + Rahul
SN 1993G   & IIL    & NGC 3690       & 53.1 & 1993 02 24 &  1 &   & 6 \\ %mfb
SN 1993J   & IIb    & M81            &  3.7 & 1993 03 28 & 29 & Y & 21 \\  % 
SN 1993N   & IIn    & UGC 5695       & 50.2 & 1993 04 15 &  2 &   & 6, 22 \\ % limit listed but no numbers in Weiler+2002
SN 1993X   & II     & NGC 2276       & 40.5 & 1993 08 22 &  1 &   & 6 \\ % limit listed but no numbers in Weiler+2002
SN 1994I   & Ic     & M51            &  7.9 & 1994 03 31 & 39 &   & 6, 19, 23  \\
SN 1994P   & II     & UGC 6983       & 19.6 & 1994 01 20 &  3 &   & 6 \\ % limit listed but no numbers in Weiler+2002
SN 1994W   & IIn-P  & NGC 4041       & 25.4 & 1994 07 30 &  3 &   & 6, 22 \\ % limit listed but no numbers in Weiler+2002
SN 1994Y   & IIn    & NGC 5371       & 46.4 & 1994 07 09 &  1 &   & 6 \\ % limit listed but no numbers in Weiler+2002
SN 1994ai  & Ic     & NGC 908        & 15.6 & 1994 12 20 &  2 &   & 6, 19 \\
SN 1994ak  & IIn    & NGC 2782       & 43.1 & 1994 12 24 &  1 &   & 6 \\ % limit listed but no numbers in Weiler+2002
SN 1995N   & IIn    & MCG-02-38-17   & 31.4 & 1994 07 04 & 18 & Y & 24 \\
SN 1995X   & II     & UGC 12160      & 25.5 & 1995 08 03 &  4 &   & 22 \\ % limit listed but no numbers in Weiler+2002; convert C->X using alph=0
SN 1995ad  & II     & NGC 2139       & 27.0 & 1995 09 22 &  1 &   & 22 \\ % limit listed but no numbers in Weiler+2002
SN 1996L   & IIn    & ESO 266-G10    & 157  & 1996 03 12 &  1 &   & 22 \\ % limit listed but no numbers in Weiler+2002
SN 1996N   & Ib     & NGC 1398       & 19.8 & 1996 03 09 &  3 & Y & 6, 19, 25 \\
SN 1996W   & II     & NGC 4027       & 12.2 & 1996 04 10 &  3 &   & 6, 22 \\ % limit listed but no numbers in Weiler+2002
SN 1996ae  & IIn    & NGC 5775       & 19.9 & 1996 01 27 &  4 &   & 6, 22 \\ % limit listed but no numbers in Weiler+2002
SN 1996an  & II     & NGC 1084       & 19.1 & 1996 05 30 &  2 &   & 22 \\ % limit listed but no numbers in Weiler+2002
SN 1996aq  & Ic     & NGC 5584       & 21.8 & 1996 08 17 &  4 &   & 6, 19, 26 \\
SN 1996bu  & IIn    & NGC 3631       & 10.3 & 1996 11 14 &  2 &   & 6  \\  % PPAA - remove "22"
SN 1996bw  & II     & NGC 664        & 79.0 & 1996 11 30 &  1 &   & 22 \\  % ?? not listed in Weiler+2002, on NRAO archive
SN 1996cb  & IIb    & NGC 3510       & 13.9 & 1996 12 12 &  3 & Y & 6, 27  \\
SN 1996cr  & IIn:   & Circinus       &  3.8 & 1995 03 01 & 11 & Y & 28 \\
SN 1997W   & II     & NGC 664        & 79.0 & 1997 02 01 &  2 &   & 6, 22 \\  % convert c->x using alpha=0
SN 1997X   & Ib/c   & NGC 4691       & 21.3 & 1997 01 25 &  3 & Y & 6, 19 \\
SN 1997ab  & IIn  & Anon J095100+2004& 53.9 & 1996 04 11 &  2 &   & 22 \\ % limit listed but no numbers in Weiler+2002
SN 1997db  & II     & UGC 11861      & 18.9 & 1997 08 02 &  3 &   & 6, 22 \\ % limit listed but no numbers in Weiler+2002; convert C->X assuming alpha=0
SN 1997dn  & II     & NGC 3451       & 27.1 & 1997 10 29 &  1 &   & 6 \\ 
SN 1997dq  & IcBL   & NGC 3810       & 15.7 & 1997 10 13 &  3 &   & 22, 19 \\  % limit listed but no numbers in Weiler+2002
SN 1997ef  & IbBL   & UGC 4107       & 55.9 & 1997 11 20 &  2 &   & 6, 19 \\ 
SN 1997eg  & IIn    & NGC 5012       & 47.6 & 1997 12 04 &  3 & Y & 29 \\ 
SN 1997ei  & Ic     & NGC 3963       & 48.8 & 1997 11 20 &  1 &   & 22 \\  % limit listed but no numbers in Weiler+2002
SN 1998S   & IIn    & NGC 3877       & 14.9 & 1998 02 28 &  8 & Y & 6, 22, 30 \\ % Weiler+2002 says "Detected"
SN 1998bm  & II     & IC 2458        & 24.7 & 1998 04 21 &  2 &   & 6 \\ 
SN 1998bw  & IcBL   & ESO 184-82     & 41.4 & 1998 04 25 & 31 & Y & 31 \\
SN 1998dl  & IIP    & NGC 1084       & 19.1 & 1998 08 02 &  2 &   & 22 \\  % not listed Weiler+2002!
SN 1998dn  & II     & NGC 337A       & 13.7 & 1998 08 19 &  2 &   & 22 \\  % not listed Weiler+2002; convert C-X alpha=0
SN 1999B   & II     & UGC 7189       & 31.2 & 1999 01 14 &  1 &   & 6 \\ 
SN 1999D   & II     & NGC 3690       & 52.6 & 1999 01 16 &  2 &   & 6, 22 \\ % limit listed but no numbers in Weiler+2002
SN 1999E   & IIn & Anon J131716-1833 & 119  & 1998 09 10 &  1 &   & 22 \\  % limit listed but no numbers in Weiler+2002
SN 1999cn  & Ic     & MCG+02-38-43   & 111  & 1999 06 14 &  1 &   & 22 \\  % not listed Weiler+2002 
SN 1999dn  & Ib     & NGC 7714       & 29.1 & 1999 08 15 &  1 &   & 6, 19 \\ % conv X-C assuming alpha=-1
SN 1999eb  & IIn    & NGC 664        & 79.0 & 1999 10 02 &  1 &   & 6 \\ 
SN 1999eh  & Ib     & NGC 2770       & 28.6 & 1999 07 26 &  2 &   & 8, 19 \\ % conv C-X assuming alpha=0
SN 1999el  & IIn    & NGC 6951       & 23.1 & 1999 10 20 &  2 &   & 6 \\  % convert C->X assuming alpha=0
SN 1999em  & IIP    & NGC 1637       & 11.5 & 1999 10 24 &  5 & Y & 6, 22, 32 \\
SN 1999ev  & IIP    & NGC 4724       & 13.9 & 1999 11 07 &  1 &   & 6 \\
SN 1999ex  & Ic     & IC 5179        & 53.3 & 1999 11 01 &  1 &   & 33 \\
SN 1999gi  & IIP    & NGC 3184       & 12.4 & 1999 12 06 &  3 &   & 6 \\  % conv C-X lim assuming alpha=0
SN 1999go  & II     & NGC 1376       & 60.4 & 1999 12 18 &  1 &   & 6 \\ 
SN 1999gq  & IIP    & NGC 4523       & 16.7 & 1999 12 23 &  1 & Y & 6 \\ % no other source found 
SN 2000C   & Ic     & NGC 2415       & 59.4 & 2000 01 01 &  1 &   & 19, 33 \\ 
SN 2000F   & Ic     & IC 302         & 86.1 & 2000 01 29 &  1 &   & 19 \\ 
SN 2000P   & IIn    & NGC 4965       & 30.2 & 2000 03 08 &  2 &   & 22 \\  % limit listed but no numbers in Weiler+2002
SN 2000S   & Ic     & MCG-01-27-20   & 138  & 1999 10 09 &  1 &   & 19 \\ 
SN 2000cr  & Ic     & NGC 5395       & 61.3 & 2000 06 21 &  1 &   & 33 \\ 
SN 2000ds  & Ib/c   & NGC 2768       & 20.5 & 2000 05 28 &  3 &   & 8, 19 \\
SN 2000ew  & Ic     & NGC 3810       & 15.7 & 2000 11 21 &  1 &   & 6 \\
SN 2000fn  & Ib     & NGC 2526       & 72.3 & 2000 11 09 &  1 &   & 33 \\ 
SN 2000ft  & ?      & NGC 7469       & 73.5 & 2000 07 19 &  7 & Y & 34 \\ % convert X->C using alpha -0.8+-0.2
SN 2001B   & Ib     & IC 391         & 27.4 & 2000 12 31 &  3 & Y & 19, 33 \\
SN 2001M   & Ic     & NGC 3240       & 57.3 & 2001 01 17 &  1 &   & 33 \\ 
SN 2001ai  & Ic     & NGC 5278       & 121  & 2001 03 24 &  1 &   & 33 \\ 
SN 2001bb  & Ic     & IC 4319        & 82.0 & 2001 04 22 &  2 &   & 19, 33\\ 
SN 2001ch  & Ic     & MCG-01-54-16   & 46.8 & 2001 03 24 &  1 &   & 19 \\
SN 2001ci  & Ic     & NGC 3079       & 16.4 & 2001 04 21 &  3 & Y & 6, 19, 33 \\
SN 2001ef  & Ic     & IC 381         & 40.2 & 2001 09 04 &  2 &   & 19, 33\\ 
SN 2001ej  & Ib     & UGC 3829       & 62.8 & 2001 09 09 &  2 &   & 19, 33\\ 
SN 2001em  & IIn\tablenotemark{f}
                    & UGC 11794      & 89.7 & 2001 09 12 &  8 & Y & 35, 36, 37\\ 
SN 2001gd  & IIb    & NGC 5033       & 17.5 & 2001 09 03 & 11 & Y & 38 \\
SN 2001ig  & IIb    & NGC 7424       &  9.3 & 2001 12 03 & 23 & Y & 39 \\
SN 2001is  & Ib     & NGC 1961       & 61.2 & 2001 12 19 &  1 &   & 19 \\  
SN 2002ap  & IcBL   & NGC 628        &  8.9 & 2001 02 28 &  9 & Y & 19, 40 \\
SN 2002bl  & IcPecBL& UGC 5499       & 77.5 & 2002 02 23 &  2 &   & 19, 33 \\ 
SN 2002cj  & Ic     & ESO 582-05     & 113  & 2002 04 16 &  1 & Y & 33 \\ 
SN 2002cp  & Ib/c   & NGC 3074       & 82.9 & 2002 04 20 &  2 &   & 19, 33 \\
SN 2002dg  & Ib & Anon J145716+0554  & 225  & 2002 05 29 &  2 &   & 19, 33 \\ % cheat X-C lim alpha -1
SN 2002dn  & Ic     & IC 5145        & 112  & 2002 06 08 &  1 &   & 19, 33 \\ 
SN 2002gy  & Ib/c:  & UGC 2701       & 107  & 2002 10 13 &  1 &   & 33 \\
SN 2002hf  & Ic     & MCG-05-03-20   & 82.9 & 2002 10 26 &  2 &   & 19, 33 \\ 
SN 2002hh  & II     & NGC 6946       &  5.6 & 2002 10 31 &  8 & Y & 22, 41\\  % cheat C-X value alpha = -0.6
SN 2002hn  & Ic     & NGC 2532       & 82.1 & 2002 10 26 &  1 &   & 33 \\ 
SN 2002ho  & Ic     & NGC 4210       & 47.2 & 2002 11 01 &  2 &   & 19, 33 \\ 
SN 2002hy  & IbPec  & NGC 3464       & 60.7 & 2002 10 28 &  2 &   & 19, 33 \\ 
SN 2002hz  & Ib     & UGC 12044      & 82.9 & 2002 11 07 &  2 &   & 19, 33 \\ 
SN 2002ji  & Ic     & NGC 3655       & 30.3 & 2002 10 19 &  3 &   & 19, 33, 42 \\ 
SN 2002jj  & Ic     & IC 340         & 60.6 & 2002 10 13 &  2 &   & 19, 33 \\ 
SN 2002jp  & Ic     & NGC 3313       & 59.6 & 2001 11 15 &  2 &   & 19, 33 \\ 
SN 2002jz  & Ic     & UGC 2984       & 22.9 & 2001 12 14 &  1 &   & 33  \\ 
SN 2003H   & IbPec  & NGC 2207       & 22.3 & 2003 01 08 &  3 &   & 6, 42 \\  
SN 2003L   & Ic     & NGC 3506       & 104  & 2001 01 01 & 40 & Y & 43 \\
SN 2003bg  & IcPecBL& MCG -05-10-15  & 19.3 & 2003 02 22 & 41 & Y & 44 \\
SN 2003bu  & Ic     & NGC 5953       & 105  & 2003 03 03 &  2 &   & 8 \\ 
SN 2003dr  & Ib/c   & NGC 5714       & 41.7 & 2004 04 10 &  3 &   & 8, 19, 42 \\
SN 2003dv  & IIn    & UGC 9638       & 33.9 & 2004 04 16 &  1 &   & 6 \\
SN 2003ed  & II     & NGC 5303A      & 25.2 & 2003 04 30 &  4 & Y & 6, 45 \\
SN 2003el  & Ic     & NGC 5000       & 93.4 & 2003 05 11 &  1 &   & 19 \\
SN 2003gd  & IIP    & NGC 628        &  8.6 & 2003 03 17 &  5 &   & 6 \\
SN 2003gk  & Ib     & NGC 7460       & 48.5 & 2003 06 15 &  1 &   & 8 \\
SN 2003ie  & IIP    & NGC 4051       & 13.7 & 2003 09 19 &  4 &   & 6, 22 \\ % cheat C-X lim assuming alpha=0
SN 2003jd  & IcPecBL& MCG -01-59-2   & 84.6 & 2003 10 10 &  4 &   & 8, 19 \\
SN 2003jg  & Ib/c   & NGC 2997       &  9.0 & 2003 10 01 &  3 &   & 8, 42 \\
SN 2003lo  & IIn    & NGC 1376       & 60.4 & 2003 12 31 &  1 &   & 6 \\
SN 2004A   & IIP    & NGC 6207       & 17.0 & 2004 01 06 &  6 &   & 6 \\ 
SN 2004C   & Ic     & NGC 3683       & 32.6 & 2003 12 23 &  3 & Y & 6, 8 \\ 
SN 2004am  & IIP    & NGC 3034       &  3.8 & 2003 11 07 &  3 &   & 22, 46 \\
SN 2004ao  & Ib     & UGC 10862      & 26.8 & 2004 02 21 &  1 &   & 8   \\
SN 2004bm  & Ic     & NGC 3437       & 24.4 & 2004 04 17 &  2 &   & 6, 42 \\
SN 2004bu  & IcBL   & UGC 10089      & 92.1 & 2004 05 14 &  1 &   & 8 \\  
SN 2004cc  & Ic     & NGC 4568       & 17.4 & 2004 05 23 &  8 & Y & 47 \\
SN 2004dj  & IIP    & NGC 2403       &  3.4 & 2004 07 13 & 40 & Y & 48 \\
SN 2004dk  & Ib     & NGC 6118       & 20.8 & 2004 07 30 & 10 & Y & 47 \\
SN 2004et  & IIP    & NGC 6946       &  5.6 & 2004 09 22 & 19 & Y & 22, 49  \\
SN 2004gq  & Ib     & NGC 1832       & 24.3 & 2004 12 08 & 21 & Y & 47 \\
SN 2004gt  & Ib/c   & NGC 4038       & 21.1 & 2004 11 27 &  2 &   & 8, 42, 60 \\  %PPAA add "60"
SN 2005E   & Ib/c   & NGC 1032       & 38.8 & 2005 01 04 &  1 &   & 8 \\
SN 2005U   & IIb    & NGC 3690       & 53.1 & 2005 01 28 &  2 &   & 6 \\
SN 2005V   & Ib/c   & NGC 2146       & 19.6 & 2005 01 01 &  5 &   & 6, 8, 42 \\  %cheat C-X lim using alpha=0
SN 2005aj  & Ic     & UGC 2411       & 41.1 & 2005 02 09 &  2 &   & 8, 42 \\  
SN 2005at  & Ic     & NGC 6744       &  7.2 & 2005 03 05 &  2 &   & 50 \\
SN 2005ay  & IIP    & NGC 3938       & 12.7 & 2005 03 21 &  4 &   & 6 \\
SN 2005cs  & IIP    & M51            &  7.9 & 2005 06 27 &  5 &   & 51 \\
SN 2005ct  & Ic     & NGC 207        & 58.6 & 2005 05 29 &  1 &   & 8 \\
SN 2005cz  & Ib     & NGC 4589       & 35.7 & 2005 06 17 &  1 &   & 8 \\
SN 2005da  & IcBL   & UGC 11301      & 74.4 & 2005 06 25 &  3 &   & 8 \\
SN 2005dl  & II     & NGC 2276       & 20.4 & 2005 08 25 &  2 &   & 6 \\
SN 2005ek  & Ic     & UGC 2526       & 73.0 & 2005 09 22 &  1 &   & 52 \\
SN 2005gl  & IIn    & NGC 266        & 68.8 & 2005 10 26 &  1 &   & 22 \\ % no other source for data found
SN 2005ip  & IIn    & NGC 2906       & 36.5 & 2005 10 27 &  3 & Y & 53 \\
SN 2005kd  & IIn    & 2MFGC 3318     & 69.4 & 2005 11 10 &  4 & Y & 6, 22, 54, 55 \\
SN 2005kl  & Ic     & NGC 4369       & 29.7 & 2005 11 01 &  1 &   & 6 \\
SN 2006aj  & IcBL   & 2XMM J032139.6+165202 
                                     & 153  & 2006 02 18 & 17 & Y & 56 \\
SN 2006be  & II     & IC 4582        & 40.6 & 2006 03 13 &  1 &   & 57 \\
SN 2006bp  & IIP    & NGC 3953       & 16.6 & 2006 04 09 &  4 &   & 22, 58 \\ 
SN 2006gy  & IIn    & NGC 1260       & 85.0 & 2005 08 20 &  8 &   & 6, 59, 60 \\
SN 2006jd  & IIn    & UGC 4179       & 83.7 & 2006 10 07 & 11 & Y & 61 \\
SN 2006my  & IIP    & NGC 4651       & 16.5 & 2006 08 01 &  2 &   & 22 \\ %convert 5->8 GHz lim alpha 0
SN 2006ov  & IIP    & NGC 4303       & 14.6 & 2006 10 26 &  2 &   & 6, 22 \\ % cheat limit 8->6 GHz alpha -1; no other source for data found
SN 2007C   & Ib     & NGC 4981       & 22.7 & 2006 12 28 &  2 & Y & 6 \\
SN 2007Y   & IbPec  & NGC 1187       & 16.8 & 2007 02 14 &  7 &   & 42, 62 \\
SN 2007ak  & IIn    & UGC 3293       & 69.6 & 2007 03 10 &  1 &   & 22 \\ % no other source for data found
SN 2007bg  & IcBL &Anon J114926+5149 & 155  & 2007 04 16 & 18 & Y & 63 \\
SN 2007gr  & Ib/c   & NGC 1058       &  5.2 & 2007 08 13 &  9 & Y & 64 \\
SN 2007iq  & IcBL   & UGC 3416       & 62.5 & 2007 08 01 &  2 &   & 8, 42  \\  %
SN 2007ke  & Ib     & NGC 1129       & 76.7 & 2007 09 02 &  1 &   & 8 \\
SN 2007kj  & Ib/c   & NGC 7803       & 79.3 & 2007 09 14 &  1 &   & 8 \\
SN 2007pk  & IInPec & NGC 579        & 73.4 & 2007 11 08 &  1 &   & 65 \\
SN 2007rt  & IIn    & UGC 6109       & 107  & 2007 09 05 &  1 &   & 66 \\
SN 2007ru  & IcBL   & UGC 12381      & 70.3 & 2007 11 25 &  2 &   & 8 \\
SN 2007rz  & Ic     & NGC 1590       & 57.1 & 2007 11 19 &  2 &   & 8, 42\\
SN 2007uy  & Ib     & NGC 2770       & 28.6 & 2007 12 27 & 16 & Y & 67 \\
SN 2008B   & IIn    & NGC 5829       & 94.5 & 2008 01 02 &  1 &   & 68 \\
SN 2008D   & Ib     & NGC 2770       & 28.6 & 2008 01 09 & 21 & Y & 69 \\
SN 2008X   & IIP    & NGC 4141       & 35.4 & 2008 01 14 &  2 &   & 6, 70 \\
SN 2008aj  & IIn    & MCG+06-30-34   & 122  & 2008 02 12 &  1 &   & 71 \\
SN 2008ax  & IIb    & NGC 4490       &  6.2 & 2008 03 03 & 24 & Y & 22, 72 \\
SN 2008be  & IIn    & NGC 5671       & 142  & 2008 03 12 &  1 &   & 73 \\
SN 2008bk  & IIP    & NGC 7793       &  3.9 & 2008 03 07 &  1 &   & 74 \\
SN 2008bm  & IIn    & Leda 45053     &  155 & 2008 03 29 &  1 &   & 75 \\
SN 2008bo  & IIb    & NGC 6643       & 19.1 & 2008 03 27 &  7 & Y & 22, 76 \\
SN 2008du  & Ic     & NGC 7422       & 72.4 & 2008 06 30 &  1 &   & 42 \\
SN 2008dv  & Ic     & NGC 1343       & 10.5 & 2008 05 26 &  2 &   & 8, 42 \\ % no other source for data found
SN 2008ew  & Ic     & IC1236         & 99.5 & 2008 08 06 &  1 &   & 8 \\
SN 2008gm  & IIn    & NGC 7530       & 53.0 & 2008 10 02 &  1 &   & 77 \\ %
SN 2008hh  & Ic     & IC 112         & 85.0 & 2008 11 04 &  1 &   & 8 \\
SN 2008hn  & Ic     & NGC 2545       & 54.2 & 2008 11 12 &  1 &   & 8 \\
SN 2008ij  & II     & NGC 6643       & 19.1 & 2008 12 19 &  1 &   & 78 \\
SN 2008im  & Ib     & UGC 2906       & 40.2 & 2008 12 15 &  1 &   & 8 \\
SN 2008in  & IIP    & NGC 4303       & 14.6 & 2008 12 22 &  2 &   & 6, 79 \\
SN 2008ip  & IIn    & NGC 4846       & 90.2 & 2008 12 31 &  1 &   & 80 \\
SN 2008iz  & ?      & M82            &  3.8 & 2008 02 20 & 25 & Y & 81  \\ 
SN 2008jb  & II     & ESO 302-14     &  9.3 & 2008 11 11 &  1 & Y & 6 \\ % no other source of data found
SN 2009E   & IIP    & NGC 4141       & 35.4 & 2008 01 01 &  1 &   & 6 \\ % 
SN 2009H   & II     & NGC 1084       & 19.1 & 2009 01 02 &  2 &   & 82 \\
SN 2009N   & IIP    & NGC 4487       & 17.2 & 2009 01 24 &  2 &   & 82 \\ 
SN 2009au  & IIn    & ESO 443-21     & 36.5 & 2009 03 07 &  1 &   & 83 \\
SN 2009bb  & IcBL   & NGC 3278       & 43.5 & 2009 03 19 & 17 & Y & 84 \\
SN 2009dd  & II     & NGC 4088       & 13.9 & 2009 04 12 &  3 &   & 6, 85 \\
SN 2009eo  & IIn    & Leda 53491     & 212  & 2009 04 13 &  1 &   & 86 \\
SN 2009fs  & IIn    & UGC 11205      & 256  & 2009 06 01 &  1 &   & 87 \\
SN 2009gj  & IIb    & NGC 134        & 16.8 & 2009 05 31 &  3 & Y & 88 \\
SN 2009hd  & II     & NGC 3627       &  9.6 & 2009 06 19 &  1 &   & 6 \\
SN 2009ip  & IIn    & NGC 7259       & 28.1 & 2009 09 13 &  5 & Y & 89 \\
SN 2009kn  & IIn    & MCG-03-21-06   & 74.6 & 2009 10 11 &  1 &   & 90 \\
SN 2009mk  & IIb    & ESO 293-34     & 20.3 & 2009 12 15 &  4 &   & 91 \\
SN 2010O   & Ib     & NGC 3690       & 53.1 & 2010 01 24 &  2 &   & 92 \\
SN 2010P   & ?      & NGC 3690       & 53.1 & 2010 01 10 &  7 & Y & 92 \\ % X->C assuming alpha=-0.6
SN 2010ah  &IcBL& Anon J114403+5541  & 230  & 2010 02 21 &  4 &   & 93 \\
SN 2010al  & IInPec & UGC 4286       & 80.6 & 2010 03 07 &  1 &   & 94 \\
SN 2010as  & IIb    & NGC 6000       & 27.4 & 2010 03 16 & 10 & Y & 60, 95 \\ %
SN 2010ay  & IcBL& Anon J123527+2704 & 314  & 2010 02 22 &  3 &   & 96 \\
SN 2010bh  & IcBL& Anon J071031-5615 & 276  & 2010 03 16 &  7 & Y & 97 \\
SN 2010br  & Ib/c   & NGC 4051       & 13.7 & 2010 04 10 &  1 &   & 98 \\
PTF10vgv   & IcBL& 2MASX J22160156+4052065
                                     & 63.8 & 2010 09 13 &  1 &   & 99 \\ %sn2010z0
SN 2010id  & II     & NGC 7483       & 74.1 & 2010 09 15 &  1 &   & 100 \\
SN 2010jl  & IIn    & UGC 5189A      & 53.3 & 2010 10 01 & 11 & Y & 101 \\
SN 2010jp  & IIn & Anon J061630-2124 & 44.8 & 2010 11 13 &  2 &   & 102 \\
SN 2010kp  & II  & Anon J040341+7045 & 22.3 & 2010 11 30 &  2 &   & 6, 103 \\
PTF10abyy  & II     & galaxy unknown & 133  & 2010 12 06 &  1 &   & 104 \\ %sn2010z2
SN 2011cb  & IIb    & IC 5249        & 36.0 & 2011 04 18 &  4 & Y & 60, 105 \\
SN 2011dh  & IIb    & M51            &  7.9 & 2011 05 31 & 16 & Y & 106 \\ 
PTF11iqb   & IIn    & NGC 151        & 55.1 & 2011 07 20 &  1 &   & 107 \\ % sn2011z0
PTF11qcj   & IcBL   & Leda 2295826   & 135  & 2011 10 08 & 20 & Y & 108 \\ % sn2011z1
SN 2011ei  & II     & NGC 6925       & 28.7 & 2011 07 25 & 11 & Y & 109 \\ 
SN 2011hp  & Ic     & NGC 4219       & 22.1 & 2011 11 04 &  1 &   & 110 \\
SN 2011hs  & IIb    & IC 5267        & 21.3 & 2011 11 06 & 10 & Y & 111 \\ 
SN 2011ja  & IIP    & NGC 4945       &  4.2 & 2011 12 12 &  2 & Y & 112 \\ 
SN 2012A   & IIP    & NGC 3239       &  9.7 & 2012 01 07 &  2 &   & 6 \\
SN 2012ap  & IcBL   & NGC 1729       & 53.6 & 2012 02 05 &  3 & Y & 113 \\ 
SN 2012au  & Ib     & NGC 4790       & 22.9 & 2012 03 03 &  3 & Y & 114 \\
SN 2012aw  & IIP    & NGC 3351       & 10.0 & 2012 03 15 &  9 & Y & 115 \\
PTF 12gzk  &Ic&SDSS J221241.53+003042.7&63.4& 2012 07 24 &  3 & Y & 116 \\ % sn2012z0
SN 2013df  & IIb    & NGC 4414       & 18.1 & 2013 06 04 &  5 & Y & 117 \\
SN 2013ej  & IIP    & NGC 628        &  8.6 & 2013 07 28 &  2 & Y & 6 \\ 
SN 2013fs  & IIP    & NGC 7610       & 53.4 & 2013 10 06 &  2 &   & 118 \\ 
SN 2013ge  & Ib/c   & NGC 3287       & 15.4 & 2013 11 07 &  3 &   & 119 \\ 
iPTF13bvn  & Ib     & NGC 5806       & 24.7 & 2013 06 16 &  2 &   & 120 \\ % sn2013z0
SN 2014C   & IIn\tablenotemark{f}
                    & NGC 7331       & 13.4 & 2013 12 30 & 14 & Y & 121 \\
SN 2014ad  & IcBL   & Mrk 1309       & 28.9 & 2014 03 09 &  6 &   & 122 \\ 
SN 2014bc  & IIP    & NGC 4258       & 14.1 & 2014 04 08 &  1 &   & 6, 123\\ 
SN 2014bi  & IIP    & NGC 4096       & 11.5 & 2014 04 22 &  2 &   & 6, 123 \\ 
SN 2014eh  & Ic     & NGC 6907       & 51.8 & 2014 10 29 &  1 &   & 124 \\ 
AT 2014ge  & Ib     & NGC 4080       & 15.5 & 2014 09 26 &  5 & Y & 125 \\ % sn2014z2;
SN 2015G   & Ibn    & NGC 6951       & 23.1 & 2015 02 27 &  3 &   & 126 \\
SN 2015J   & IIn & Anon J073505-6907 & 24.1 & 2015 04 26 &  5 & Y & 60, 127 \\
iPTF15eqv  & IIb/Ib & NGC 3430       & 26.5 & 2015 08 18 &  4 &   & 128 \\ % sn2015z0 
ASASSN-15oz& IIL    & HIPASS J1919-33& 34.6 & 2015 08 27 &  2 & Y & 129 \\ % sn2015z1
PSN J22460504-1059484&Ib& NGC 7371   & 41.4 & 2015 07 10 &  1 & Y & 130 \\ % sn2015z2 -> more data avail
PSN J14102342-4318437&Ib& NGC 5483   & 18.5 & 2015 12 03 &  1 & Y & 131 \\ % sn2015z3 
SN 2016aqf & II     & NGC 2101       & 16.1 & 2016 02 24 &  2 & Y & 132 \\ 
SN 2016bas & IIb    & ESO 163-11     & 42.4 & 2016 03 02 &  8 & Y & 60, 133 \\
SN 2016bau & Ib     & NGC 3631       & 10.3 & 2016 03 12 &  2 & Y & 6, 134 \\
SN 2016coi & IcBL   & UGC 11868      & 18.1 & 2016 05 24 &  7 & Y & 135 \\
SN 2016cvk & IIn-pec& ESO 344-21     & 50.4 & 2016 06 13 &  1 &   & 136 \\
SN 2016gfy & II     & NGC 2276       & 20.4 & 2016 09 10 &  1 &   & 6 \\ 
Spirits 16tn& ?     & NGC 3556       & 10.0 & 2016 05 05 &  2 &   & 137 \\ %sn2016zz0 
SN 2017ahn & II     & NGC 3318       & 39.8 & 2017 02 08 &  1 &   & 138 \\
SN 2017eaw & IIP    & NGC 6946       &  5.6 & 2017 05 12 &  4 & Y & 139 \\ 
SN 2017gax & Ib/c   & NGC 1672       & 11.8 & 2017 08 12 &  1 &   & 140 \\
SN 2018ec  & Ic     & NGC 3256       & 40.3 & 2017 12 27 &  1 &   & 60 \\
SN 2018ie  & IcBL   & NGC 3456       & 70.6 & 2018 01 05 &  1 &   & 141 \\
SN 2018if  & IcBL   &SDSS J091423.85+493533.4 & 141  & 2018 01 19 &  1 &   & 141 \\
SN 2018bvw & IcBL   &SDSS J115244.11+254027.1 & 258  & 2018 04 25 &  4 & Y & 142 \\ % ZTF18aaqjovh
SN 2018cow & Icpec  & CGCG 137-068   & 72.7 & 2018 06 16 &  7 & Y & 143 \\
SN 2018gep &IcBL&SDSS J164348.22+410243.3&144&2018 09 09 &  3 & Y & 144 \\
SN 2018lab & II     & IC 2163        & 21.0 & 2018 12 29 &  1 &   & 145 \\
SN 2019eez & II     & NGC 2207       & 22.3 & 2019 04 26 &  1 &   & 146 \\
SN 2019ehk & Ib     & NGC 4321       & 16.2 & 2019 04 28 &  5 &   & 147 \\
SN 2019ejj & II     & ESO 430-20     & 11.5 & 2019 04 29 &  1 &   & 146 \\
SN 2019esa & IIn    & ESO 035-18     & 25.9 & 2019 05 05 &  1 &   & 146 \\
SN 2019fcn & II     & ESO 430-20     & 11.5 & 2019 05 03 &  1 &   & 146 \\
SN 2019mhm & IIP    & NGC 6753       & 50.6 & 2019 10 09 &  1 &   & 148 \\
SN 2019qar & Ib/c-pec& NGC 7083      & 48.5 & 2019 09 10 &  1 &   & 149 \\
SN 2020ad  & II     & IC 4351        & 28.8 & 2019 12 03 &  1 &   & 150 \\
SN 2020oi  & Ic     & NGC 4321       & 16.2 & 2020 01 07 &  9 & Y & 151 \\
SN 2020bvc & IcBL   & UGC 09379      & 122  & 2020 02 04 &  2 & Y & 152 \\
SN 2020fqv & Ib/c   & NGC 4568       & 21.0 & 2020 03 31 &  1 &   & 153 \\ 
SN 2020fsb & II     & ESO 515-04     & 35.2 & 2020 04 02 &  1 &   & 153 \\
SN 2020llx & II     & NGC 7140       & 46.7 & 2020 05 29 &  1 &   & 154 \\ 
\enddata
\tablenotetext{a}{The Type of the SN. ``BL'' stands for
  ``broad-lined'', ``Pec'' for ``peculiar'', a ``:'' means the
  Type is somewhat uncertain, and ``?'' means the SN Type is unknown,
  because no optical spectrum was available.  We do not include the
  unknown-Type SNe in either our I b/c or II groups.}
\tablenotetext{b}{The (luminosity) distance to the SN, derived from the
  NED database (see text for details).}
\tablenotetext{c}{The explosion date, $t_0$, is taken from the
  literature.  If the maximum-light time is known, but there is no
  other estimate of the explosion date, we take $t_0$ to be two weeks
  prior to maximum light.  If maximum light time is also not known we
  use the discovery date for $t0$, in most of these cases, the radio
  observations occur only several months later and the exact value of
  $t0$ will have relatively little effect.}
\tablenotetext{d}{The number of measurements refers to those used
  in this work.  For each SN we picked one of 4-8 GHz (C-band) or 8-12
  GHz (X-band), whichever had more or better measurements, with the
  exception of SN~1987A where we picked 2.3 GHz.}
\tablenotetext{e}{References:
  1 \citet{Weiler+1986};
  2 \citet{Weiler+1991, Montes+2000};
  3  \citet{Weiler+1992b};
  4 \citet{Montes+1998};
  5 \citet{Weiler+1989};
  6 re-reduced archival data; 
  7 \citet{Yin1994};
  8 \citet{SoderbergPhD2007};
  9 \citet{vDyk+1996};
  10 \citet{SramekPW1984};
  11 \citet{PanagiaSW1986};
  12 \citet{vDyk+1998};
  13 \citet{Montes+1997};
  14 \citet{WeilerPS1990};
  15 \citet{SN86J-1, SN86J-3};
  16 \citet{Turtle+1987}; 
  17 \citet{vDyk+1993b, Williams+2002};
  18 \citet{vDyk+1993a};
  19 \citet{Soderberg+2006b};
  20 \citet{vDyk+IAUC6378};
  21 \citet{SN93J-2};
  22  measurements retrieved from the website of the late Kurt W. Weiler;
  23 \citet{Weiler+2011}; 
  24 \citet{Chandra+2009a};
  25 \citet{vDyk+IAUC6375};
  26 \citet{Stockdale+CBET1714};
  27 \citet{vDyk+IAUC6528};
  28 \citet{Bauer+2008};
  29 \citet{Lacey+1998};
  30 \citet{vDyk+IAUC7322};
  31 \citet{Kulkarni+1998, WieringaKF1999}; 
  32 \citet{Lacey+IAUC7336};
  33 \citet{Berger+2003};
  34 \citet{Alberdi+2006, Perez-Torres+2009b};
  35 \citet{Schinzel+2009}, interpolated between the measured 22~GHz and 5~GHz values;
  36 \citet{Stockdale+2004};
  37 \citet{SN2001em-1, SN2001em-2};
  38 \citet{Stockdale+2007, ChandraRB_IAUC7982};
  39 \citet{Ryder+2004};
  40 \citet{BergerKC2002};
  41 \citet{Beswick+IAUC8572};
  42 \citet{SN2003gk-VLBI};
  43 \citet{Soderberg+2005};
  44 \citet{Soderberg+2006e};
  45 \citet{Stockdale+IAUC8153};
  46 \citet{Beswick+IAUC8332};
  47 \citet{WellonsSC2012};
  48 \citet{NayanaCR2018};
  49 \citet{Marti-Vidal+2007};
  50 \citet{Kankare+2014};
  51 \citet{Stockdale+IAUC8603};
  52 \citet{Drout+2013}; 
  53 \citet{Smith+2017}, and Charles Kilpatrick, private communication;
  54 \citet{ChandraS_ATel1182};
  55 \citet{Dwarkadas+2016}; 
  56 \citep{Soderberg+2006c};
  57 \citet{Argo2007};
  58 \citet{Kelley+CBET495};
  59 \citet{Argo+ATel1084, SN2006gy-1, sn2006gy-2, sn2006gy-3};
  60 {\em this paper};
  61 \citet{Chandra+2012a};
  62 \citet{Stritzinger+2009};
  63 \citet{Salas+2013};
  64 \citet{SN2007gr-Soderberg};
  65 \citet{ChandraS_ATel1271};
  66 \citet{ChandraS_ATel1359};
  67 \citet{vdHorst+2011, Roy+2013};
  68 \citet{ChandraS_ATel1366};
  69 \citet{SN2008D-Nature,SN2008D-VLBI};
  70 \citet{ChandraS_ATel1410};
  71 \citet{ChandraS_ATel1409};
  72 \citet{Argo+ATel1469, Stockdale+ATel1439, Roming+2009};
  73 \citet{SoderbergC_ATel1470}; 
  74 \citet{Stockdale+ATel1452};
  75 \citet{ChandraS_ATel1869};
  76 \citet{Stockdale+ATel1477, Stockdale+ATel1484};
  77 \citet{Soderberg_ATel1811};
  78 \citet{Stockdale+ATel1915};
  79 \citet{Stockdale+ATel1883,   Stockdale+ATel1912};
  80 \citet{ChandraS_ATel1891};
  81 \citet{Marchili+2010, Brunthaler+2010b, Kimani+2016};
  82 \citet{Stockdale+ATel1915, Stockdale+ATel1925};
  83 \citet{ChandraS_ATel2351};
  84 \citet{SN2009bb-VLBI};
  85 \citet{Stockdale+ATel2016}; 
  86 \citet{ChandraS_ATel2358};
  87 \citet{ChandraS_ATel2070};
  88 \citet{Stockdale+IAUC9056};
  89 \citet{Margutti+2014a};
  90 \citet{ChandraS_ATel2335};
  91 \citet{Ryder+ATel2450};
  92 \citet{Romero-Canizales+2014};
  93 \citet{Corsi+2011};
  94 \citet{Chandra+ATel2532};
  95 \citet{Ryder+CBET2242};
  96 \citet{Sanders+2012};
  97 \citet{Margutti+2013b};
  98 \citet{vdHorst+ATel2612};
  99 \citet{Corsi+2012};
  100 \citet{Kasliwal+ATel2864};
  101 \citet{Chandra+2015};
  102 \citet{Smith+2012b};
  103 \citet{Kasliwal+ATel3090};
  104 \citet{Kasliwal+ATel3093};
  105 \citet{Ryder+ATel3370};
  106 \citet{Krauss+SN2011dh-II, Horesh+2013b, SN2011dh_Alet};
  107 \citet{Horesh+ATel3512};
  108 \citet{Palliyaguru+2019};
  109 \citet{Milisavljevic+SN2011ei};
  110 \citet{Ryder+ATel3764};
  111 \citet{Bufano+2014};
  112 \citet{Chakraborti+2013};
  113 \citet{Chakraborti+2015_SN2012ap};
  114 \citet{Kamble+2014c};
  115 \citet{Yadav+2014};
  116 \citet{Horesh+2013a};
  117 \citet{Kamble+2016_SN2013df, Perez-Torres+ATel8452};
  118 \citet{Yaron+2017};
  119 \citet{Drout+2016};
  120 \citet{KambleS_ATel5154,Horesh+ATel5198};
  121 \citet{Margutti+2017_SN2014C, SN2014C_VLBI};
  122 \citet{Marongiu+2019};
  123 \citet{SNE2014_ATel};
  124 \citet{Kamble+ATEL6724};
  125 \citet{Chandra+2019};
  126 \citet{Shivvers+2017};
  127 \citet{Ryder+ATel7762};
  128 \citet{Milisavljevic+2017};
  129 \citet{Bostroem+2019};
  130 \citet{Kamble+ATel7845};
  131 \citet{HancockHATel8504};
  132 \citet{Ryder+ATel8815};
  133 \citet{Ryder+ATel8836};
  134 \citet{Kamble+ATel8911};
  135 \citet{Argo+ATel9147, Terreran+2019};
  136 \citet{Ryder+ATel9475};
  137 \citet{Jencson+2018};
  138 \citet{Ryder+ATel10147};
  139 \citet{Argo+ATel10421, Argo+ATel10472};
  140 \citet{Bannister+ATel10660};
  141 \citet{CorsiHK_ATel11295};
  142 \citet{Ho+2020b};
  143 \citet{Dobie+2018a, Dobie+2018b, Dobie+2018c, Margutti+2019_AT2018cow}
  144 \citet{Ho+2019c};
  145 \citet{Ryder+ATel12373};
  146 \citet{Ryder+ATel12820};
  147 \citet{Jacobson-Galan+2020};
  148 \citet{KunduR_ATel13040};
  149 \citet{Ryder+ATel13136};
  150 \citet{Kundu+ATel13477};
  151 \citet{Horesh+2020};
  152 \citet{Ho+2020c};
  153 \citet{Ryder+ATel13642};
  154 \citet{Kundu+ATel13805}}
\tablenotetext{f}{SN~2001em and SN~2014C were initially classified as Type Ic 
  and Ib, respectively, but both developed the spectral characteristics of 
  a Type IIn later in their evolution.  Since the bright radio
  emission occurred at later times corresponding to the IIn spectra, we 
  classify both as IIn}
\end{deluxetable*}

\subsection{Observed Radio Lightcurves}
\label{slightcurve}

\begin{figure*}
  \centering
\includegraphics[width=\linewidth]{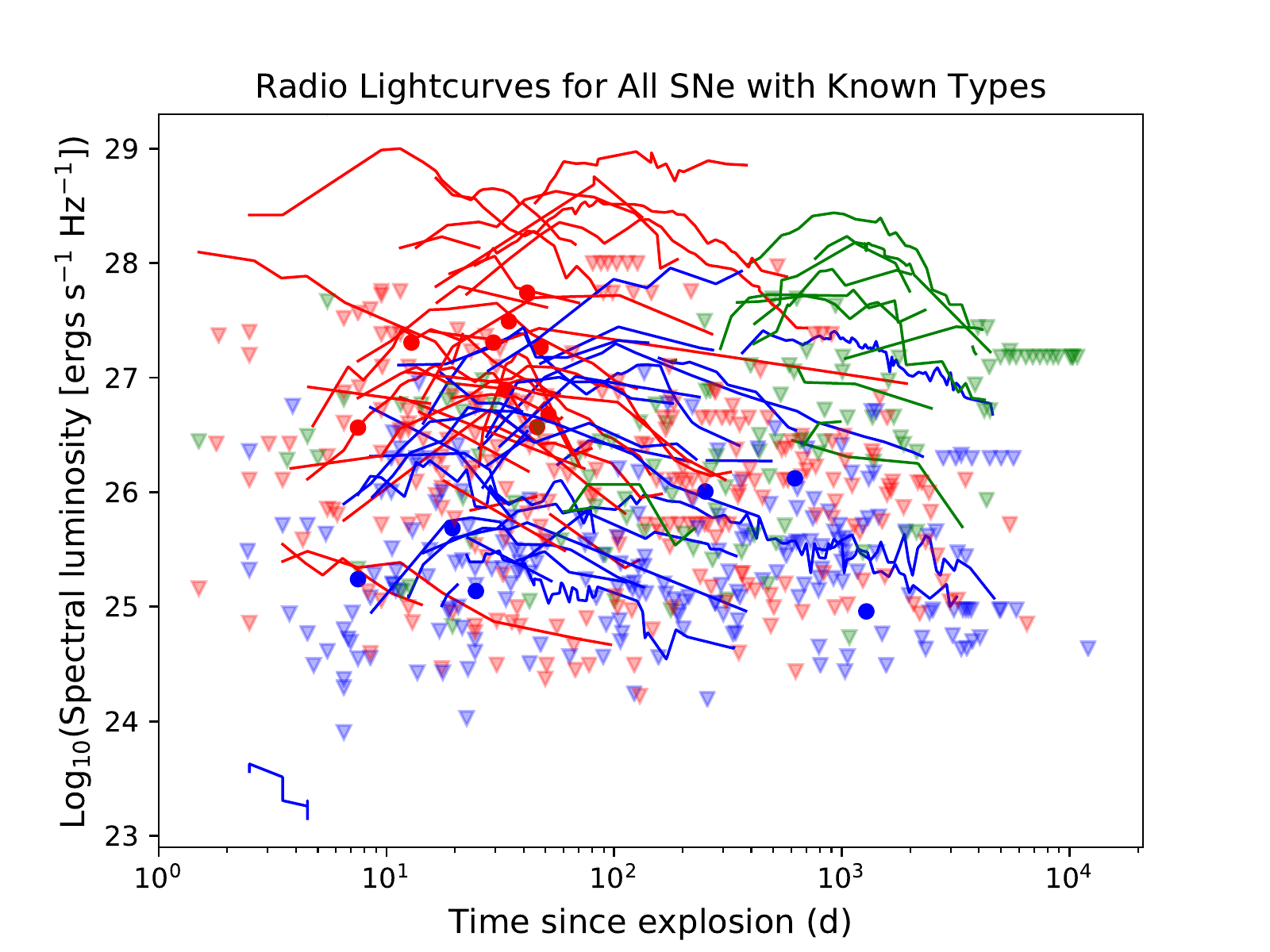}
\caption{A plot of the measurements and lightcurves for all the 289 SNe
  in our sample with known SN Types.  We plot the
  spectral luminosity against the time since the explosion, with both
  axes being logarithmic.  The 129 Type I b/c SNe are plotted in red.
  The 53 SNe of Types IIn are plotted in green, and 107 remaining Type
  II's in blue.  The lines are lightcurves in the case of multiple
  detections, while the round points are single detections, and the
  pale triangles show upper limits. The lines do not show any fit,
  they just connect the available measurements.  All measurements were
  between 4 and 10~GHz with the exception of SN~1987A.  For SN~1987A,
  which is the lowest-luminosity curve in the plot, there were only
  very few early measurements available above 2.3-GHz, and we
  therefore use the more complete 2.3~GHz lightcurve.}
\label{falllc}
\end{figure*}

\begin{figure*}
\centering
\includegraphics[width=\linewidth]{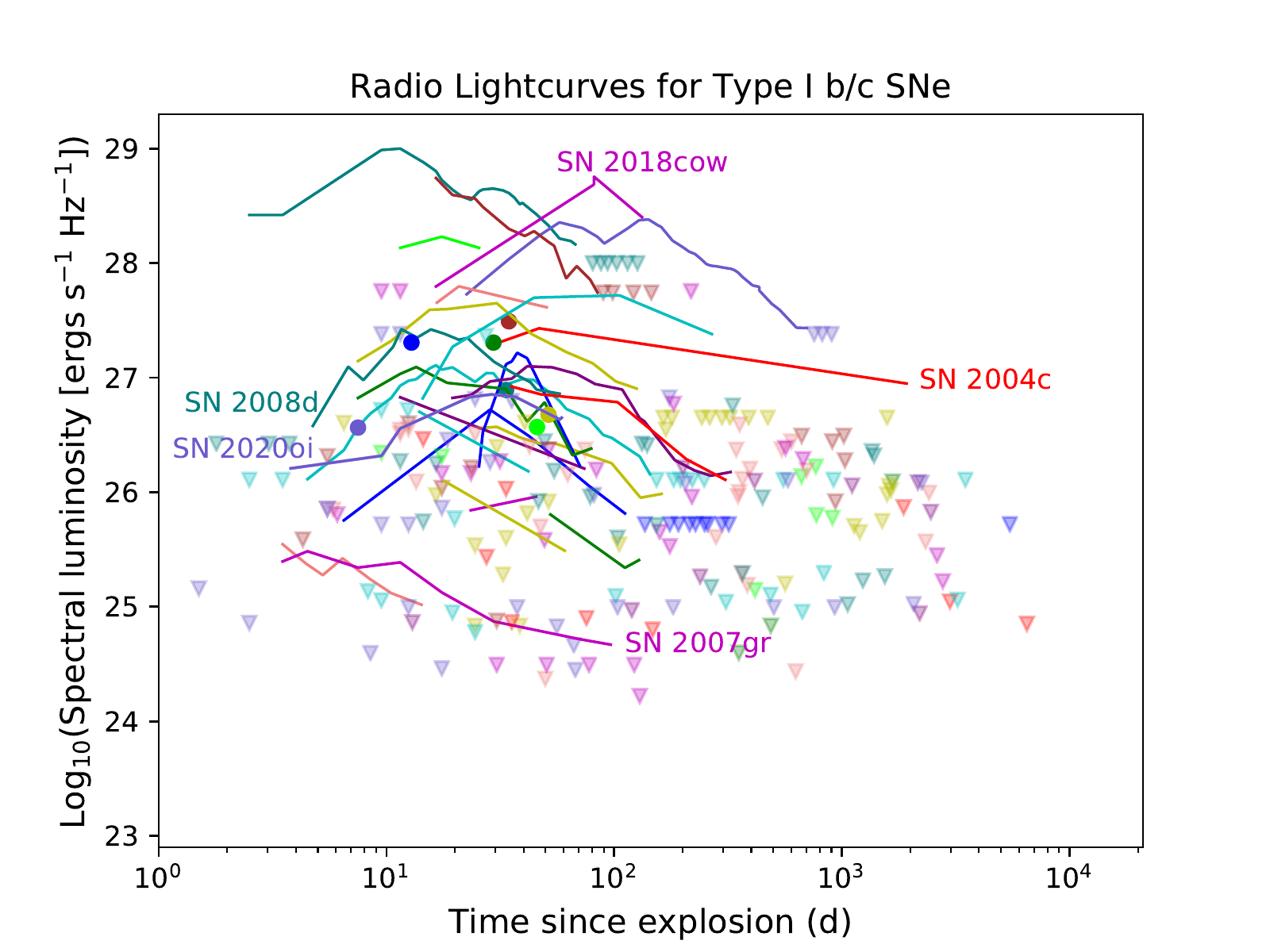}  
\caption{As Figure~\ref{falllc}, but showing only the 110 SNe of Type
  I b/c and showing only those at $D < 100$~Mpc.  Different colors are
  used for the different SNe.  The lightcurves for several SNe are
  labeled in the corresponding colors.}
\label{falllc_type1}
\end{figure*}

\begin{figure*}
\centering
\includegraphics[width=\linewidth]{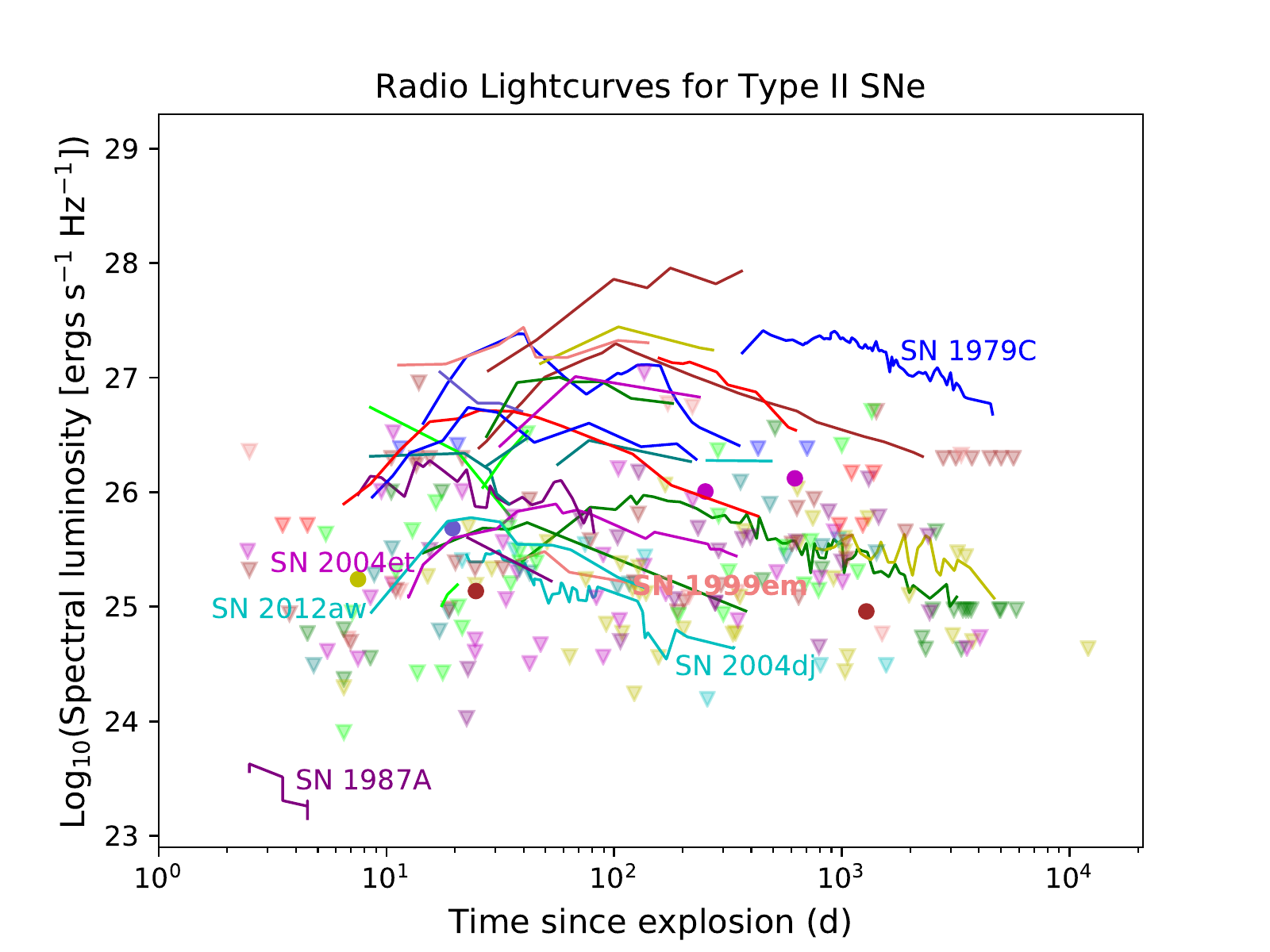} 
\caption{As Figure~\ref{falllc}, but showing only the 106 SNe of Type
  II (excluding Type IIn) at $D < 100$~Mpc.  Different colors are used
  for the different SNe.  The lightcurves for several SNe, including
  SN~1987A, are labeled in the corresponding colors.  The lightcurve
  for SN~1987A is at 2.3~GHz, unlike all the others which are at 4 to
  10~GHz. }
\label{falllc_type2}
\end{figure*}

\begin{figure}
\centering
\includegraphics[trim=0.2in 0 0.4in 0.2in, clip, width=\linewidth]{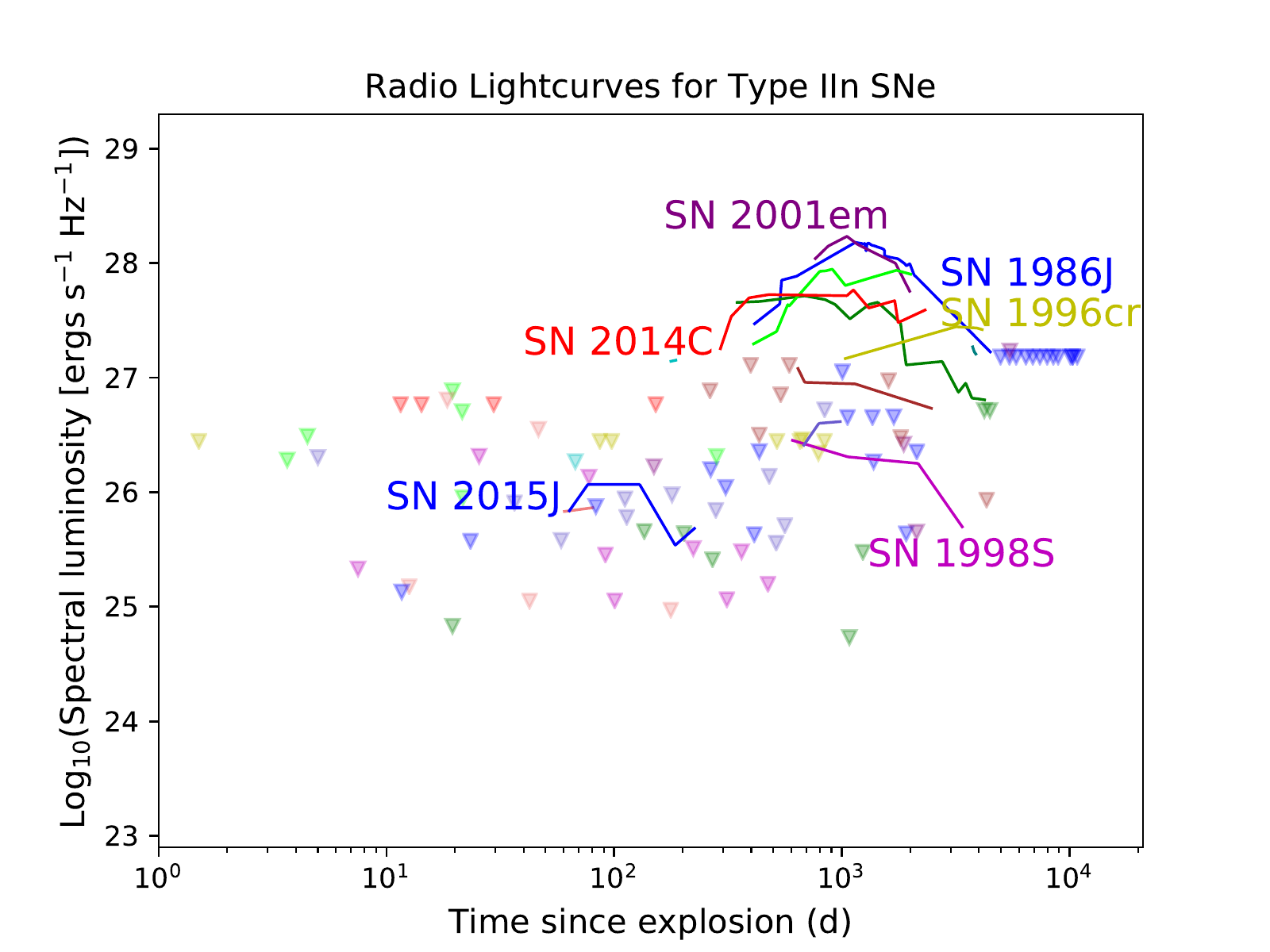} % checked, is RSN_master_v5.6 (no change in v5.63)
\caption{As Figure~\ref{falllc}, but showing only the 41 SNe of Type
  IIn at $D < 100$~Mpc.  Different colors are used for the different
  SNe.  The lightcurves for a few SNe are labeled in the corresponding
  colors.  Since we treat values below 10\% of the peak as upper
  limits (see Section~\ref{sdata}), the trailing part of the SN~1986J
  lightcurve is indicated here as upper limits, though in fact the
  spectral luminosities are well measured \citep{SN86J-4}.}
\label{falllc_IIn}
\end{figure}

\begin{figure}
\centering
\includegraphics[trim=0.2in 0 0.4in 0.2in, clip, width=\linewidth]{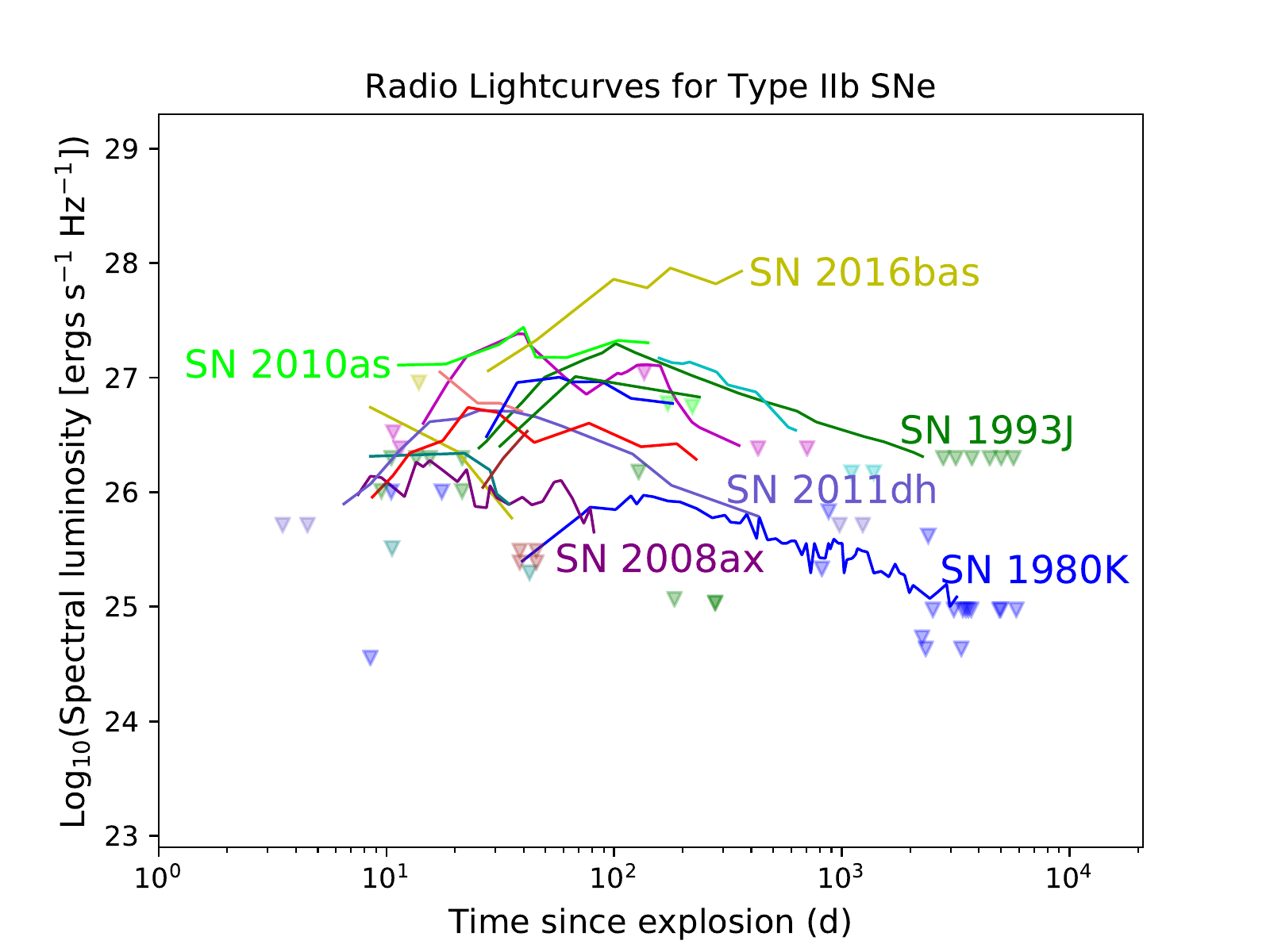} % 
\caption{As Figure~\ref{falllc}, but showing only the 19 SNe of Type
  IIb at $D < 100$~Mpc.  Different colors are used for the different
  SNe.  The lightcurves for several SNe are labeled in the
  corresponding colors.}
\label{falllc_IIb}
\end{figure}

\begin{figure}
  \centering
  \includegraphics[trim=0.2in 0 0.4in 0.2in, clip, width=\linewidth]{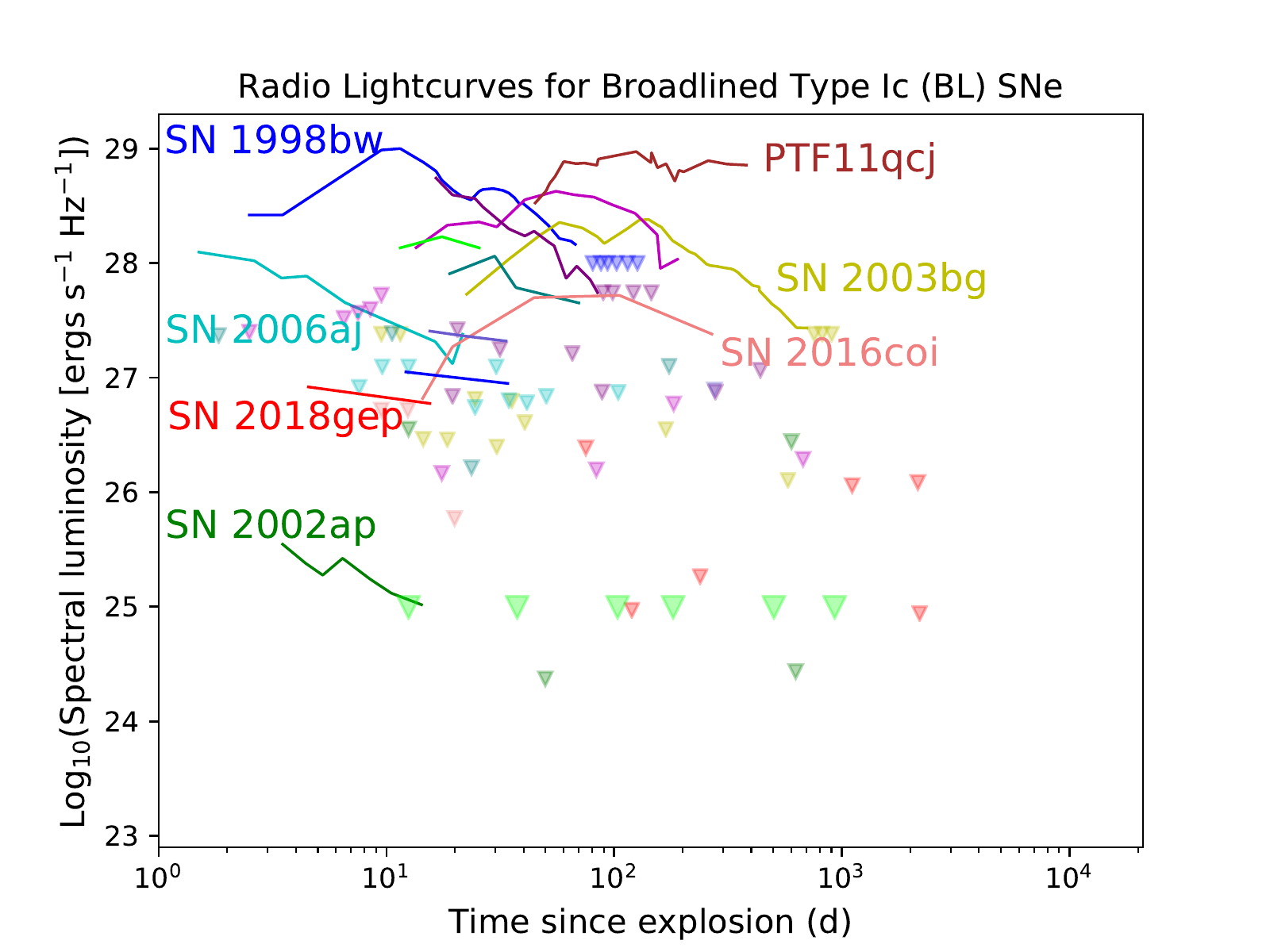}
  \caption{As Figure~\ref{falllc}, but showing only the 27 broad-lined
    (BL) SNe (including those at $D > 100$~Mpc).  Different colors are
    used for the different SNe.  We highlight the limits for SN~2014ad
    \citep{Marongiu+2019} with larger, lime-green triangles, since for
    that SN the measurements constrain the radio emission to low
    levels over a wide range of times.}
\label{falllc_BL}
\end{figure}

We plot the observed values in the form of radio lightcurves (i.e.,
spectral luminosity curves), including any upper limits, for all our
SNe with known Types in Figure~\ref{falllc}.  \newstuff{We then also
  separate the SNe by Type, and restrict our sample to those SNe at
  $D < 100$~Mpc (except as noted below), and plot values for Type~I
  b/c SNe in Figure~\ref{falllc_type1}, \newstuff{those for Type II
    SNe (excluding IIn's) in}\ Figure~\ref{falllc_type2}, and those
  for Type IIn SNe in Figure~\ref{falllc_IIn}. 

  The subtype IIb seem to have brighter radio emission than
  the remainder of the Type II's, and we plot the Type IIb's
  separately from the other Type II's in Figure~\ref{falllc_IIb}.
  
  Finally we plot the values for the ``broad-lined'' (BL) TYpe Ic SNe
  separately from the remainder of the Type I b/c in
  Figure~\ref{falllc_BL}.  Since SNe-BL are rare, and we have only 6
  detected examples at $D < 100$~Mpc, we plot all 27 BL SNe in our sample,
  regardless of $D$.}

A number of things are apparent from these figures.  First, it can be
seen that the lightcurves vary over a large range.  \Lpk\ can vary
over more than 5 orders of magnitude, from $\sim \! 10^{29}$~\ergsHz\
for SN~1998bw \citep{Kulkarni+1998}, which is
associated with GRB~980425, and SN~2009bb \citep{SN2009bb-VLBI}, to
$<10^{24}$~\ergsHz\ for SN~1987A \citep{Turtle+1987}.  Similarly, some
SNe, such as SN~1987A peak at \tpk\ $\lesssim 2$~d, while others such
as SN~1986J have $\tpk > 1000$~d \citep{SN86J-1}, almost 3 orders of
magnitude larger.

It can also be seen that the lightcurves exhibit a wide variety of
forms.  While generally they do show an initial rise and a subsequent
decay of approximately power-law form, various ``bumps'' and changes
in the slope of the power-law decay are seen.

\newstuff{Figure~\ref{falllc} shows that Type I b/c (red) reach the
  highest peak luminosities, followed by the Type IIn (green), while
  those of Type~II SNe (blue) are lower.  Type~I b/c's are more likely
  to peak earlier, while the Type~II's are likely to peak later and
  the Type IIn even later}.  This pattern has been noted earlier, for
example in \citet{Weiler+2002}, but with only a relatively small
sample of SNe.  While we only have a single example detected at a low
value of $\Lpk < 10^{24}$~\ergsHz\ (SN~1987A, at $D$ only
$\sim$50~kpc), which was of Type II, the distribution of upper limits
for Type~I b/c SNe is not obviously different than that for Type~II's,
implying that low values of $\Lpk < 10^{25}$~\ergsHz\ likely occur for
both Type I b/c and II SNe.  \newstuff{Type IIb SNe tend to have
  higher values of \Lpk\ than the remainder of the Type II's, and are
  therefore more likely to be detected.  The Type Ic-BL SNe also tend
  to have high values of \Lpk\, but note that some Ic-BL SNe, such as
  SN~2002ap and SN~2014ad have fairly low values of
  $L_\nu \lesssim 10^{25.5}$ \ergsHz.}

\section{Lightcurve Modeling}
\label{sfit}

\subsection{The Model}
\label{smodel}

As mentioned, in our model for the lightcurves, the spectral
luminosity, $L_\nu$, of the SN rises to a peak, and then decays
in a power-law fashion with $L_\nu \propto t^\beta$, where we take
$\beta = -1.5$.  Our model has only two free parameters,
\tpk, the time from the explosion to reach the peak, and
\Lpk, the peak spectral luminosity.

The rise in the lightcurve is caused by an optical depth, $\tau$,
which decreases as a function of time. \newstuff{This optical depth
  could be due to either external free-free absorption or internal
  synchrotron self-absorption, or a combination of the two}.  The peak
in the lightcurve occurs approximately when $\tau = 1$. We take
$\tau \propto t^{-\delta}$ and $\delta = 1$, which is a value which
fits most SNe moderately well, although we explore different rise
parameterizations in Section \ref{sdiscussMdot} below.

We fixed the slope of the power-law decay at $\beta = -1.5$ for all
SNe.  Different well-observed SNe do in fact show different values of
$\beta$: For example, SN~1993J has a flatter decay particularly during
the first $\sim$1000~d \citep[Figure~\ref{fsn93j} and][]{SN93J-2}, while
SN~1986J shows a steeper decay \citep{SN86J-4}.  However, for our
purposes, an average value of $\beta = -1.5$ gives a reasonable fit
near the peak of the lightcurve.

Our model lightcurve, normalized so that it reaches \Lpk\ at \tpk,
therefore has the form
$$ L(t) = \Lpk \cdot 4.482 \cdot e^{-1.5 (t_{\rm pk}/t)} \cdot (t/\tpk)^{-1.5}.$$

As can be seen in Figure~\ref{falllc}, the lightcurves of individual
SNe are often more complex than our simple model.  However, our model
gives an adequate fit to the peak in the lightcurve, and thus
serves our purpose here of providing an approximate, but sufficient,
parameterization of SN lightcurves in general.

While more complex models are certainly warranted for studying
individual SNe, and would likely yield more accurate values for \tpk\
and \Lpk, our purpose here is to examine the distribution of \tpk\ and
\Lpk\ over all SNe, so the approximate values obtained from our simple
model are adequate.  In particular, the fitted distributions of \tpk\
and \Lpk\ depend only very weakly on the choice of parameterization
for the rise and fall of the lightcurve, so even in cases where the shape
of the actual lightcurve differs from the model, our fitted values for
\tpk\ and \Lpk\ should be adequate to our purpose.

In cases where we have many measurements, clearly those near to \tpk\
provide the best constraints on \tpk\ and \Lpk\@.  Values that are
either much earlier or much later than \tpk\ and well below \Lpk\
provide little additional constraint on \tpk\ and \Lpk, and could
drive the fitted values to deviate from the peak in the actual
lightcurve in cases where our model is not a good match for the actual
lightcurve shape.  To minimize this effect, for any given SN, we
downweight any measurements that are at $<10$\% of the observed peak
by treating them as upper limits.  Note that we downweight
measurements in this way only in cases where we have better
measurements available for the same SN, that is those with $>10\,\times$
higher flux density.  The effect of this is two-fold: firstly any
``bumps'' in the lightcurve that happen well below the peak have
little effect on our fitted values of \tpk\ and \Lpk, and secondly, it
serves to smooth the likelihood function in the \tpk-\Lpk\ plane
slightly, which reduces the effect of our relatively coarse sampling
in this plane.

An example of this can be seen in the case of SN~1993J, where the
slope of the decay changes.  Figure~\ref{fsn93j} shows the full set of
8.4-GHz measurements for SN~1993J, while the left panel of
Figure~\ref{flcmany} below shows the values that we used to fit \tpk\
and \Lpk\ in this case, with the flux-densities $< 10$\% of the peak
treated as upper limits.

\subsection{Estimates of \texorpdfstring{\tpk\ and \Lpk}{}}
\label{stpklpk}

For many of our SNe, particularly if only upper limits were obtained,
the measurements do not determine a unique set of values of \tpk\ and
\Lpk\@.  Instead, some ranges of values are allowed and others
excluded.  In order to establish the {\em distribution}\/ of \tpk\ and
\Lpk\ over our sample, we proceed in a Bayesian fashion as follows.
We define a 2-dimensional array of possible values of \tpk\ and
\Lpk. We choose logarithmically spaced values of \tpk\ and \Lpk\ in
view of the large range these quantities can take on.  Then, for each
SN, we calculate the likelihood of obtaining the flux-density
measurements for that SN as a function of \tpk\ and \Lpk\ (assuming
the distance given in Table~\ref{tsne}).  If the likelihood is
high for some particular pair of values \tpk\ and \Lpk, then a
lightcurve characterized by those values of \tpk\ and \Lpk\ represents
a good fit to the measurements of the spectral luminosity.

Some values of \tpk\ and \Lpk\ are un-physical: the frequency at which
the spectrum turns over due to synchrotron self-absorption (SSA)
depends only on the luminosity and size of the source.  A lower limit
on the size of the source can therefore be estimated from the
observing frequency and value of \Lpk\ \citep[see][]{ChevalierF2006}.
Assuming a spherical source, this size can be expressed as a radius,
which we call the SSA-radius, $r_{\rm SSA}$.  In the case that
absorbing mechanisms other than SSA are active, for instance free-free
absorption (FFA) in the CSM, the turnover frequency could be higher,
so the source could be larger, but not smaller than calculated
assuming only SSA, so $r_{\rm SSA}$, is a lower limit on the physical
radius.  The speed, $\vSSA = r_{\rm SSA}/\tpk$ is therefore a lower
limit on the source's expansion speed.  \newstuff{Projection effects
  do allow apparent velocities somewhat larger than $c$ in the case of
  relativistic SNe, as were observed in SN~2003dh / GRB 030329
  \citep{Pihlstrom+2007}, but highly superluminal values are not
  expected.  To exclude physically unlikely cases where highly
  relativistic expansion would be required, we therefore assign a
  likelihood of 0 to all points in the \tpk, \Lpk\ plane for which
  $\vSSA > 2c$.  Although we use a non-relativistic calculation for
  \vSSA, which will not provide accurate values when
  $\vSSA \gtrsim c$, our cut at $\vSSA > 2c$ should nonetheless serve
  to exclude the majority of the physically unlikely combinations of
  \tpk\ and \Lpk}\@.  (Indeed, there are no well-determined values of
\tpk\ and \Lpk\ in this part of the plane.)

\begin{figure*}
  \centering
\includegraphics[trim=1.2in 0 1.1in 0, clip, width=0.31\linewidth]{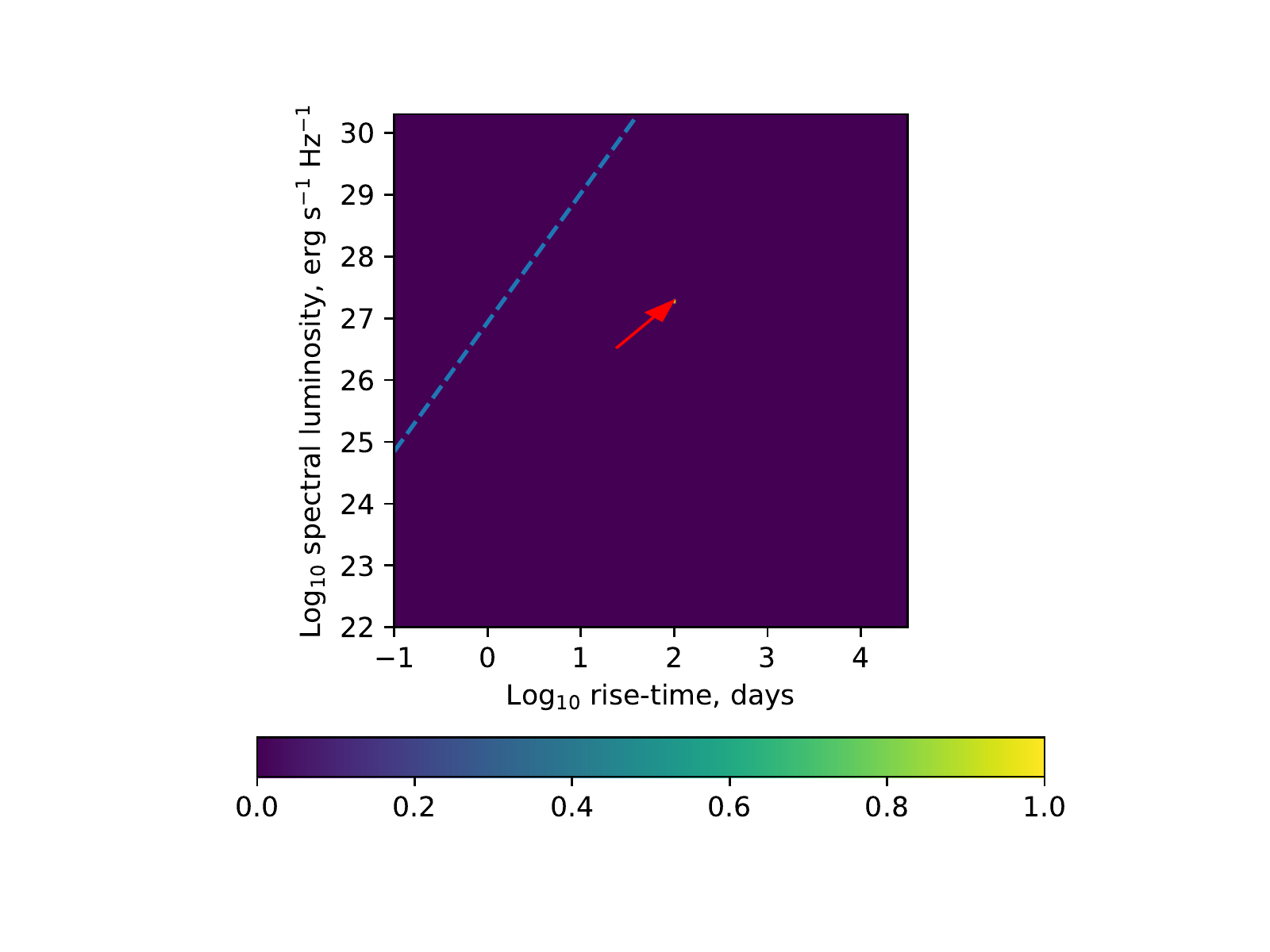}
\includegraphics[trim=1.2in 0 1.1in 0, clip, width=0.31\linewidth]{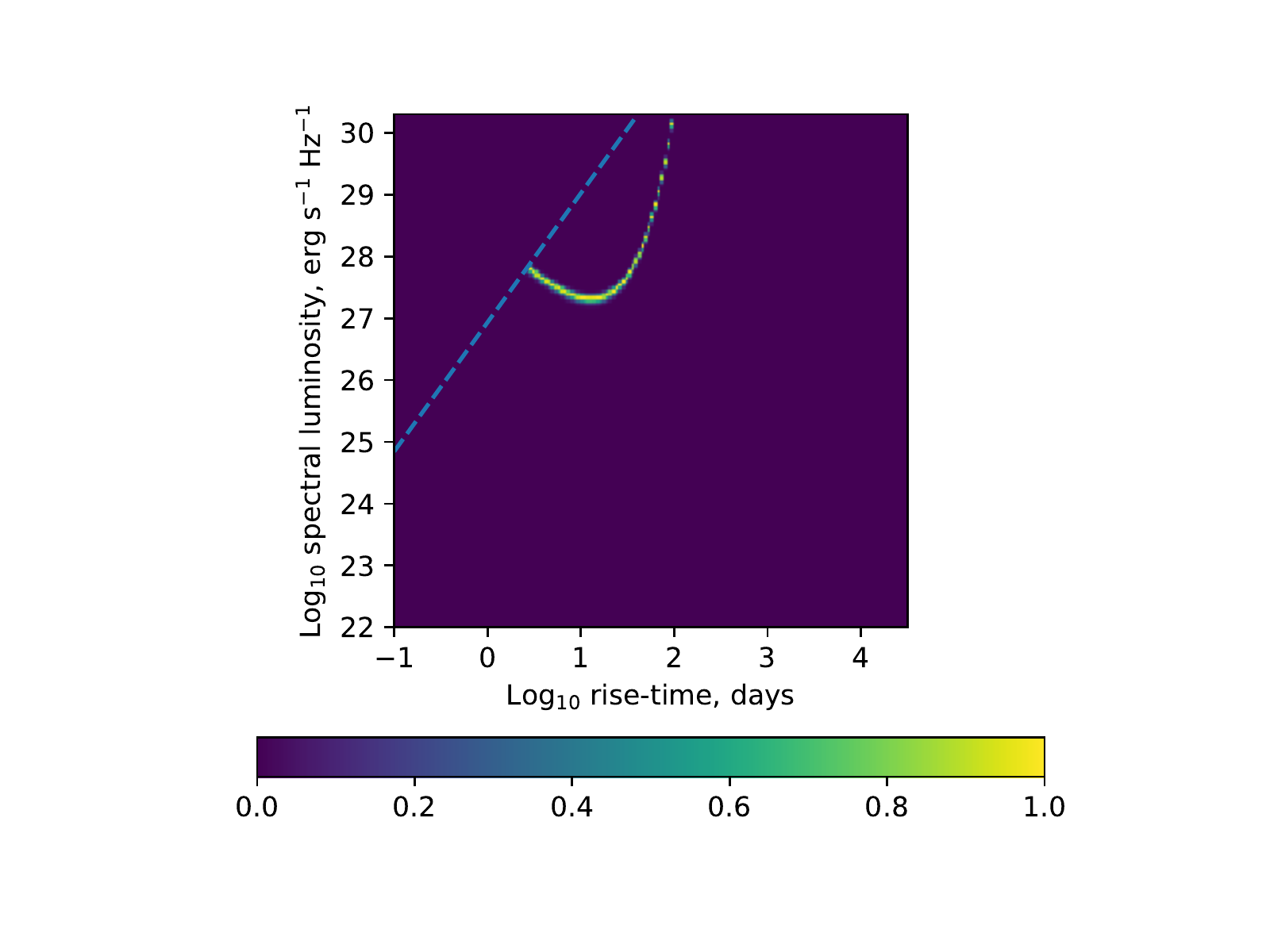}
\includegraphics[trim=1.2in 0 1.1in 0, clip, width=0.31\linewidth]{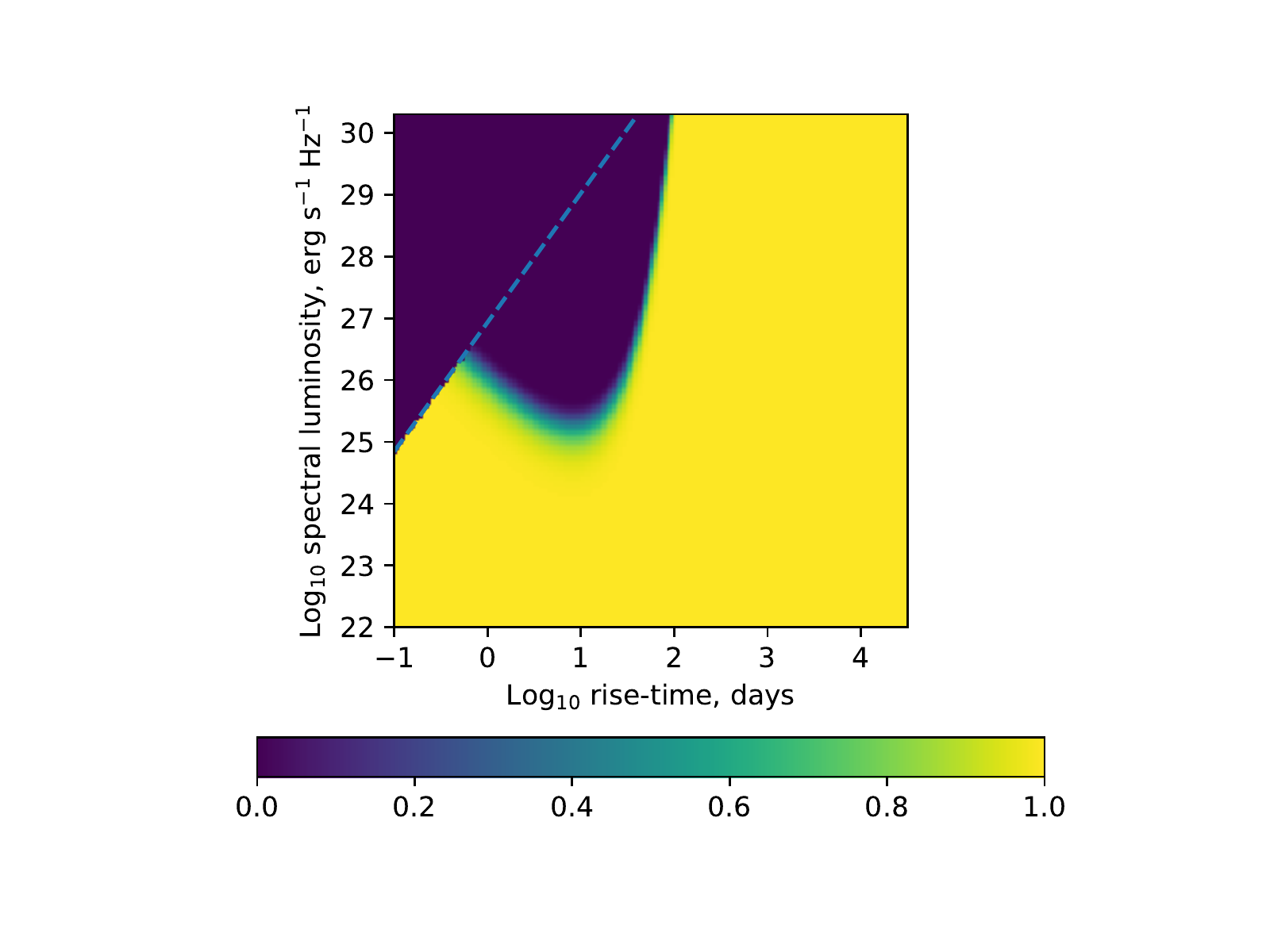}
\caption{Plots of the likelihood of pairs of \tpk\ and \Lpk\ values for three example SNe.
  The likelihoods are normalized to a maximum value of unity.
  The horizontal axis is \tpk, while the vertical one is \Lpk,
  and the likelihood is shown in color. The values of \tpk\ and \Lpk\
  that imply an apparent expansion speed $>2c$ are excluded, which
  results in the region above the dashed line, at the top left of the
  plots, always having zero likelihood.
  {\bf Left:} SN~1993J, for which many measurements tightly constrain
  the possible values of \tpk\ and \Lpk\ to a region smaller than
  our resolution in the \tpk, \Lpk\ plane, and thus to a single
  pixel in the image, which is indicated by the red arrow.
  {\bf Middle:} PSN J22460504-1059484, for which there was only a single measurement,
  but the SN was detected, thus constraining the possible locations in the
  \tpk-\Lpk\ plane to the thin curved line, occupying only a small part of the plane.
  The pixellation of the
  curved region is an artefact of our relatively low resolution in the
  \tpk, \Lpk\ plane, but should not significantly affect our results.
  {\bf Right:} SN~2017gax, for which a single measurement yielded only
  an upper limit to the flux density.  Many parts of the \tpk-\Lpk\
  plane are therefore almost equally likely.}
\label{fSNprob}
\end{figure*}

\begin{figure*}
  \centering
\includegraphics[trim=0.2in  0 0.5in 0, clip, height=1.9in]{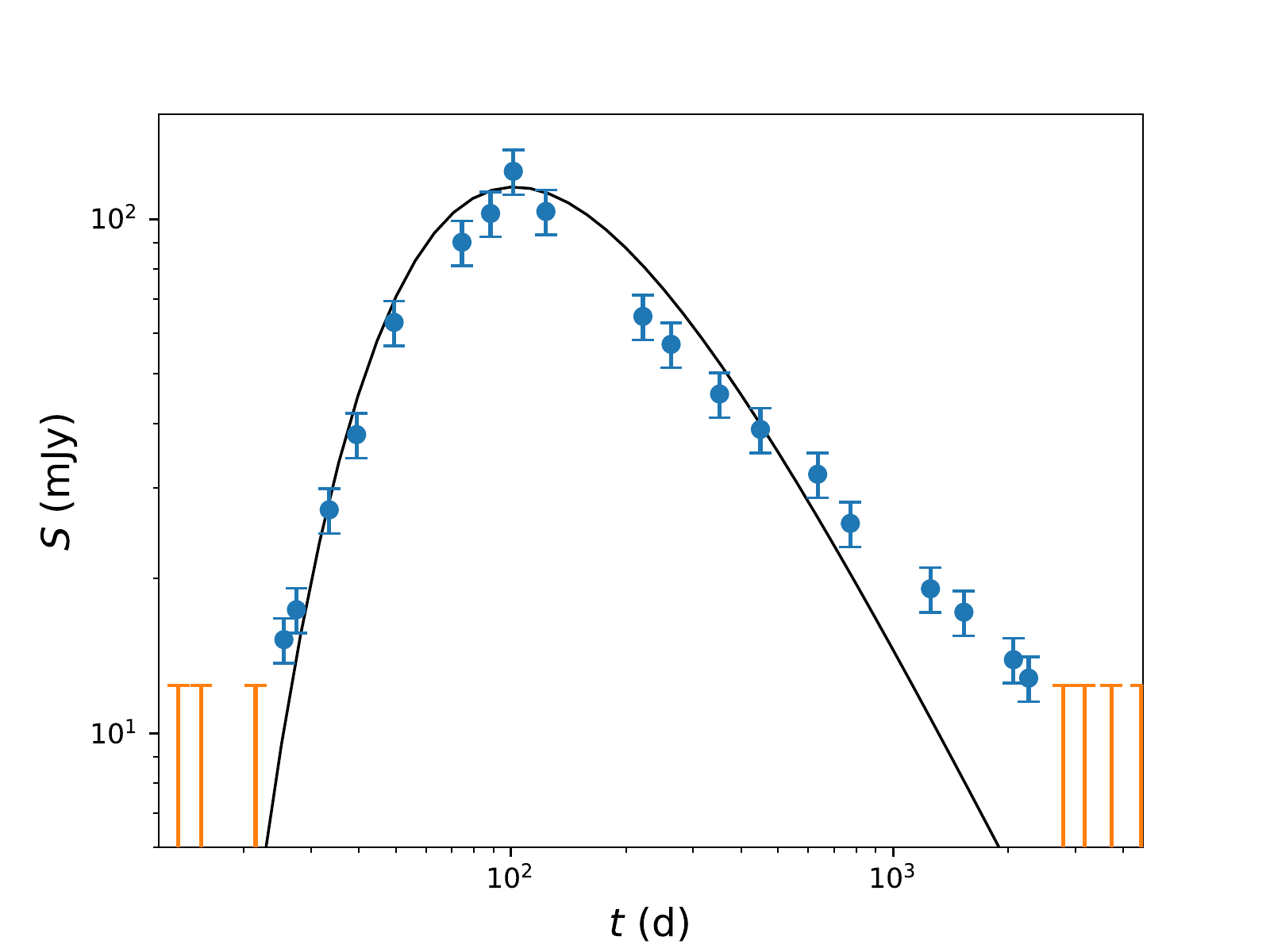}
\includegraphics[trim=0.39in 0 0.5in 0, clip, height=1.9in]{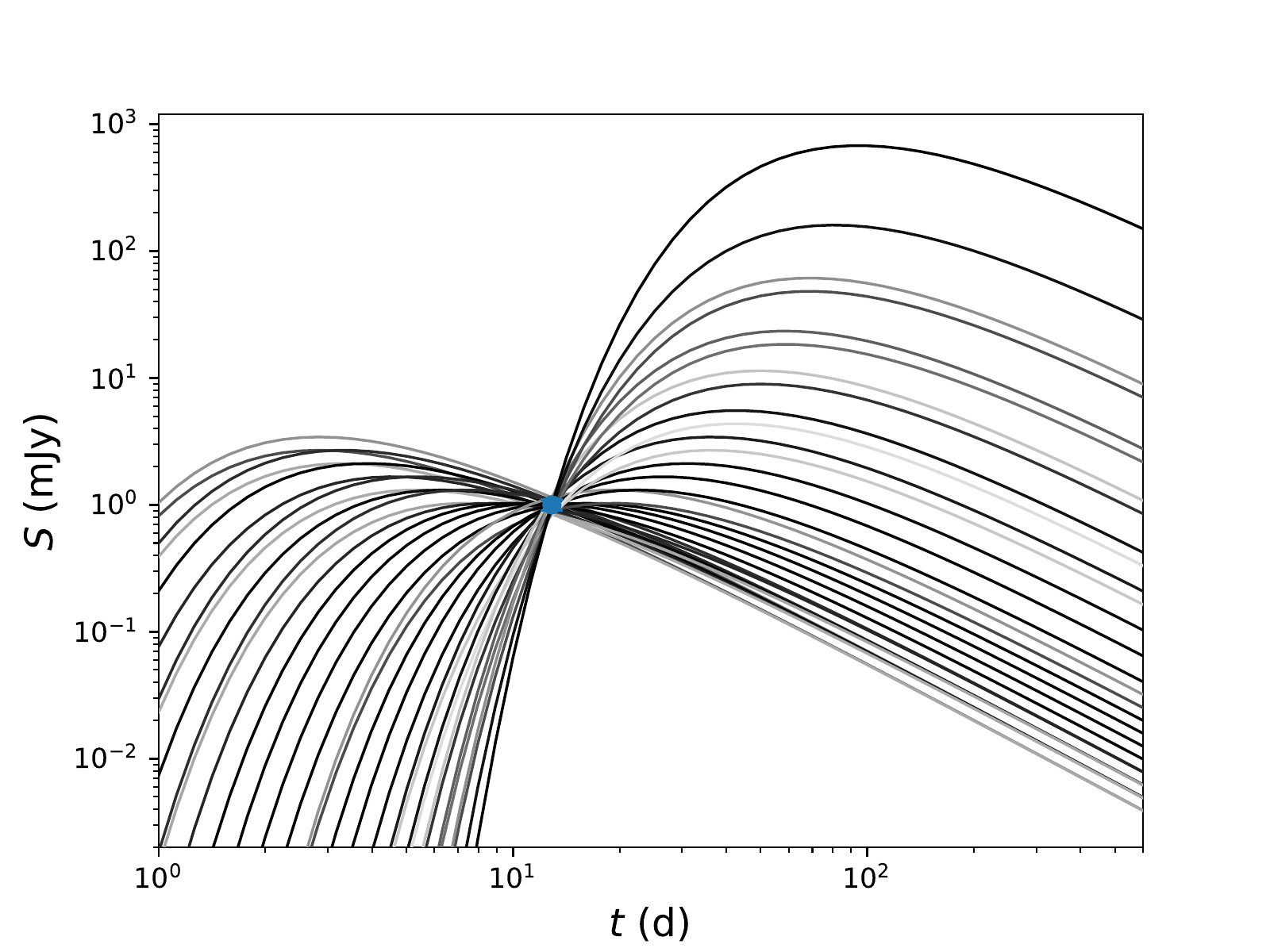}
\includegraphics[trim=0.39in 0 0.5in 0, clip, height=1.9in]{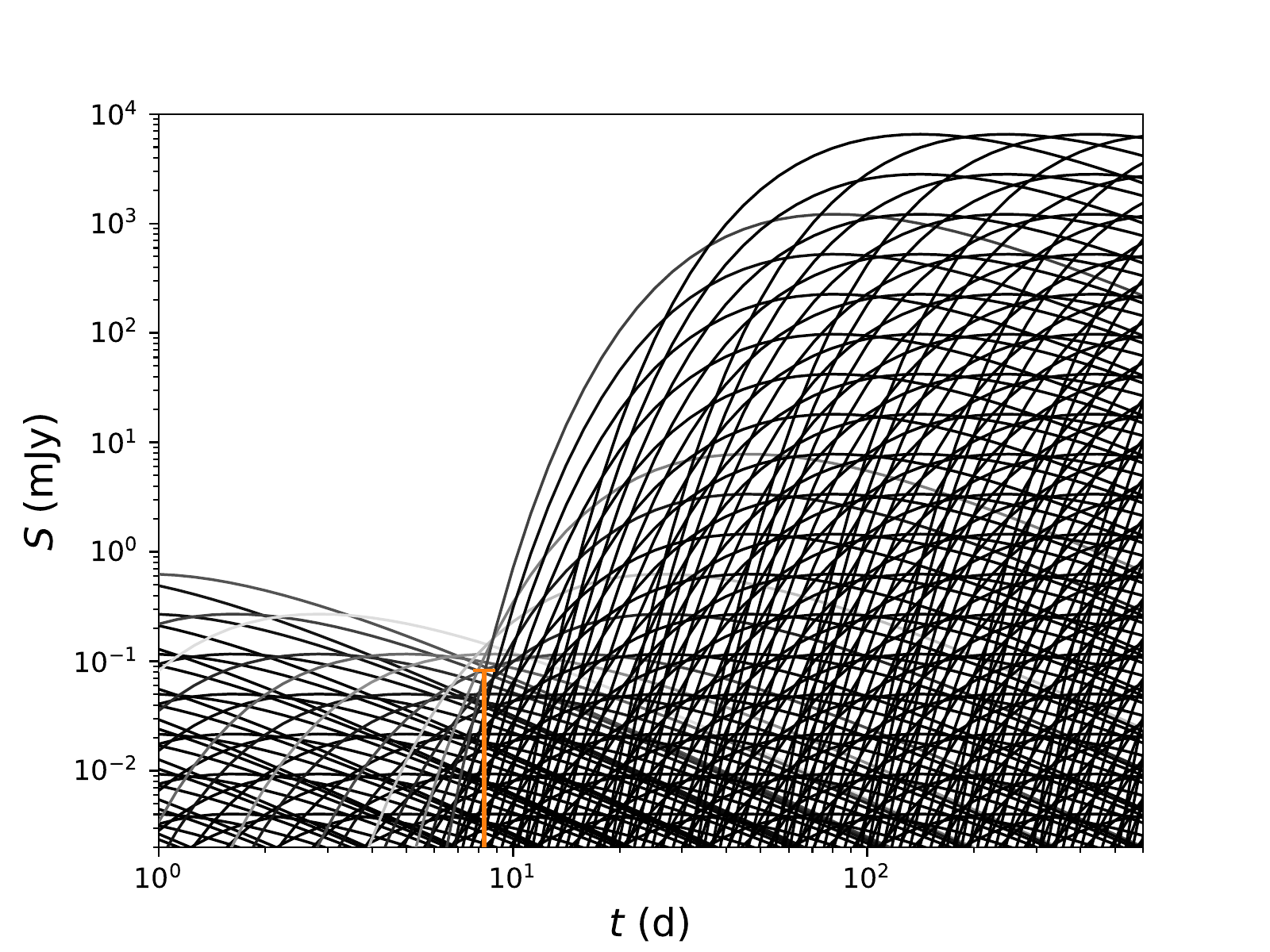}

\caption{Plots of the possible lightcurves for the three example SNe.
  The darkness of the line indicates the relative likelihood of the
  measurements for that particular lightcurve.  The errorbars indicate the
  $p = 68$\% ($1\sigma$) confidence limits in the case of both measured values (blue)
  and limits (orange).  The lightcurves (in mJy) are
  calculated using the distances given in Table~\ref{tsne}.
  {\bf Left:} SN~1993J\@.  The measurements (including the limits) have a high likelihood only
  for a single lightcurve defined by a particular set of \tpk, \Lpk\ values.
  In this case, one can see that modeled lightcurves do not
  match the measurements precisely, with the measurements suggesting a
  slightly slower rise, as well as a flatter decay, especially at
  $t > 1000$~d, than our simple two-parameter model.  However, the
  model reasonably reproduces the peak of the lightcurve.  Note also
  that the measurements plotted as lower limits here were in fact
  detections (see Figure~\ref{fsn93j}). As we explain in
  Section~\ref{sdata}, we treat all values below 10\% of the brightest
  observed value as upper limits so as to not unduly influence the
  fits near the peak.
  {\bf Middle:} PSN J22460504-1059484.  The measurements have a high likelihood
  for a range of related lightcurves, in some cases (with small
  values of \tpk) placing the single measurement during the rise, and
  in others (with larger values of \tpk) placing it during the decay.
  {\bf Right:} SN~2017gax. The measurements have a high likelihood for
  a wide range of lightcurves, but nonetheless some lightcurves,
  e.g., those having $S_{\rm peak} \gtrsim 0.1$~mJy and \tpk
  $\sim 5$~d, are excluded by the measurements.}
\label{flcmany}
\end{figure*}

We show three examples of these likelihood arrays in
Figure~\ref{fSNprob}, and three examples of the possible lightcurves
in Fig~\ref{flcmany}.

The first example is for a well-sampled case like SN~1993J
\citep[e.g.,][]{SN93J-2}, Figure~\ref{fSNprob} left.  The many
luminosity measurements allow for only one specific fit of our model,
which narrowly constrains the possible pairs of values of \tpk\ and
\Lpk\ and only one specific pair, corresponding to a single pixel in
the \tpk, \Lpk\ plane, has a significantly non-zero likelihood.  Only
a single lightcurve fits the measurements in Figure~\ref{flcmany} left.

The second example is for a supernova with only a single detection
like PSN J22460504-1059484 \citep{Kamble+ATel7845}, shown in
Figure~\ref{fSNprob} center.  In this case many lightcurves are
possible, all of them going through the sole luminosity measurement
but some having the measured luminosity on the rising part and some on
the falling part of the model lightcurve. In this case the allowed
pairs of values of \tpk\ and \Lpk\ are constrained to a thin curve.
A family of related lightcurves, all passing through the single
measurement, fit in Figure~\ref{flcmany} center.

The third example is for a case where only one single upper limit of a
luminosity measurement is available, like for SN~2017gax
\citep{Bannister+ATel10660}, shown in Figure~\ref{fSNprob} right.  Here
the range of lightcurves with high likelihood is the largest, with
many points in the \tpk-\Lpk\ plane having almost the same high
likelihood, but still a portion of the plane is excluded.  A range of
lightcurves, constrained only by having to go below the observed
limit, fit in Figure~\ref{flcmany} right.

\section{The Radio Luminosity-Risetime Function, or the Distribution of \texorpdfstring{\tpk\ and \Lpk}{}}
\label{sdistrib}

\subsection{The Distribution of the Observed Values of \texorpdfstring{\tpk\ and \Lpk}{}}
\label{sobsdist}

\begin{figure}
  \centering
  \includegraphics[width=\linewidth]{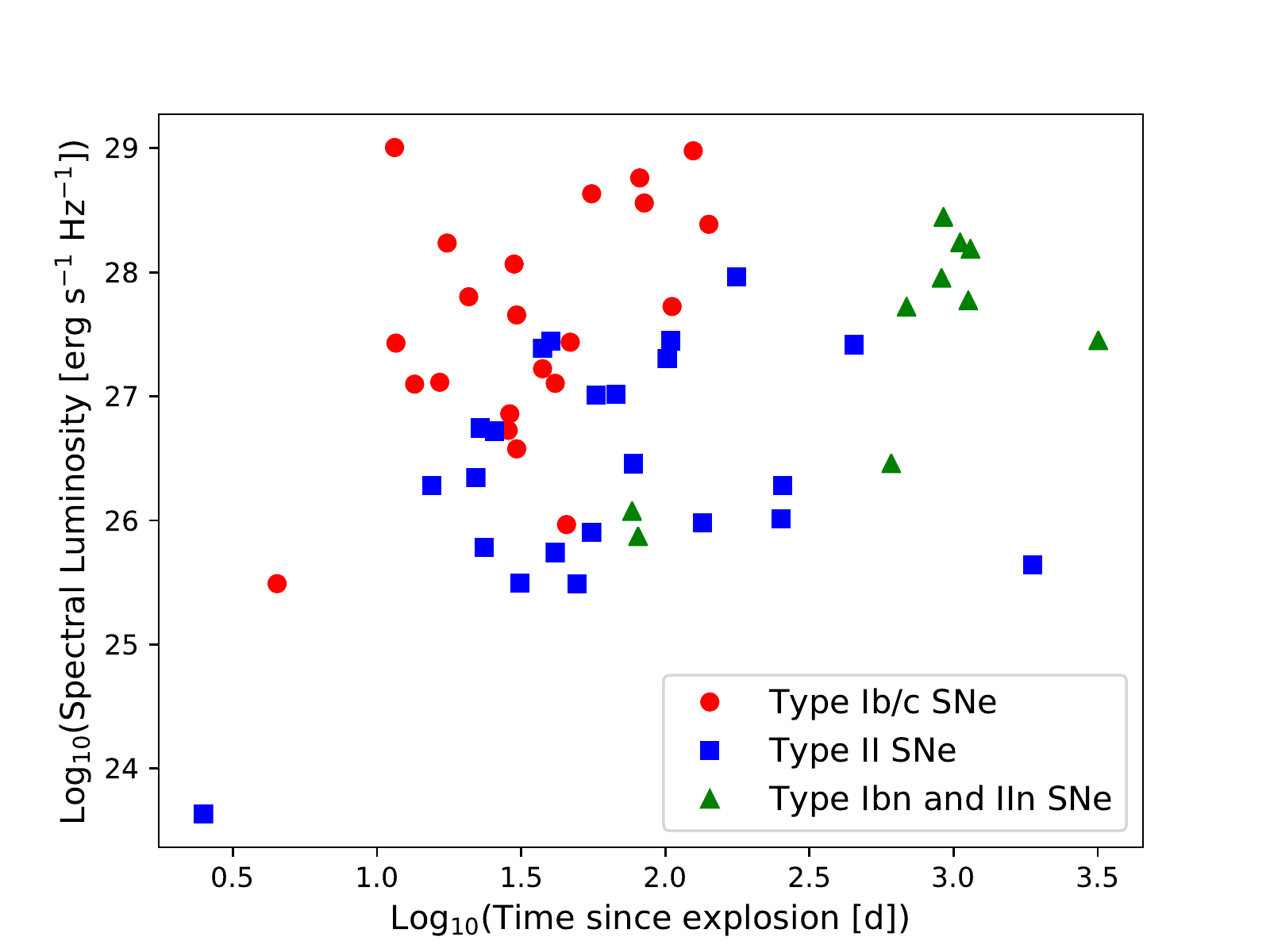}
  \caption{The observed values of the log$_{10}$ of the peak spectral
    luminosity, \Lpkobs, against the time in days at which it occurs,
    \tpkobs, for all 54 SNe for which these values were determined.
    These values are not derived from any lightcurve fit to the
    measurements, but for each SN are merely the largest value of
    $L_\nu$ that was observed and the time it was observed.  We do not
    plot SNe for which the highest observed value was the earliest (or
    only) one, since in those cases the peak cannot be determined.
    The interacting SNe, of Type IIn are shown as green
    triangles, while the remainder of the Type I b/c SNe are shown as
    red circles and the remainder of the Type II SNe are shown as
    blue squares.  The isolated square at the lower left corner is
    SN~1987A\@.  Many SNe for which only upper limits on the flux
    density could be determined would likely fall in the range below
    \Lpk\ $< 10^{25}$~\ergsHz.}
  \label{fpeakplot}
\end{figure}

We want to determine the distribution of \tpk\ and \Lpk, which is the
radio luminosity-risetime function for core-collapse SNe.  To guide
our investigation, we start first with the subset of SNe that have
well-determined values of \tpk\ and \Lpk, which is the subset of
examples similar to SN~1993J in Figures~\ref{fSNprob} and
\ref{flcmany}.  We adopt simple observational values of \tpk\ and
\Lpk\ here, where \Lpkobs\ is the $L_\nu$ corresponding to the highest
measured flux density, provided that the highest value was not either
the first or the last measurement, and \tpkobs\ is the time since the
explosion of that measurement.  Note that these observational values
of \tpkobs\ and \Lpkobs\ will generally not be identical to the values
of \tpk\ and \Lpk\ that have the highest likelihood from the previous
section, since the latter are influenced by all the measured values,
not just the single highest measurement.  However, the maximum
likelihood values of \tpk\ and \Lpk\ should be similar to \tpkobs\ and
\Lpkobs. We will return below to the fitted values of \tpk\ and \Lpk,
which are required for the majority of SNe for which \tpkobs\ and
\Lpkobs\ are not determined. First however, we plot a scattergram of
the observed values of \tpkobs\ and \Lpkobs\ in
Figure~\ref{fpeakplot}.  As already noted in Figure~\ref{falllc}, SNe
of Type I b/c (shown in red) tend to have higher values of \Lpk\ and
lower values of \tpk\ than do Type II\@.

\begin{figure}
  \centering
  \includegraphics[width=\linewidth]{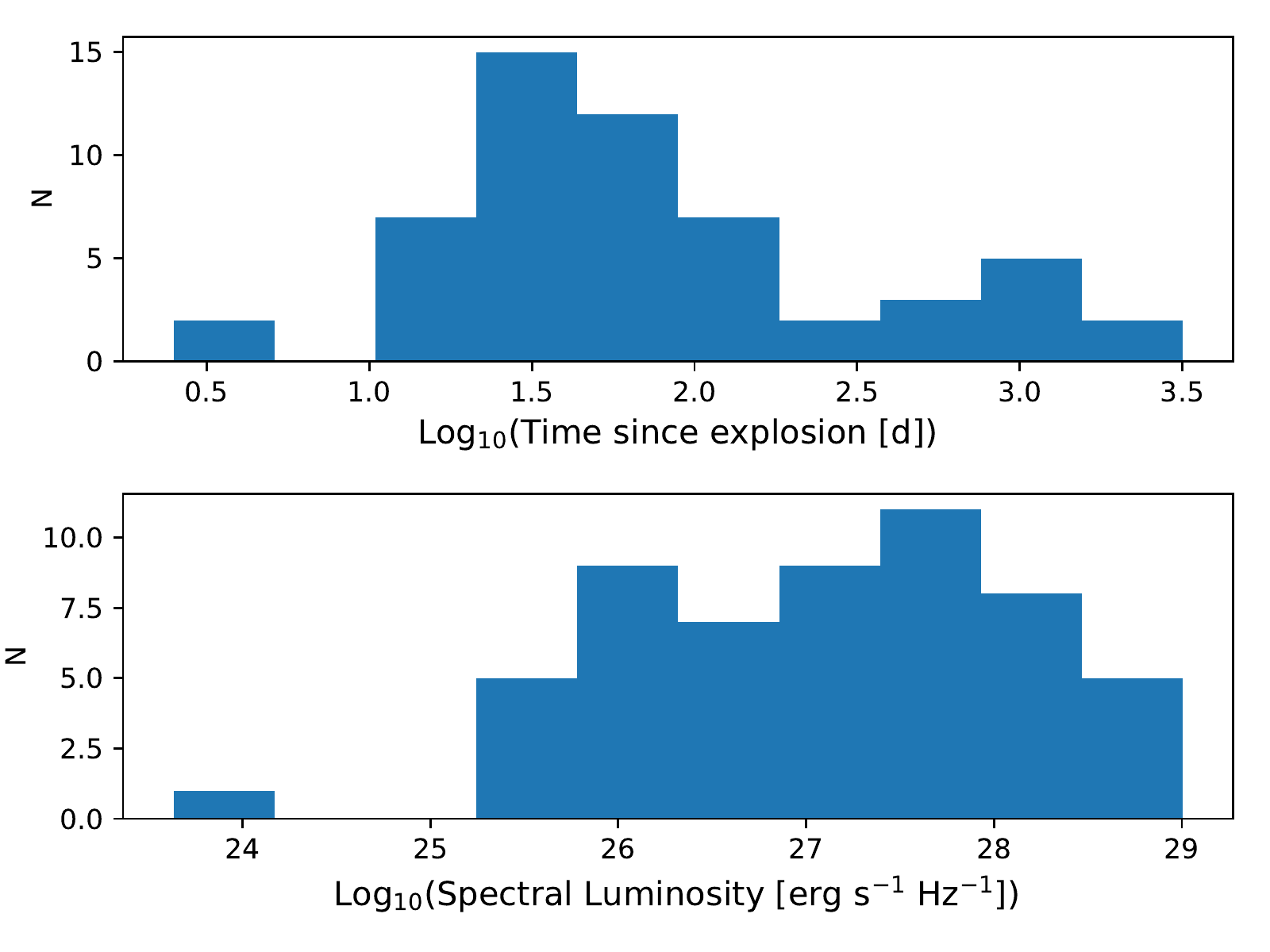}
  \caption{Histograms of the observed values of the log$_{10}$ of the
    time in days at which the observed peak occurred, \tpkobs, (top),
    and the spectral luminosity of that peak, \Lpkobs\ (bottom) for
    all 57 SNe for which these values were determined.  See
    Figure~\ref{fpeakplot} for the description of \tpkobs\ and
    \Lpkobs\@. These histograms represent only the population of
    detected, well-observed, SNe, and are not representative of the
    overall population, of which 69\% is not detected.  In particular,
    the distribution of \Lpkobs\ is strongly biased by exclusion of
    SNe for which only limits on $L_\nu$ were obtained.  The mean of
    \tpkobs\ was 1.88, and the standard devotion was 0.67, while the
    corresponding values for \Lpkobs\ were 27.09 and 1.09.}
  \label{fhistoobs}
\end{figure}

In Figure~\ref{fhistoobs} we plot the histograms showing distributions
of \tpkobs\ and \Lpkobs\@.  For both, the values are scattered
relatively uniformly in logarithmic space, suggesting that
parameterizing the distributions of \tpk\ and \Lpk\ in logarithmic
space. Only for 57 SNe, (19\% of our total of \NSNe), can the values
of \tpkobs\ and \Lpkobs\ be determined.

It is important to note that the histograms in Figure~\ref{fhistoobs}
represent only the population of well-observed, detected SNe, and are
not representative of the overall population at $D < 100$~Mpc, of
which 69\% was never detected and 80\% do not have well-defined
values of \tpkobs\ and \Lpkobs.

The most obvious bias is in the distribution of \Lpkobs: If one were
to take into account the 69\% of SNe for which only upper limits on
$L_\nu$ were ever obtained, many of them would be at
$\Lpk < 10^{25}$~\ergsHz, and the distribution of \Lpkobs\ must
therefore be biased towards higher values than the distribution of
\Lpk\ over all SNe.  Indeed, only for SN~1987A could a value of
$\lLpk < 25$ have been observed.

As far as \tpk\ is concerned, very few SNe are observed at all at
times $<1$~week, therefore many SNe could lie in the range
$\tpk < 10$~d, and the distribution in Figure~\ref{fhistoobs} may be
significantly biased here also.

\subsection{The Distribution Function for \texorpdfstring{\Lpk\ and \tpk}{} From All SNe}
\label{sdistall}

We now turn to incorporating the 80\% of our sample for which \tpkobs\
and \Lpkobs\ were not defined, which includes the 69\% of SNe for
which only upper limits on $L_\nu$ could be determined.  Although the
observations for these SNe do not determine \tpk\ or \Lpk\ uniquely,
they do provide some constraints on their possible values.  We
incorporate them by examining the likelihood of various values of
\tpk\ and \Lpk\ given the observations.

\begin{figure}
  \centering
  \includegraphics[trim=1.3in 0.5in 1.0in 0.3in, clip, width=\linewidth]{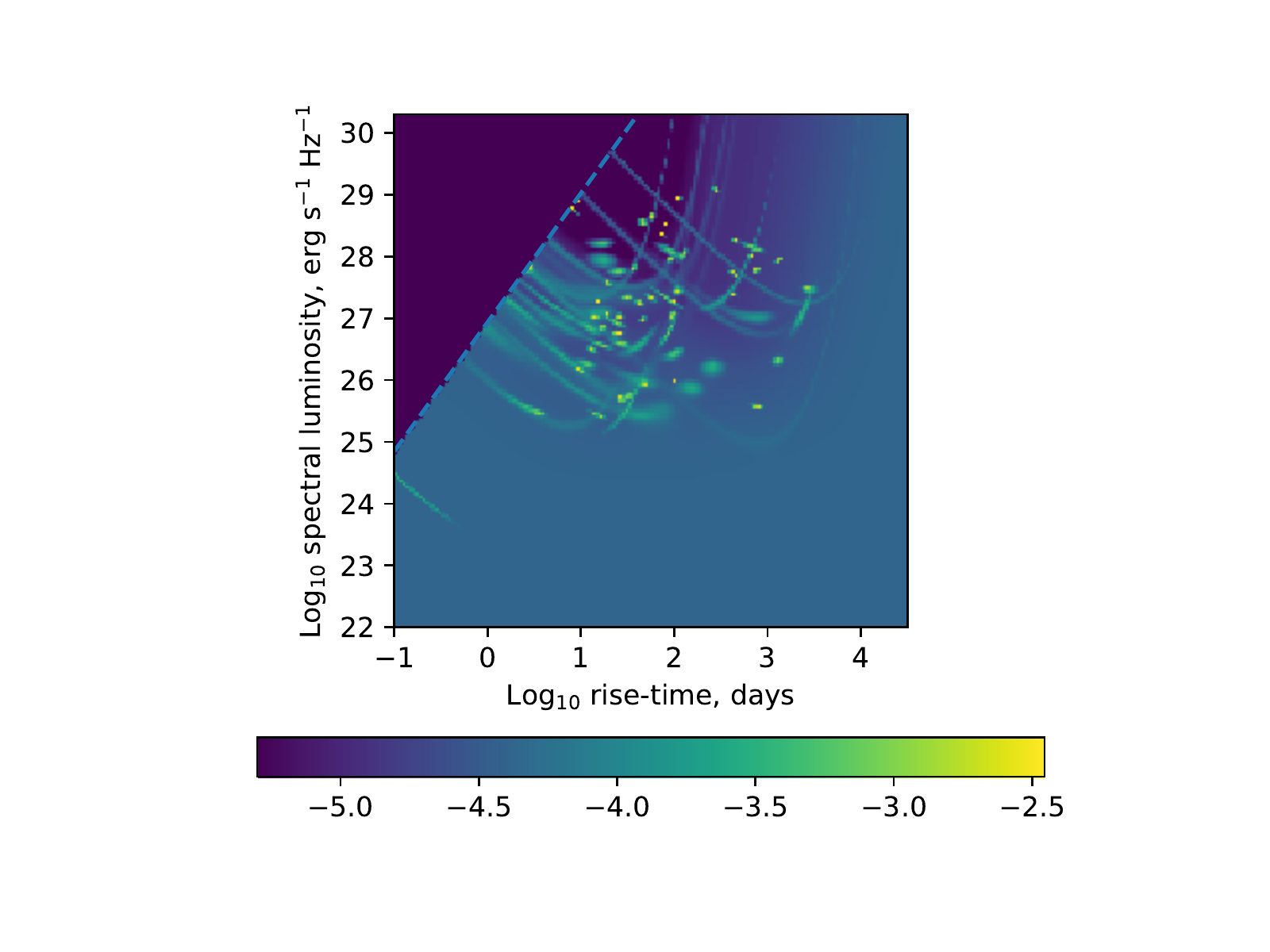}
  
  \caption{The logarithm of the probability,
    $\log_{10}[\ptot(\tpk, \Lpk)]$ of the measurements as a function
    of \tpk\ and \Lpk\ over all our SNe.  The maximum value of \ptot\
    is 0.0039, which is approximately $1/(N_{\rm SNe} = \NSNe)$.  For
    a small number of SNe (e.g., SN~1993J, Figure~\ref{fSNprob} left),
    the values of \tpk\ and \Lpk\ are well determined, and therefore
    some particular pair of values has $p_i \simeq 1$, thus
    contributing $1/N_{\rm SNe}$ to \ptot.  For other SNe, allowed
    values of \ptot\ lie on curved lines, whose peak values will be
    lower than $1/N_{\rm SNe}$ (since $p_i$ must sum to 1 over the
    whole image).  Finally, for SNe for which only upper limits are
    available, the maximum $p_i$ is lower still since it is spread out
    more or less uniformly over the bottom part of the \tpk-\Lpk\
    plane.  Although the probability for any particular \tpk\ and
    \Lpk\ in this region is low, the integral of \ptot\ below, say,
    $\Lpk = 10^{25}$~\ergsHz\ is substantial, so that the probability
    of $\Lpk < 10^{25}$~\ergsHz\ is not negligible. Again, the region
    with $\ptot = 0$ at the top left above the dashed line is excluded
    because it would require strongly superluminal expansion.}
  \label{fsumprob}
\end{figure}

In Section~\ref{stpklpk}, we calculated the likelihoods for each SN
for different pairs of values of \tpk\ and \Lpk, with examples being
shown in Figure~\ref{fSNprob}.  If we normalize these likelihood
functions, they become the probability, $p_i(\tpk, \Lpk)$, of SN
number $i$, having some particular pair of \tpk\ and \Lpk\ values (in
Bayesian terms, this is equivalent to incorporating a flat prior for
\tpk\ and \Lpk\ to form the posterior probability).  If we then sum
these arrays over all of our SNe and divide by our total number of SNe
(\NSNe), we arrive at the probability for particular pairs of values
of \tpk, \Lpk\ over all of our SNe, $p_{\rm tot}(\tpk, \Lpk)$. We show
$p_{\rm tot}(\tpk, \Lpk)$ in Figure~\ref{fsumprob}.

The probability of different values of \tpk\ and \Lpk\ is hard to
interpret from Figure~\ref{fSNprob}. On the one hand, there are a
small number of SNe that have well-determined \tpk\ and \Lpk\ (those
in Figure~\ref{fpeakplot} that produce a small number of
high-probability pixels in Figure~\ref{fSNprob}).  As mentioned, these
constitute an almost certainly biased subset of only 19\% of our
sample.  On the other hand, there are many SNe for which the sparse
measured values or limits mean that  large areas of the \tpk-\Lpk\
plane have low, but significantly non-zero probability.  Pairs of
\tpk-\Lpk\ values which are physically unlikely, such as \ltpk = 4,
\lLpk = 30, have non-zero probability because for many SNe they are
not excluded by the measurements.

To proceed we want to impose some reasonable constraints on the
distributions of \tpk\ and \Lpk, for example considering extreme
values unlikely even if they are allowed by our measurements.  So,
instead of attempting to estimate the probability distributions of
\tpk\ and \Lpk\ from Figure~\ref{fSNprob}, we will proceed by {\em
  hypothesizing}\/ some functional forms for the distributions.
Although there is no physical reason to expect that the values of either
\Lpk\ or \tpk\ are in fact drawn from any distribution with a
simple functional form, determining approximate forms of the
distributions of \tpk\ and \Lpk\ should prove useful until more
physically-motivated versions can be found, for example, for
estimating the likelihood of detecting future SNe in the radio.  It
also allows us to compare the distributions across different types of
SNe, and may also provide some insight into the physics of radio
emission from SNe.

We have noted in Section~\ref{sobsdist} that the values of both \tpk\
and \Lpk\ seem relatively uniformly scattered in logarithmic space. A
normal distribution in, for example, \tpk\ therefore seems
incompatible with the measurements, whereas a normal distribution in
$\log(\tpk)$, that is, a lognormal distribution in \tpk, could provide
a reasonable fit.  We will therefore mostly work with the logarithms
of \tpk\ and \Lpk, which we denote by
$\ltpk = \log_{10}(\tpk/{\rm d})$ and
$\lLpk = \log_{10}(\Lpk /[\ergsHz])$.

Since there is no strong correlation between the more probable values
of \tpk\ and \Lpk\ in Figure~\ref{fsumprob}, we consider only separate
distributions for \tpk\ and \Lpk, so the hypothesized joint
probability for \tpk\ and \Lpk\ can be obtained by multiplying their
respective hypothesized probability distributions.  This product would
be the anticipated radio luminosity-risetime function for
core-collapse SNe.

\subsection{Finding the Most Likely Distribution Function for \texorpdfstring{\Lpk\ and \tpk}{}}
\label{sfinddist}

We try therefore the following three forms for the distribution functions
for \ltpk\ and \lLpk:
\begin{trivlist}
\item{1}. A uniform distribution in $\log_{10}(x)$ where $x$ is either
  \tpk\ or \Lpk\@.  This distribution has two free parameters, namely
  the low and high limits, $x_{\rm low}$ and $x_{\rm high}$.  (With
  well determined values of \tpk\ and \Lpk\ the highest probability
  would be achieved by placing these limits at just below the smallest
  and just above the highest observed values.  However, given that our
  measurements do not uniquely determine \tpk\ or \Lpk\ in the
  majority of cases, the boundaries are flexible, and we determine the
  values of the limits that give the highest probability.)

\item{2}. A lognormal distribution in $x$, which is a normal
distribution in $\log_{10}(x)$.  This has also two free parameters, the
mean, $\mu$, and the standard deviation, $\sigma$, so the probability,
$p(x) = \frac{1}{\sigma \sqrt{(2\pi)}} e^{-0.5
  (\frac{x-\mu}{\sigma})^2}$.

\item{3}. A power-law distribution, where $p(x) = K x^q$ if
  $x > x_{\rm min}$ and $p = 0$ otherwise.  This distribution also has
  two free parameters, namely $q$ and $x_{\rm min}$.  Given that
  Figures~\ref{fpeakplot} and \ref{fhistoobs} suggest that both very
  small and large values of \tpk\ are unlikely, we consider the
  power-law distribution only for \Lpk, where the many lower limits
  means small values of \Lpk\ could be likely.
\end{trivlist}

In all cases we normalize the distributions over the ranges
$-1 < \ltpk < 10^{4.5}$ (0.1~d to 86 yr) and $22 < \lLpk < 30.3$
($10^{22}$ to $2\times10^{30}$~\ergsHz).

For each SN, we then multiply the likelihood function for \tpk\ and
\Lpk\ (Figure~\ref{fSNprob}) by the hypothesized joint distribution of
\tpk\ and \Lpk\@.  The integral of this product over all possible
values of \tpk\ and \Lpk\ then gives the likelihood of the
measurements for this SN for this particular hypothesized \tpk, \Lpk\
distribution. The likelihood of the measurements for all SNe given the
hypothesized distributions of \tpk\ and \Lpk\ is then the product of
the likelihoods for the individual SNe.

Our first goal is to determine which functional form, i.e.,
lognormal, uniform, or power-law, is most appropriate for \tpk\ and
\Lpk\@.  Since our sample is almost certainly notably incomplete at
larger distances, we use here only those SNe at $D < 100$~Mpc, where
our sample is more complete, retaining 262 SNe from our total of \NSNe.

We evaluate in a brute-force fashion the likelihood for each possible
value of the four free parameters over the two distributions (two in
\tpk\ and two in \Lpk; for example $\mu$ and $\sigma$ in the case of a
lognormal distribution). 
We find that the highest likelihood occurs for lognormal distributions
in both \tpk\ and \Lpk\@.  The maximum-likelihood estimate of the
lognormal distribution function for \ltpk\ has mean, $\mu = 1.7$ and
standard deviation $\sigma = 0.9$, while that for \Lpk\ has
$\mu = 25.5, \sigma=1.5$.

We give the values of the maximum likelihoods for other combinations
of distribution functions relative to that for the best-fitting case
of lognormal distributions in both \tpk\ and \Lpk, along with the
associated parameter estimates in Table~\ref{tdistribfun}\@.  A
lognormal distribution in both \tpk\ and \Lpk\ results in a
significantly higher likelihood than any other combination of the
three functions (lognormal, power-law, uniform) that we tried.

\begin{deluxetable*}{l l@{\hspace{0.5in}}L L}
\tablecaption{Distribution Functions for \tpk\ and \Lpk \label{tdistribfun}}
\tablewidth{0pt}
\tablehead{
  & \\  % need extra space in  here for tablenotemarks
\multicolumn{2}{c}{Distribution functions\tablenotemark{a}}
& \multicolumn{2}{l}{Maximum likelihood} \\
\colhead{\protect \tpk} & \colhead{\protect \Lpk}  
& \colhead{$\Delta \log_{10}p$\tablenotemark{b}} & \colhead{best-fit parameters} }
\startdata
lognormal & lognormal  & 0     & \log_{10}\tpk: \mu=1.7, \sigma= 0.9;\phn \log_{10}\Lpk: \mu=25.5, \sigma=1.5 \\  % v5.63,1b
lognormal & power-law  & -4.34 & \log_{10}\tpk: \mu=1.7, \sigma= 0.9;\phn \log_{10}\Lpk: {\rm min}=23.9, {\rm exponent}=-1.24 \\ %v5.63, 3
uniform   & lognormal  & -5.11 & \log_{10}\tpk:  {\rm min}$-0.3$, {\rm max}=3.5;\phn \log_{10}\Lpk: \mu=25.6, \sigma=1.5  \\ % v5.63, 4b
lognormal & log-uniform& -5.63 & \log_{10}\tpk: \mu=1.8, \sigma= 0.9;\phn \log_{10}\Lpk: {\rm min}=22.0, {\rm max}=29.1 \\ % v5.63, 2c
\enddata
\tablenotetext{a}{The functional form of the distribution
  functions. Lognormal is a Normal (Gaussian) distribution in
  $\log_{10}(x)$, characterized by the mean, $\mu$, and standard
  deviation, $\sigma$.  Log-uniform is a uniform distribution in
  $\log_{10}(x)$.  \tpk\ is in days, and \Lpk\ is in \ergsHz.}
\tablenotetext{b}{We give the log$_{10}$ maximum likelihood values
  relative to that for the best-fitting case where both distribution
  functions were lognormal.
}
\end{deluxetable*}

\subsection{The Lognormal Distributions for Different SN types}
\label{sdistribbytype}

Thus guided towards the use of lognormal distributions, we proceed to
determine the distributions of \tpk\ and \Lpk\ for various groups of
SNe, to study whether different kinds of SNe are characterized by
different distributions of \tpk\ and \Lpk.  In addition to the maximum
likelihood estimates of the means and standard deviations of the
lognormal distributions, we also obtain the $p=68$\% points, being
the points where the overall likelihood is 68\% of that associated
with the best-fit values.  We give our results in
Table~\ref{tdistribvals}.

\begin{figure}
  \centering
  \includegraphics[trim=0.2in 0 0.4in 0.4in, clip, width=\linewidth]{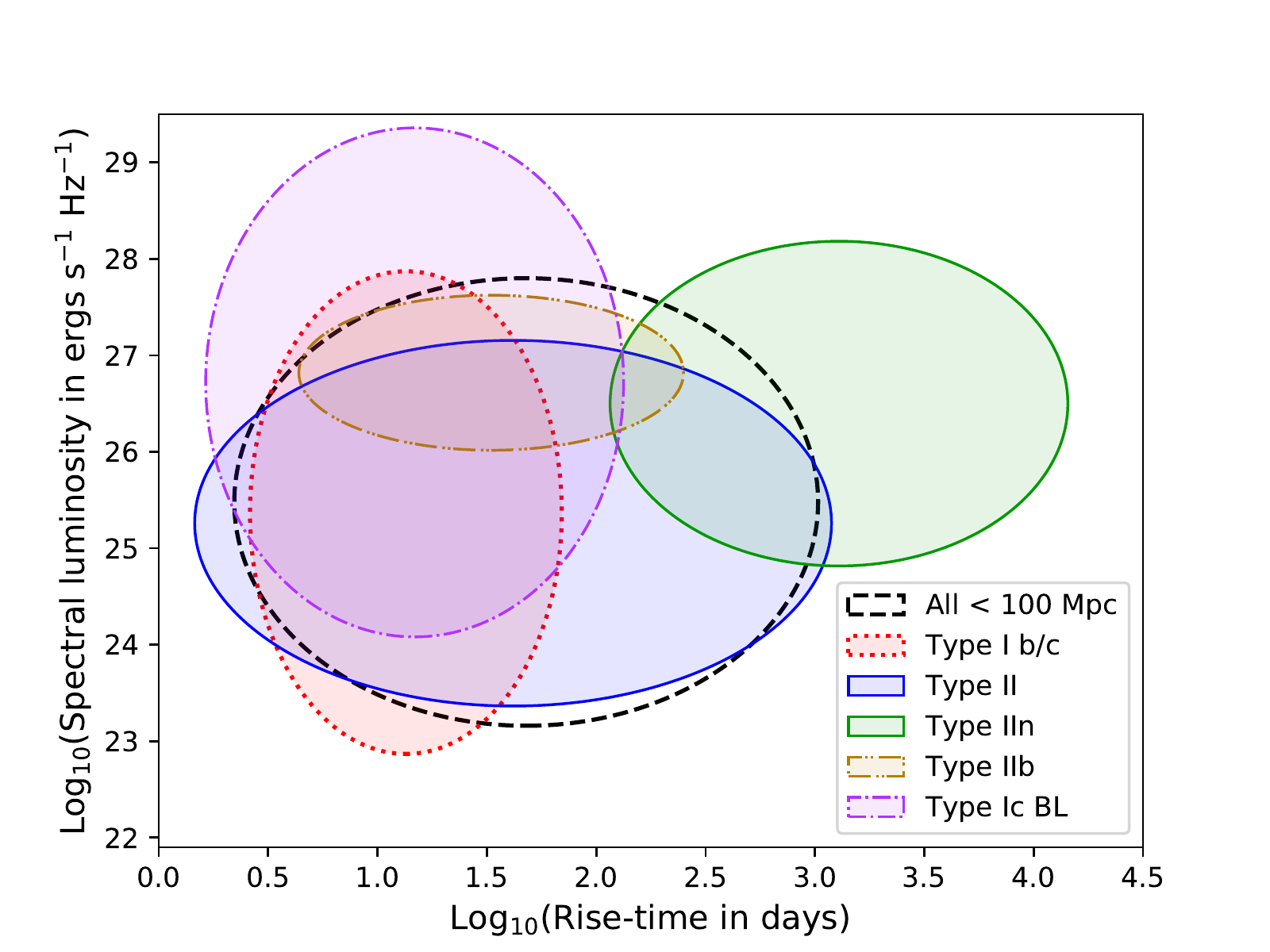}
  \caption{The distributions of different Types of SNe in the \tpk,
    \Lpk\ plane.  The ellipses show the $p=68$\% contour, i.e., the
    contour containing 68\% of the SNe, 
    of the relevant most-likely lognormal distributions for the
    various kinds of SNe (see Table~\ref{tdistribvals}).
    We illustrate the distributions for different samples of SNe as:
    All (dashed, black, $n=262$), Type I b/c (red, $n=110$), all Type
    II (except Type IIn; blue, $n=106$), Type IIn (green, $n=41$), 
    Type IIb (mustard, $n=19$), and Type Ic-BL SNe (magenta, $n=27$).  We
    use the samples at $D < 100$~Mpc, except for the rare Type Ic BL
    where we take all examples regardless of $D$. Note that the Type
    IIb and Ic BL distributions are rather uncertain because of the
    small number of SNe.}
  \label{fellipses}
\end{figure}

Are different Types of SNe characterized by different distributions of
\tpk\ and \Lpk?  We have already seen from Figure~\ref{falllc} that
Type I b/c SNe tend to have higher \Lpk\ and shorter
\tpk. \newstuff{We split our set of SNe by Type as discussed in our
  introduction, and fit the distributions of \tpk\ and \Lpk\
  separately for the different Types.  We use the following three main
  classes: Type I b/c, Type IIn, and the remainder of the Type II's.
  We also examine the subset of Type I b/c SNe that are broad-lined
  Type Ic (Ic-BL) and the Type IIb subset of the Type II SNe.  The
  results are given in Table~\ref{tdistribvals}, and we plot the
  distributions in Figure~\ref{fellipses}.

  Because of the completeness considerations mentioned earlier, we
  again consider only subsamples of SNe at $D < 100$~Mpc, with the
  exception of the rare BL subclass, where we include all examples
  regardless of $D$\@.  Note that the first line of
  Table~\ref{tdistribvals} represents the same fit as the first line
  of Table~\ref{tdistribfun}.}

\newstuff{We find that the 110 Type~I b/c SNe are characterized by
  values of \tpk\ $\sim\!3\times$ lower and values of \Lpk\
  $\sim\!1.3\times$ higher than are the 106 Type~II SNe.  The range of
  \Lpk\ values is higher for Type I SNe ($\sigma$ of \lLpk = 1.7) than
  for Type~II's ($\sigma$ of \lLpk = 1.3).

  Type I b/c are over-represented in our sample, they form 42\% of our
  sample at $D < 100$~Mpc}, while they represent only 26\% of all the
SNe in the Lick Observatory Supernova Search
\citep[LOSS;][]{Smith+2011a} and 19\% of a complete nearby sample of
175 SNe from LOSS \citep{Li+2011}. The reason for the
over-representation is that Type I b/c's were more actively observed
because of the potential association with GRBs.

We ask whether the presence of the Type II SN~1987A, which was clearly
unusual in a number of respects, and which, because of its low radio
luminosity, could be detected because of its nearness, biases our
derived distributions of \tpk\ and \Lpk? \newstuff{We redid the fit
  for the 105 Type II SNe excluding SN~1987A, and found that the
  best-fit distribution of \tpk\ and \Lpk\ (Table~\ref{tdistribvals})
  changed only slightly}, so we can conclude that our derived
distributions are not overly sensitive to the presence of the unusual
SN~1987A\@.

\newstuff{We examined the 41 Type IIn SNe (at $D < 100$~Mpc) for which
  we have measurements.  Type IIn SNe are associated with particularly
  strong radio emission.  They are characterized by $\sim\!31\times$
  longer \tpk, and $\sim\!17\times$ higher values of \Lpk\ than the
  remainder of the Type~II population.  We find that Type IIn SNe are
  also over-represented in our sample, they are 16\% of our sample at
  $D < 100$~Mpc, while they represent only 9\% of the core collapse
  SNe in the whole LOSS sample \citep{Smith+2011a} and 5\% of the
  nearby complete LOSS subsample \citep{Li+2011}.  The reason for the
  over-representation is that IIn's were probably more actively
  observed in the radio because of their strong association with radio
  emission.

  Our Type II sample contained 19 SNe of Type IIb (none at
  $D > 100$~Mpc).  Although this number is too low to permit a very
  reliable determination of the $\tpk$ and $\Lpk$ distributions, we
  did find some interesting trends.  Type IIb's were much more likely
  to be detected than other types of SNe, with 79\% being detected.
  The Type IIb's have values of \tpk\ in between those of Type I b/c
  and Type II, but closer to those of Type II\@.  They have a high
  mean value of \Lpk, $36\times$ higher than that of the remainder of
  the Type II's (excluding IIn's), and about $2\times$ higher even
  than that of Type IIn's.  The spread in the values of $\Lpk$ is
  considerably smaller than for other Types, $\sigma$ in \lLpk\ being
  only 0.5\@.}

Finally, we examined Type Ic SNe classified as broad-lined, BL, which
are the type associated with gamma-ray bursts.  BL SNe are relatively
rare, and only 6 were detected within $D < 100$~Mpc, so we take all
\newstuff{27 BL SNe in our database, regardless of $D$\@.  They are
  characterized by a relatively short rise-time, with a mean \ltpk\ of
  only 1.2 ($\sigma = 0.6$), and a fairly high \lLpk, with a mean of
  26.7 with $\sigma = 1.7$, with the mean \lLpk\ being
  $\sim \! 20\times$ higher than that for all Type I b/c's.  However,
  since there were only 13 detected BL SNe in our sample, the
  distribution of \tpk\ and \Lpk\ must be regarded as rather
  uncertain.  We note that two unusually nearby BL SNe, SN~2002ap
  \citep{BergerKC2002, Soderberg+2006b} and SN~2014ad
  \citep{Marongiu+2019}, were observed over a wide range of times and
  had $L_{8.4 \rm GHz} \lesssim 10^{25.5}$~\ergsHz.  Since our sample
  is probably biased in favour of radio-bright examples,
  it seems likely that $\sim$10\% of BL SNe have radio luminosities
  $<10^{25.5}$~\ergsHz, unless they have very short $\tpk$ less than a
  few days.}

\begin{deluxetable*}{l r r c c c c}
\tablecaption{Lognormal Distributions of \tpk\ and \Lpk}
\label{tdistribvals}
\tablehead{
  \colhead{Set of SNe} & \colhead{$N_{\rm SNe}$} &
  \colhead{$\log_{10}(p)$}  & \multicolumn{2}{c}{Distribution of \ltpk} & \multicolumn{2}{c}{Distribution of \lLpk} \\
  & & \colhead{per measurement\tablenotemark{a}} &
  \colhead{$\mu$\tablenotemark{b}} &
  \colhead{$\sigma$\tablenotemark{c}} &
  \colhead{$\mu$\tablenotemark{b}} &
  \colhead{$\sigma$\tablenotemark{c}} }
\startdata
  All ($D < 100$~Mpc)&262 &$ -1.09 $& 1.7 (1.6, 1.8) & 0.9 & 25.5 (25.2, 25.7) & 1.5 \\  % v5.63 Prod.I PPAA
  All ($D < 50$~Mpc)& 189 &$ -1.15 $& 1.6 (1.5, 1.7) & 0.8 & 25.5 (25.2, 25.7) & 1.5 \\  % v5.63 Prod.I PPAA
  Type I b/c     &    110 &$ -1.00 $& 1.1 (1.1, 1.3) & 0.5 & 25.4 (24.8, 25.7) & 1.7 \\  % v5.63 Prod.I PPAA
  Type II         &   106 &$ -1.05 $& 1.6 (1.4, 1.9) & 1.0 & 25.3 (25.0, 25.6) & 1.3 \\  % v5.63 Prod.I PPAA
  Type II w/o SN 1987A&105&$ -1.05 $& 1.7 (1.5, 2.0) & 1.0 & 25.4 (25.1, 25.7) & 1.2 \\  % v5.63 all vals within 1 sigma of II with 87A
  IIn               &  41 &$ -1.20 $& 3.1 (2.8, 4.1) & 0.7 & 26.5 (25.9, 27.0) & 1.1 \\  %ck v5.60-F almost identical to Ibn+IIn
  IIb             &    19 &$ -1.55 $& 1.5 (1.3  1.7) & 0.6 & 26.8 (26.7, 27.0) & 0.5 \\  %ck v5.63 Prod.I PPAA
  Broad-lined(BL)\tablenotemark{d}
                  &    27 &$ -1.00 $& 1.2 (0.9, 1.4) & 0.6 & 26.7 (25.9, 27.2) & 1.7     %ck v5.63 
\enddata
\tablenotetext{a}{The average $\log_{10}$ of the probability per
  measurement if \tpk\ and \Lpk\ are distributed with the most
  probable log-Gaussian distribution.  This is more comparable over
  different numbers of SNe than the probability for all the
  measurements, which is expected to be lower the larger the number of
  measurements.}
\tablenotetext{b}{The mean, $\mu$, of the normal distributions in
  \ltpk\ and \lLpk, i.e.\ the lognormal distributions in \tpk\ and
  \Lpk, with the $p=68$\% confidence range in parenthesis
  following.}
\tablenotetext{c}{The standard deviation, $\sigma$, corresponding to
the mean values $\mu$ in the preceding column.}
\tablenotetext{d}{Due to the rarity of broad-lined (BL) SNe, we relax
  our restriction on $D$ to include $D > 100$~Mpc for these SNe.}
\end{deluxetable*}

\section{Mass-Loss Rates}
\label{sMdot}

Massive stars lose a significant fraction of their mass before
exploding as SNe.  This mass-loss is still poorly understood.  An
exploding SN provides a probe of this mass-loss, since the medium into
which the SN shock expands is the circumstellar medium (CSM) which
consists of the star's wind during the period before it exploded.  The
radio emission from the SN is due to the interaction of the SN ejecta
with the CSM, and its brightness depends in part on the CSM density,
which is a function of the mass-loss rate, \Mdot, of the
progenitor. Although the flux-density measurements provide useful direct
constraints on \Mdot\ of only a small fraction of well-observed SNe, we
can use the distributions of \tpk\ and \Lpk\ obtained in
Section~\ref{sdistribbytype} to constrain the distribution of \Mdot\ over
our sample of SNe.

The SN shock is expected to both amplify the magnetic field and
accelerate some fraction of the electrons to relativistic energies.
The amount of synchrotron radio emission depends on the energy in the
magnetic field as well as that in the relativistic electrons.  In the
absence of any absorption, the amount of synchrotron radiation can be
estimated by assuming that constant fractions of the post-shock
thermal energy density are transferred to magnetic fields and
relativistic electrons \citep[see, e.g.,][]{Chevalier1982b,
  ChevalierF2006}.  The spectral luminosity, $L_\nu$, at a given time
will therefore depend on the CSM density at the corresponding shock
radius.  $L_\nu$ will also depend on the square of the shock speed
and the volume of the emitting region.  Although the shock speed and
radius are measured using VLBI for some SNe (e.g., SN~1993J,
\citealt{SN93J-2}; SN~2011dh \citealt{SN2011dh_Alet}; for a review see
\citealt{SNVLBI_Cagliari}) they are not measured for the great
majority of SNe.

The post-shock energy density, at time, $t$ when the shock has radius
$r$, will be $\propto \rho_{\rm CSM}(r) v^2(t)$, or in the case of a
steady wind, $\propto \Mdot v^2(t) r(t)^{-2}$.  If there is
equipartition between the relativistic electrons and the magnetic
field, then a measurement of the spectral luminosity, $L_\nu$, can be
used to estimate $\Mdot$, provided that a number of things are known
or, in our case, can be assumed.

The first thing we need to assume is the wind speed of the
progenitor, \vwind.  $L_\nu$ actually depends on the density, which is
proportional to $\Mdot / \vwind$, rather than depending directly on
\Mdot. Type I b/c SNe generally have Wolf-Rayet progenitors, with
fast, low-density winds, with $\vwind \sim 1000$~\kms.  Type II SNe,
on the other hand, have supergiant progenitors, which generally have
slow, dense winds with $\vwind \sim 10$~\kms. In calculating $\Mdot$,
we will assume $v_{\rm wind} \sim 1000$~\kms\ for the Type I b/c's,
and $\sim 10$~\kms\ for the Type II's.

The next thing that we need to assume is the efficiency of the
conversion of thermal energy to both magnetic field and relativistic
particle energies.  These efficiencies are usually expressed as the
ratio between the energy density of the magnetic field and the
relativistic particles to the post-shock thermal energy density, and
we will denote the two ratios with $\epsilon_B$ and $\epsilon_e$,
respectively.  Although the values are not well known, it is
\newstuff{often assumed}\ that $\epsilon_B \simeq \epsilon_e$
(equipartition), and that both are $\sim 0.1$.  \newstuff{We will here
  also assume $\epsilon_B = \epsilon_e = 0.1$, and we note that our
  values of \Mdot\ must remain somewhat speculative, but we hope
  nonetheless instructive.  We discuss the uncertainty in deriving
  \Mdot\ from radio lightcurves further in Sec.~\ref{sdiscussMdot}
  below.}

Finally, the volume of the emitting region and the speed of the shock
also needs to be known or assumed.  In the case of Type I b/c SNe, the
absorption producing the rising part of the lightcurve is most often
synchrotron self-absorption (SSA).  In this case the absorption is
internal to the emitting region, and \tpk\ and \Lpk\ allow an estimate
of the radius at time \tpk\ (as noted already in
Section~\ref{stpklpk}).  If we assume the emitting region to be a
spherical shell with outer radius 26\% larger than the inner one, then
the filling factor is 0.5, which is considered typical.  We will
assume $f = 0.5$\@.  With these assumptions, \citet{ChevalierF2006}
and \citet{Soderberg+SN2011dh-I} find that
\begin{equation}  % no empty lines in here
  \begin{multlined}
  \label{eqMdotSSA}
\Mdot = 1.1 \times 10^{-7} 
\left( \frac{0.1}{\epsilon_B} \right)
\left( \frac{\epsilon_e}{\epsilon_B} \right) ^{-8/19}  \; \cdot \\
\left( \frac{\Lpk}{10^{26} \; \ergsHz} \right) ^{-4/19} 
\left( \frac{ \tpk}{\rm d} \right)^2  \; \Msolyr 
\end{multlined}
\end{equation}
where we have recast the equation given in \citet{Soderberg+SN2011dh-I}
for our nominal frequency of 8.4 GHz, and taken $\vwind = 1000$~\kms.

This equation, however, is only applicable if the spectral energy
distribution (SED) is dominated by SSA\@.  As can be seen in
Figs.~\ref{falllc_type1} and \ref{fpeakplot}, Type I b/c SNe show a
wide range of lightcurve behaviors.  In particular, for ones which are
slow-rising and faint, the rise cannot be reproduced by SSA without
assuming expansion velocities too low to be believable.  In those
cases, therefore, there is likely significant FFA absorption.  In the
presence of FFA, the radius and velocity implicit in the above
calculation are only lower limits.  Most Type I b/c SNe show expansion
velocities of $> 30,000$~\kms\ \citep{Chevalier2007}.
For any Type I b/c SN where \tpk\ and \Lpk\ imply
$\vSSA < 20,000$~\kms, the assumption of an SSA-dominated lightcurve
is problematic, and eq.~\ref{eqMdotSSA} therefore not applicable, and
the speed of the shock volume of the emitting region must be estimated
in some fashion other than from SSA.

Eq.~\ref{eqMdotSSA} is also not applicable for Type II SNe, where the
absorption is generally dominated by FFA\@.  However, Type II SNe seem
to be characterized by a relatively narrow range in expansion
velocity: \citet{deJaeger+2019} found that for the 51 Type II SNe from
the Berkeley sample, the standard deviation of expansion velocity
measured from the H$\alpha$ line at $t = 10$~d was only 19\%,
suggesting a fairly narrow range of velocities \footnote{Expansion velocities
  from radio are expected to be somewhat higher than those from optical,
  e.g., from H$\alpha$, because the former usually relate
  to the forward shock and the latter to expanding areas interior to
  it. However, the optical and radio velocities are
  expected to be well correlated, so a narrow range in H$\alpha$
  suggests a correspondingly narrow range in forward shock velocities.
  See discussion in \citet{SN93J-4}.}.

Since the spread in the velocity for Type II SNe is not large,
we assume a single representative value for all Type II SNe.
\citet{Weiler+2002} give expressions in this case, which are based on
assuming a self-similar evolution of the SN, with $r \propto t^m$,
where $m$ is called the deceleration parameter, as well as assuming a
single characteristic initial expansion velocity for all SNe of a
particular Type (I or II).  Although the latter assumption is
demonstrably poor for rapidly-expanding Type I b/c SNe, which have
speeds ranging up to $c$, it is probably reasonable for the slower
Type I b/c's, i.e., those with \vSSA $< 20,000$~\kms, as well as for
Type II's, which generally do not show high expansion velocities.

Although \citet{Weiler+2002} take $m = 1$ for Type II SNe, the
observations do not necessarily bear this out, with some Type II SNe
showing substantial deceleration, for example, $m=0.781$ for SN~1993J
between ages of $\sim$1.5 and $\sim$5~yr \citep{SN93J-2} and
$m = 0.69$ observed for SN~1986J \citep{SN86J-2}.
Overall, the Type II SNe show a similar range of values of $m$ as do
the few Type I b/c's for which we have reliable estimates of $m$, and
therefore we assume $m=0.8$ for both Type I and Type II SNe.
Following \citet{Chevalier1982b}, we find in this case that
\begin{equation}
  \label{eqLnuboth}
  \begin{multlined}
L_\nu  \propto (1-m) (\Mdot/\vwind)^{(p - 7 + 12m)/4} \; \cdot \\
m^{(5+p)/2} t^{-(p + 5 - 6m)/2} \nu^{-(p-1)/2},
\end{multlined}
\end{equation}
where $p$ is the energy index of the relativistic electron population.
Representative values of $p$ for Type I b/c and II SNe are 3 and 2.4,
respectively \citep{Weiler+2002}.  Setting $m = 0.8$, this equation
simplifies to
\begin{equation}
  \label{eqLnuType1}
  {\rm Type\;I\; b/c:} \; L_\nu  \propto (\Mdot/\vwind)^{1.4} \, t^{-1.6} \, \nu^{-1}
\end{equation}
and
\begin{equation}
  \label{eqLnuType2}
  {\rm Type\;II:} \; L_\nu \propto (\Mdot/\vwind)^{1.25} \, t^{-1.3} \, \nu^{-0.7}
\end{equation}

Now we need to determine the constant of proportionality in eqs.\
\ref{eqLnuType1} and \ref{eqLnuType2}, to use them to obtain
$\Mdot/\vwind$.  For Type I b/c SNe, we determine the constant by
requiring that \Mdot\ have the same value as that calculated using
eq.~\ref{eqMdotSSA} for a representative value of
$\Lpk = 2\times10^{26}$~\ergsHz, and our rounded mean value of $\tpk$
for Type I b/c SNe of 20~d (Table \ref{tdistribvals}), which
corresponds to $v_{\rm SSA} = 20,000$~\kms.  We obtain
\begin{equation}
  \begin{multlined}
  \label{eqMdotTypeI}
  {\rm Type\;I\;b/c:} \; \Mdot = 7.2 \times 10^{-7} \; \times \\
   \left( \frac{\Lpk}{10^{26}~\ergsHz} \right)^{0.71}
   \left( \frac{\tpk}{\rm 1 d} \right)^{1.14} \; \Msolyr
   \end{multlined}
 \end{equation}
for \vwind\ = 1000~\kms.

For Type II SNe, we determine the constant of proportionality by using the
mean values of $\log_{10}(\Mdot)$ determined from the absorption for
four well observed SNe (SN 1970G, SN 1979C, SN 1980K, SN
1981K)\footnote{We omit SN~1982aa from this calculation, because no optical
  spectrum was ever obtained and its Type is therefore uncertain (see
  Section~\ref{sfindtype} below).} given in \citet{Weiler+2002}.
We obtain
\begin{equation}
  \begin{multlined}
  \label{eqMdotType2}
  {\rm Type \; II:} \; \Mdot = 1.1 \times 10^{-7} \; \times \\
   \left( \frac{\Lpk}{10^{26}~\ergsHz} \right)^{0.80}
   \left( \frac{\tpk}{\rm 1 d} \right)^{1.04} \; \Msolyr
   \end{multlined}
 \end{equation}
for \vwind\ = 10~\kms.

Given the distributions of \tpk\ and \Lpk\ we obtained in
Table~\ref{tdistribvals}, we can now calculate the corresponding
distribution of \Mdot\ for Type I b/c and Type II SNe.  Since we found
that a lognormal distribution was appropriate for \tpk\ and \Lpk, we
determine the distribution of $\log_{10}(\Mdot)$.

\newstuff{We find that the mean of
  $\log_{10} (\Mdot \; {\rm in} \; \Msolyr)$ for Type I b/c SNe is
  $-5.6 \pm 1.1$, assuming \vwind\ = 1000~\kms.  For Type II SNe,
  (excluding IIn), \lMdot\ = $-6.8 \pm 1.4$}, assuming \vwind =
10~\kms. The progenitors and wind velocity of the Type IIn's are not
well known and their \Mdot\ rates are likely strongly time-variable,
and therefore equation \ref{eqMdotType2} will be poorly calibrated for
them, so we do not extend this analysis to the Type IIn SNe.

\section{Discussion}
\label{sdiscuss}

Our large compilation of 1475 radio measurements of \NSNe\ SNe shows % 1475 is for v5.63 PPAA
that the radio lightcurves of SNe are extremely varied.  With our
simple characterization of the lightcurves with only two parameters,
\tpk\ (rise-time) and \Lpk\ (peak spectral luminosity), we find that
both \tpk\ and \Lpk\ can vary over large ranges, at least 3 and 5
orders of magnitude, respectively (Figure~\ref{fpeakplot}).  We showed
that a lognormal distribution was appropriate for both \tpk\ and \Lpk.

We find that the normal distribution of \ltpk\ has a mean of 1.7,
corresponding to 50~d, with $\sigma$ (standard deviation) of 0.9 (line
2 in Table~\ref{tdistribvals}, \newstuff{using only SNe at
  $D < 100$~Mpc})\@.  Both quite short risetimes of 7~d and quite long
ones of $> 1$~yr are within the range of $\pm 1 \sigma$ and thus not
uncommon.

We find that many SNe must be fairly faint in the radio.  Indeed, to
date, only about 31\% of the SNe at $D < 100$~Mpc
that have been observed in the radio were detected at all. The
results published so far tend to be biased in favour of the detections or
towards higher radio luminosities.  If we include the many
non-detections, we find that the most probable distribution of \lLpk\ in \ergsHz\
has mean of 25.5, corresponding to $3\times10^{25}$~\ergsHz, with
$\sigma = 1.5$. % PPAA

This distribution has a significantly lower mean \lLpk, as well as a
wider range, than the mean of 27.3, corresponding to $2\times10^{27}$
\ergsHz, with $\sigma = 1.25$ given by \citet{Lien+2011}, which was
based on only 20 {\em detected}\/ SNe.  In fact, if we repeat the
calculation from Table~\ref{tdistribvals} for only those SNe with at
least 3 detections, we find that the mean \lLpk\ = 27.1, close to that
found by \citet{Lien+2011}.  The inclusion of the many limits is
crucial for obtaining the distribution of {\em all}\/ radio SNe, not
just the well-studied radio-bright ones.

We find that more than half of all SNe will have peak luminosities
$<10^{26}$~\ergsHz\ (at 4 to 8 GHz), corresponding to $\sim$1 mJy at
10 Mpc
Although SN~1987A is at the faint end of the distribution with \Lpk\
$\lesssim 10^{24}$~\ergsHz, we expect $\sim$11\% of all SNe, or
$\sim$6\% of Type II SNe, will be comparably faint in the radio.

\subsection{Differences Between SNe of Type I b/c and II}
\label{s12diffs}

It has long been accepted that Type Ib/c SNe tend to have more
rapidly-evolving radio lightcurves, characterized by shorter values of
\tpk, than do Type II's.  However, until the present work, this has
only been asserted on the basis of relatively small numbers of SNe
\citep[e.g.,][]{Weiler+2002, Weiler+2010}.  While we find the
assertion to be true, with the values of \ltpk\ being characterized by
a mean of \newstuff{1.1 (13~d)}\ for Type I b/c SNe, in comparison
\newstuff{to 1.6 (40~d) for Type II SNe excluding Type IIn's}, the
caveat that must be stated here is the standard deviations in \ltpk\
were large for both Types, being \newstuff{0.5}\ for I~b/c and 1.0 for
Type II\@.  Therefore, as can also be seen in Figure~\ref{falllc},
there is considerable overlap, with some Type I b/c SNe having very
slow rise times up to several years, while some Type II SNe have short
rise times of $<1$ month, and SN~1987A has one of $< 2$~d.

\newstuff{We further find that the Type I b/c and Type II SNe reach a
  similar range of \Lpk\ values.  For our sample of SNe at
  $D < 100$~Mpc, the mean value of
  $\log_{10} (\Lpk \; {\rm in}\; \ergsHz)$ for Type I b/c's was 25.4,
  while that for Type II's (excluding IIn) was marginally lower at
  25.3.  The standard deviations for \lLpk\ were large, being 1.6 for
  Type~I b/c's, and 1.3 for Type II's, so there is very significant
  overlap in the distribution of \lLpk\ (see also
  Figure~\ref{falllc}).  Some Type I b/c SNe have quite low values of
  $\lLpk \lesssim 25.5$, while many Type II SNe have high values of
  $\lLpk > 25$.  We note, however, that for Type I b/c's, the standard
  deviation of the \lLpk\ distribution is notably higher than it is
  for Type II's, so both extreme high and low values of \Lpk\ are more
  likely for Type I b/c's.

  While it had been suggested on the basis of only four examples that
  Type I b/c SNe could be approximate radio standard candles
  \citep{Weiler+1998}, our data (Figure~\ref{falllc_type1}) make
  clear that this is very much not the case, with the variation in
  \Lpk\ extending over several orders of magnitude.}

Our best-fit distributions of \tpk\ and \Lpk\ are illustrated in
Figure~\ref{fellipses}\@.  The 41 SNe of Type IIn (at $D < 100$~Mpc)
have higher and later radio peaks than the remainder of the Type II's,
with mean values of \ltpk\ and \lLpk\ being \newstuff{3.1
  (corresponding to 3.5~yr) and 26.5,
  respectively, but the standard deviations in \ltpk\ and \lLpk\ are
  large, being 0.7 and 1.1, respectively}, thus overlapping with the
other SN Types.  We note that \citet{Stockdale+2007} suggested a much
higher \lLpk\ of 28 for Type IIn's, but again this result was biased
by not including non-detections.  There is a possibility that Type IIn
SNe have similar radio lightcurves to other Type II's initially, i.e.,
with risetimes on the order of \tpk\ = 40~d, and relatively low values
of \Lpk, but are characterized by luminous late-time radio emission,
since there are relatively few observations of IIn's at earlier times
(Figure~\ref{falllc_IIn}).

We note again that the values of \Lpk\ of the few SNe that have many
measurements are notably higher than the mean (except for SN~1987A),
being around $\Lpk \gtrsim 10^{27}$~\ergsHz.  The reason is that the
radio SNe that have attracted the most attention are the most
luminous examples, but our many upper limits show that the majority of
SNe are in fact relatively faint.

\subsection{The Synchrotron-Self-Absorption Expansion Velocity}
\label{svSSA}

As mentioned in Section \ref{stpklpk}, if SSA (synchrotron self-absorption)
is the dominant absorption mechanism, the emitting volume, and thus
the radius, can be deduced from the frequency at which the SED
peaks. Equivalently, for some particular frequency, $\nu$, \Lpk\
allows calculation of the source volume or radius at the time \tpk.  We
call this radius $r_{\rm SSA}$, and the corresponding velocity
$\vSSA = r_{\rm SSA}/\tpk$.  Both $r_{\rm SSA}$ and \vSSA\ are just
calculated from $\nu$, \tpk, and \Lpk, regardless of whether SSA is in
fact the dominant absorption mechanism.  Only if SSA {\em is}\/
dominant do $r_{\rm SSA}$ and \vSSA\ correspond to the physical radius
and speed.

For each of our SNe the measurements provide some constraint on \vSSA,
to the degree to which the measurements constrain \tpk\ and \Lpk\@.  In
Figure~\ref{fSNprob}, we showed the likelihood of various values of
\tpk\ and \Lpk\ given our measurements for three example SNe.  The
lines of constant \vSSA\ are parallel to the dotted line in the top
left corner showing $\vSSA = 2c$ in the \tpk-\Lpk\ plane.  Integrating along
lines of constant \vSSA, we can therefore determine the
probability of particular values of \vSSA\ given our measurements.

Referring again to the three example SNe shown in
Figure~\ref{fSNprob}, for some SNe, such as SN~1993J, \tpk, \Lpk\ and
thus \vSSA\ are well determined, and only a single value of \vSSA\ is
allowed by the measurements, while for others such as SN~2017gax, we
have only very weak constraints on \vSSA, and virtually any value of
\vSSA\ can be accommodated by the (single) measurement.  If we
normalize the probability for each SN, over the range of \vSSA =
1~\kms\ to $2c$, we can determine a probability distribution of \vSSA\
over our collection of SNe by summing over all our SNe, giving each
equal weight.

\begin{figure}
\centering
\includegraphics[width=\linewidth]{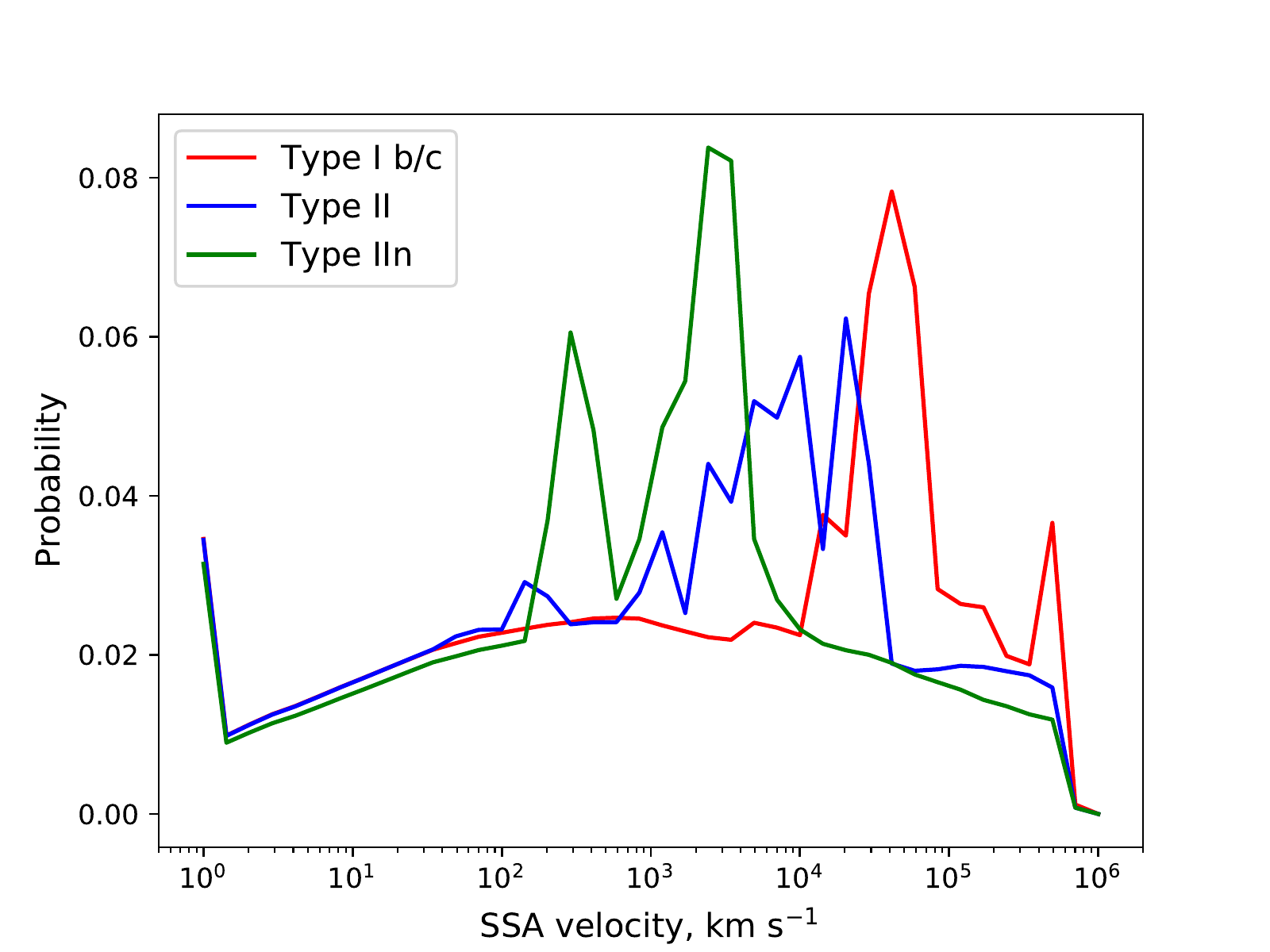}
\caption{The normalized likelihood of different values of the SSA
  velocity, \vSSA, over different samples of SNe.  \vSSA\ is the
  velocity calculated from \tpk\ and \Lpk\ assuming that the dominant
  absorption mechanism is synchrotron self-absorption.  The likelihood
  is that of all our measurements arising from SNe characterized by a
  particular value of \vSSA.  Values of \vSSA\ which may never occur
  in any actual SN can have non-zero likelihood if they are not
  disallowed by our flux-density measurements.  Note that \vSSA\ is a
  {\em lower}\/ limit to the shock speed.  We weight each SN equally.
  We show $p(\vSSA)$ for SNe of Type I b/c SNe (red), II (blue), and
  IIn (green).}
\label{fvSSAdist}
\end{figure}

We show this distribution in Figure~\ref{fvSSAdist}, showing
separately the distributions for Type I b/c, Type II, and Type IIn SNe.
We note that the probability we show is that of particular values of
\vSSA\ given all our observations.  A non-zero probability for some
value of \vSSA\ means that value is allowed by the observations for
some fraction of our SNe, but does not require that there exist any SN
characterized by that value of \vSSA.  This is particularly true of
the very low values of $\vSSA < 100$~\kms, which are allowed by the
measurements for a significant number of our SNe, but which likely do
not occur in any real SNe.  Nonetheless, in the absence of concrete
measurements of \vSSA, (or better, the actual shock speed) for a large
number of SNe, Fig~\ref{fvSSAdist} will give some insight into what
values of \vSSA\ are allowed by the currently existing measurements.

Recall also that \vSSA\ is only a lower limit to the shock velocity
(see Section \ref{stpklpk}): if free-free absorption dominates and the
peak in the SED is not due to SSA, then both $r_{\rm SSA}$ and
\vSSA\ are only lower limits to the physical $r$ and $v$.  In fact,
given that the shock speeds observed in SNe are almost always larger
than a few thousand \kms, much of the portion of
Figure~\ref{fvSSAdist} below $10^4$~\kms\ is likely due to cases where
in fact FFA dominates, and the shock speed is larger than \vSSA.

Comparing the distributions of \vSSA\ for different types of SNe, we
find the following: For Type I b/c SNe, high values of
$\vSSA > 20,000$~\kms\ are the most probable, with values up to and
even exceeding $c$ occurring.\footnote{As mentioned earlier,
  $\vSSA > c$ was seen in SN~2003dh, and the superluminal
    apparent expansion was confirmed directly by VLBI observations
    \citep[e.g.,][]{Pihlstrom+2007}.}
It is generally accepted that Type I
b/c SNe tend to have higher speeds than Type II's \citep[see,
e.g.,][]{Chevalier1998, Chevalier2007}, but this has only been
concluded on the basis of much smaller numbers of SNe in the past, and
we can now confirm this pattern on the basis of a much larger sample.
Note also that some fraction of Type I b/c can have low \vSSA
$< 10,000$~\kms, so even in the case of Type I b/c SNe, \vSSA\
can be significantly lower than the shock speed.

The most probable values of \vSSA\ for Type II SNe are
$\sim$3000~\kms.  Since this is lower than the expected shock speeds,
we conclude that for the majority of Type II SNe, FFA is the dominant
absorption mechanism rather than SSA, and \vSSA\ therefore is lower
than the shock speed.  Again our conclusion is in agreement with
statements made earlier \citep[e.g.,][]{Chevalier1998, Chevalier2007},
but which had in the past been made only on the basis of a far smaller
sample of SNe.  Type IIn SNe are characterized by even lower values of
\vSSA.

\subsection{Identifying the SN Type on the Basis of the Radio Lightcurve}
\label{sfindtype}

Can the radio emission be used to determine the Type of an SN, for
example in cases where there is no optical detection?  Because of the
large overlap in the lightcurves of different SN types, the radio
lightcurve for any particular SN generally does not reliably indicate
the SN type.

There are two exceptions where the radio lightcurve nevertheless can
give a fairly reliable indication of the SN Type.  First, a 8.4-GHz
spectral luminosity $L_{\rm 8.4 GHz} >10^{28}$~\ergsHz\ in the first
month seems to occur only for Type I b/c SNe.  Such high, early
luminosities therefore strongly suggest a Type I b/c SN\@.  In
general, very high radio luminosities of
$L_{\rm 8.4 GHz} > 10^{28.5}$~\ergsHz\ seem to occur predominantly for
Type I b/c SNe regardless of age.  Second, a very high luminosity at
late times with $\Lpk >10^{27}$~\ergsHz\ at $\tpk>1000$~d, strongly
suggests a Type IIn supernova.

Third, a high value of $\vSSA\ > 30,000$~\kms\ (see
Section~\ref{svSSA}) suggests that the SN is much more likely to be of
Type I b/c, while values of $\vSSA\ < 10,000$~\kms\ are more likely in
Type II SNe.

\begin{figure}
  \centering
  \includegraphics[width=\linewidth]{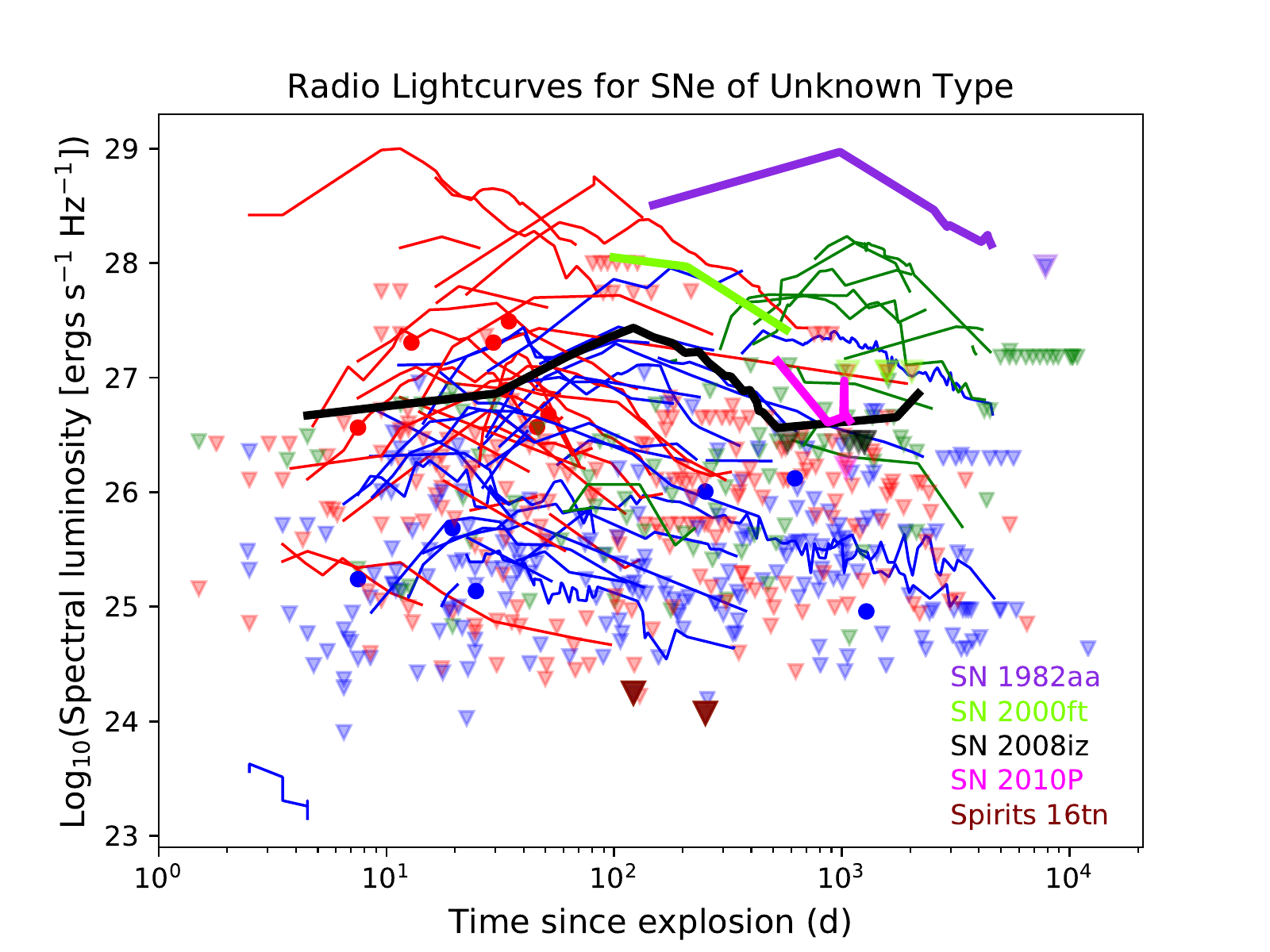}
  \caption{The radio lightcurves or limits of five SNe which were not
    spectroscopically identified, compared to those of the
    spectroscopically confirmed SNe of Types I b/c (red) and II (blue)
    and IIn (green) in our sample.  SNe~1982aa, 2000ft, 2008iz, 2010P
    and \newstuff{Spirits~16tn are highlighted in violet, light green,
      black, magenta and maroon, respectively.  Spirits 16tn was not
      detected, but two fairly low limits were obtained, shown as
      larger maroon triangles between $t = 100$ and 300~d\@.
      SN~1982aa is most likely of Type IIn.  For SNe 2000ft, 2008iz
      and 2010P, and Spirits 16tn, the radio lightcurves do not permit
      a conclusive}\ identification of the SN Type.}
  \label{funknowntype}
\end{figure}

In our database, there are \newstuff{five}\ SNe of which no optical
spectrum was obtained, and of which therefore the SN Type (I or II) is
unknown: SNe~1982aa,
2000ft, 2008iz, and 2010P \newstuff{and Spirits 16tn. Can the Type be
  determined from radio observations alone on the basis of our
  distributions of \tpk\ and \Lpk}? We show the radio lightcurves of
these SNe in comparison to the remainder of the SNe in our sample in
Figure~\ref{funknowntype}.

SN~1982aa in NGC 6052 was detected in the radio and reached a very
late and high peak \citep{Yin1994}.  Although the explosion date is
uncertain, the values of \tpk\ and \Lpk\ are fairly
well determined at $\sim 10^{2.5}$~d and $\sim 10^{29.0}$~\ergsHz,
respectively.  These values of \tpk\ and \Lpk\ strongly suggest a
Type~IIn (Figures~\ref{falllc_IIn}, \ref{fellipses}), although the SN
is exceptional regardless of Type.

SN~2000ft in NGC~7469 was detected only after the radio peak
\citep{Alberdi+2006}, and the explosion time is again uncertain.
Although it was optically detected \citep{Colina+2007}, no spectrum
was obtained.  The values of \tpk\ and \Lpk\ are fairly well
determined at $\sim \! 10^{2.0}$~d and $\sim 10^{28.1}$~\ergsHz,
respectively.  It is quite luminous compared to the majority of SNe,
but the lightcurve and the values of \tpk\ and \Lpk\ are equally
compatible with either Type I b/c or II, so its SN Type remains
unknown.

SN~2008iz was detected in the radio in M82, and was never detected
optically despite the close distance (3.8~Mpc), presumably because of
very strong optical extinction.  It has a very unusual radio
lightcurve \citep{Marchili+2010, Brunthaler+2010b}.  The values of
\tpk\ and \Lpk\ are fairly well determined at $\sim \! 10^{1.8}$~d and
$\sim \! 10^{27.3}$~\ergsHz, respectively.  It showed both an unusually
slow rise and a relatively shallow decay, and seems to be
showing a late-time rise after $t \simeq 1000$~d.  Although a Type II
has been suggested, the radio lightcurve is equally compatible with
either Type I b/c or II, although \Lpk\ was higher than the average
for either Type.  Its SN Type therefore also remains unknown.

SN~2010P, in Arp~299, was discovered in the infrared and subsequently
detected in the radio \citep[and references
therein]{Kankare+2014, Romero-Canizales+2014}. Infrared observations
and an optical spectrum were obtained by \citet{Kankare+2014}.  The
spectrum had relatively low signal-to-noise ratio due to the high
extinction, and was compatible with an SN of either Type Ib or a IIb.
In this case the peak of the radio lightcurve is not well determined,
and the first measurement occurred only at $t = 523$~d, so a wide range
of \tpk\ and \Lpk\ are compatible with the measurements.  The likely
values of \ltpk\ are between 1.2 and 2.5 and those of \lLpk\ between
27 and 29, with the higher values of \Lpk\ occurring in conjunction
with earlier values of \tpk.
While \citet{Kankare+2014} suggest that the radio evolution precludes
a Type Ib, we find (see Figure~\ref{funknowntype}) that, when compared
to our broad sample, SN~2010P's radio evolution is not inconsistent
with that seen in some Type I b/c's.  It is, however, more luminous
than the mean of any of our SN Types.  Although the optical spectrum
rules out a normal Type II, whether SN~2010P was of Type Ib or IIb
must remain uncertain.

\newstuff{Spirits 16tn was a heavily obscured SN, detected in the
  infrared, for which spectroscopic classification was not possible
  \citep{Jencson+2018}.  Fairly low limits of
  $L_{\rm 6 GHz} \lesssim 10^{24.3}$~\ergsHz\ were placed on the radio
  luminosity \citep{Jencson+2018}.  However, as can be seen from
  Figs.\ \ref{fellipses} and \ref{funknowntype}, such low values can
  occur for either Type I b/c or Type II SNe, therefore its SN Type must
  also remain uncertain.}

\subsection{Determining Mass Loss Rates from Radio Emission}
\label{sdiscussMdot}

From the distribution of \tpk\ and \Lpk, a distribution of mass-loss
rates, \Mdot, for the progenitors can be estimated.  For both SN Type
I b/c and Type II the mean values of \Mdot\ are low compared to the
published values for well-studied SNe.  This is because the
well-studied SNe tend to be substantially brighter than the average,
and thus tend to have denser CSM to produce the stronger radio
emission than the average.  We found that the average value of
$\log_{10}(\Mdot)$ for a Type I b/c SNe was \newstuff{$-5.6 \pm 1.1$
  (in \Msolyr;}\ assuming \vwind\ = 1000~\kms),
while the equivalent value for Type II's \newstuff{(excluding IIn's)
  was lower at $-6.8 \pm 1.4$ (assuming \vwind\ = 10~\kms)}.  However,
we caution against over-interpreting this difference, since our values of
\Mdot\ rely on a number of assumptions (see Section \ref{sMdot}), and
there may be systematic biases dependent on the type of the SN since
the mass-loss rates are calculated differently for Type I b/c's than
for Type II's.  Also, as with \tpk\ and \Lpk, the range of values of
\Mdot\ is quite large, with the standard deviations of
$\log_{10}(\Mdot)$ over our sample being 1.1 and 1.4, respectively, so
here also there is considerable overlap between Type I b/c's and Type
II's.

To indicate the uncertainty in estimating the mass loss rate we give
as an example SN~1993J, whose 8.4-GHz lightcurve we showed in
Figure~\ref{fsn93j}. Despite being one of the most intensely studied
SNe with extensive, multi-frequency radio lightcurves as well as X-ray
data, estimating \Mdot\ seems to be far from straightforward, and
various authors have reported a considerable range of
$\log_{10}(\Mdot)$ for its
\newstuff{progenitor\footnote{We note that part of
      the variation in \Mdot\ derives from different assumptions about
      the poorly-known values of the efficiencies $\epsilon_B$ and
      $\epsilon_e$ (Sec.~\ref{sMdot}).  However, some of the values of
      \Mdot, e.g., those of \citet{Weiler+2007}, are derived from
      absorption only, and independent of any assumed values of
      $\epsilon_B$ and $\epsilon_e$, so the variation in derived
      values of \Mdot\ cannot be ascribed entirely to the use of
      different values of $\epsilon_B$ and $\epsilon_e$.}}.  In
$\log_{10}$(\Mdot/\Msolyr), for \vwind = 10 \kms, \citet{Weiler+2007}
reports values between $-6.3$ and $-5.2$, \citet{Bjornsson2015}
reports $-5.0$, \citet{Weiler+2002} report $-4.6$, and
\citet{FranssonB1998} report $-4.3$\@.  We found that the distribution
of $\log_{10}(\Mdot)$ for all Type II SNe excluding Type IIn's had a
mean of $-6.8$ and standard deviation of 1.4\@.  The cited values for
SN~1993J, are higher than the mean of the distribution, although not
outlandish, with, for example, $\sim$4\% of SNe having \Mdot\ higher
than even the highest of the values reported for SN~1993J
($\log_{10}(\Mdot) = -4.3)$\@.  Since SN~1993J was exceptionally radio
luminous, it is not surprising that it has a denser CSM, and thus that
its progenitor had a higher \Mdot\ than the population average.

In the standard self-similar model of an SN, the rise of the radio
lightcurve to \Lpk\ is relatively rapid, regardless of whether FFA or
SSA is the dominant absorption mechanism, and is followed by a slower decay.
For external FFA by a uniform wind medium, with density
$\propto r^{-2}$, the optical depth, $\tau$, is expected to decay with
time as $\tau \propto t^{-\delta}$ with $\delta \simeq 2.5$
\citep[see, e.g.][]{Weiler+2002}.  Inspection of Figure~\ref{falllc}
shows that such steeply rising lightcurves are not common.  We used
$\delta = 1$, which is more representative, for our lightcurve
model. To check the dependence of our results on the assumed
$\delta=1$, we tried a model with $\delta=2.5$ but obtained a much
lower likelihood than with $\delta=1$.  We note that
$\delta = 0.9$ produces a marginally higher likelihood than
$\delta = 1$, but the difference is small, and our results should not
be compromised by our use of $\delta=1$.

If the rising part of the lightcurve (in a self-similar model) were
due to SSA within the emitting region, rather than FFA, then the rising
part would be expected to be a power-law,
$L_\nu \propto t^a$, rather than exponential as seen with FFA\@.
In the case of pure SSA, values of $a \sim 2.1$ are expected
\citep{Chevalier1998}.
This expected value again is much steeper than the majority of the
observed lightcurves in Figure~\ref{falllc}.  We fitted our data with
SSA-like model lightcurves with a power-law, rather than an
exponential rise.  We found that a fit with $a = 2.1$ produces much
lower likelihood than our standard model (exponential rise with
$\delta = 1$).  Smaller values of $a$ produced higher likelihoods,
although over the whole sample, an exponential rise with
$\tau \propto t^{-1}$ produced a somewhat higher likelihood than a
power-law rise for any value of $a$.

\citet{Weiler+2002} also found that absorption by a uniform wind
medium cannot fit the rising portions of many SN radio lightcurves.
They appeal to geometrical effects from a clumpy absorbing medium to
flatten the rise and produce a better fit to the rising part of the
lightcurves. \newstuff{\citet{BjornssonK2017} model the effect of an
  inhomogeneous synchrotron-emitting region in SSA-dominated SNe, and
  find the effect is to flatten the part of the SED below the peak,
  which would also tend to make the rise in the lightcurve less
  steep.

  From our much larger collection of measurements we can conclude that
  the slowly rising lightcurves are a fairly general phenomenon and
  that therefore some form of geometrical effect, such as clumpiness
  in the CSM or inhomogeneity in the synchrotron-emitting region,
  are common.}

\newstuff{
\subsection{Structure of a Comprehensive Radio SN Observing Program}
\label{sprogram}

Our sample is of necessity heterogeneous and incomplete, with only a
fraction of SNe being observed at all in the radio, and even if
observed, often with very sparse sampling in time. While a census of
southern SNe is one of the goals of the Variables and Slow Transients
survey \citep[VAST;][]{Murphy+2013}, to be conducted with the
Australian Square Kilometre Array Pathfinder at 1.4~GHz, as well as of
the ThunderKAT transients programme underway at MeerKAT
\citep{Fender_ThunderKAT2016}, a systematic multi-frequency program of
observing SNe in the radio would be desirable to obtain a clearer
picture of the radio SN luminosity-risetime function.  What would such
a program entail?

We found the mean \lLpk\ for our sample of all Types of core-collapse
SNe was 25.5 (in \ergsHz) with a standard deviation of 1.6  % PPAA
(Table~\ref{tdistribvals}).  
This mean \lLpk\ corresponds to a flux density of 30~$\mu$Jy at 30
Mpc.

To achieve a $5\sigma$ detection of 30~$\mu$Jy requires $\sim$30~min
at 10~GHz or $\sim$3~h at 1.5 GHz with the VLA, $\sim$4~h at 10~GHz or
$\sim$7~h at 2~GHz with ATCA, and $\sim$2~h at 1.3 GHz with MeerKAT\@.
Note that this is only the {\em mean}\/ \lLpk, so at this sensitivity
level $\sim$50\% of SNe would remain undetected.  On the Transient
Name Server\footnote{\url{https://wis-tns.weizmann.ac.il}; we looked
  at the SNe listed between 2020 Jan 1 and June 30, of which 8 had
  $D < 30$~Mpc.}, the rate of classified SNe with $D < 30$~Mpc is
$\sim$16~yr$^{-1}$ over the whole sky.
Observations to fully sample the luminosity distribution of
radio SNe will therefore be challenging with current instrumentation.

Given the wide range of \tpk\ that we have found, with the 1-$\sigma$
range being from 7~d to 1~yr, an observing program with at least 7
logarithmically-spaced observations of each SN starting after about
one week and extending to at least $t=1$~yr would be required to get
reasonably complete sampling and provide accurate constraints on \tpk\
and \Lpk.  Obviously such a program will miss the $\sim$17\% of SNe
with $\tpk < $ 1 week but observations on a shorter timescale would be
hard to schedule.

A systematic program to provide more robust statistics with a more
complete sample than we have been able to do with the existing ad-hoc
sample would therefore be a challenging and long-term project with
current instrumentation, but would certainly be an important project
with the Square Kilometre Array, whose sensitivity will greatly
surpass that of current instruments \citep{Perez-Torres+2015}.
Notwithstanding the difficulty of obtaining a complete sample with
current instrumentation, it is still well worthwhile to observe nearby
or unusual SNe on a case-by-case basis, and we encourage observers to
publish non-detections.}

\section{Conclusion}
\label{sconclude}

We examined a large number of radio flux density measurements for
\NSNe\ SNe at between 5 and 10~GHz.  We parameterize the radio
lightcurves by a simple model consisting of an optically-thick
rise over time \tpk\ from the explosion, to a maximum value of the spectral
luminosity, \Lpk, followed by a power-law decay with
$L_\nu \propto t^{-1.5}$.  We concentrate here only on the part of the
lightcurve near the initial peak, and disregard any late-time rises in
flux density, such as observed in, e.g., SN~1987A.

We find that both \tpk\ and \Lpk\ vary over large ranges.  In the case
of \tpk, some SNe (such as SN~1987A) had \tpk\ of a couple of days or even
less, while others (such as SN~1986J) do not reach the peak until
$\tpk \gtrsim 1000$~d.

The range in \Lpk\ is even larger: 
SN~1987A reached \Lpk\ of only
$\sim \! 10^{23.6}$~\ergsHz, while that of SN~1998bw was
$\sim \! 10^{29}$~\ergsHz.

\begin{trivlist}
  
\item{1.} We find that, over our sample of SNe, lognormal
  distributions of \tpk\ and \Lpk\ provide a reasonable fit to the
  measurements, including the many upper limits.

\item{2}. \newstuff{Many SNe in our sample have low values of \Lpk}.
  At 8.4~GHz, 50\% of all SNe have $\Lpk < 10^{25.5}$~\ergsHz\ or flux
  densities $<30 \; \mu$Jy at $D = 30$~Mpc.

\item{3}. The median value of \Lpk\ is $\sim$30 times lower than that
  obtained if one does not consider the many upper limits in addition
  to the detections.

\item{4}. For Type I b/c SNe at $D < 100$~Mpc, the mean value and
  standard deviation of \newstuff{\tpk\ were $10^{1.1 \pm 0.5}$~d, and
    those of \Lpk\ were $10^{25.4 \pm 1.7}$~\ergsHz}.  Type I b/c SNe
  are characterized by more rapid rises than are Type II's, but they
  reach similar values of \Lpk.

\item{5}. For Type II SNe, \newstuff{at $D < 100$ Mpc and excluding
    Type IIn's, the mean value and standard deviation of \tpk\ were
    $10^{1.6 \pm 1.0}$~d, and those of \Lpk\ were
    $10^{25.3 \pm 1.3}$~\ergsHz.

    \item{6}. Type IIn SNe are characterized by long rise\-times,
    $10^{3.1 \pm 0.7}$~d but high values of \Lpk\ of
    $10^{26.5 \pm 1.1}$~\ergsHz.}

  \newstuff{\item{7}.  Type IIb SNe seem to be characterized by
    considerably higher \Lpk\ than the remainder of the Type II's of
    $10^{26.8 \pm 0.5}$~\ergsHz, and also a narrower range of \Lpk\
    than other Types.  However, our sample contained only 19 Type IIb
    SNe, so this distribution is somewhat uncertain.}

\item{8}. In general, given the wide distributions, the values of
  \tpk\ and \Lpk\ for any particular SN do not reliably indicate
  whether the SN is of Type I b/c or II\@.

\item{9}. The exception to item 8.\ above is that $L_\nu > 10^{28}$ \ergsHz\
  in the first month strongly suggests a Type I b/c SN.

\item{10}. From the distribution of \tpk\ and \Lpk\ values we estimated
  also the distribution of mass-loss rates, $\Mdot$.  We found that
  for Type~I b/c SNe, \newstuff{$\Mdot = 10^{-5.6 \pm 1.1}$ \Msolyr,
    while for Type~II SNe excluding Type IIn's,
    $\Mdot = 10^{-6.8 \pm 1.4}$ \Msolyr}, for assumed \vwind\ of
  1000~\kms\ and 10~\kms, respectively.  We caution, however, that the
  determination of \Mdot\ from \tpk\ and \Lpk\ is very imprecise, and
  possibly subject to biases that could be dependent on the SN Type.

\item{11}.  We find that the rising part of the lightcurves is in most
  cases too shallow to be described either by synchrotron
  self-absorption (SSA) or free-free absorption in a uniform medium.
  This relative flatness suggests that geometrical effects, such as a
  clumpy CSM or non-spherically symmetric structure in the ejecta or
  the CSM, are likely common among SNe.

\end{trivlist}

\section*{Acknowledgements }

This research was supported by both the National Research Foundation
of South Africa and the National Sciences and the Engineering Research
Council of Canada.
In addition to the authors, Mekuanint Hailemariam and Nceba Mhlalo
reduced some archival VLA data, which has contributed to this paper.
The first author thanks Richard Grumitt for helpful discussions, and
we thank the anonymous referee for his or her helpful suggestions.
The National Radio Astronomy Observatory is a facility of the National
Science Foundation operated under cooperative agreement by Associated
Universities, Inc.
The Australia Telescope Compact Array (ATCA) is part of the Australia
Telescope National Facility, which is funded by the Australian
Government for operation as a National Facility managed by CSIRO. We
acknowledge the Gomeroi people as the traditional owners of the ATCA
Observatory site.
(e-)MERLIN is a National Facility operated by the University of
Manchester at Jodrell Bank Observatory on behalf of the Science and
Technology Facilities Council of the UK.
We have made use of NASA's Astrophysics Data System Abstract Service.

\software{AIPS \citep{AIPS},
  CASA \citep{CASA},
  e-Merlin Pipeline \citep{e-Merlin_Pipeline}}

\bibliographystyle{apj}
\bibliography{mybib1,rsn}

\clearpage
\end{document}